\documentclass[twocolumn,eqsecnum,nofootinbib,showpacs,preprintnumbers,widetext]{revtex4}

\usepackage{amsmath}
\usepackage{amssymb}

\usepackage{latexsym}
\usepackage{amsmath,amsfonts}
\usepackage{times}

\newcommand{\rf}[1]{(\ref{#1})}

\def\be{\begin{equation}}
\def\ee{\end{equation}}
\def\beq{\begin{eqnarray}}
\def\eeq{\end{eqnarray}}

\def\parline{\,\partial\kern -0.55em /\,\,}

\def\half{{\frac{1}{2}}}

\def\AA{{\cal A}}

\def\DD{{\cal D}}

\def\LL{{\cal L}}

\def\Ybf{{\bf Y}}

\def\phik{|\phi\rangle}

\def\phikst#1{|\phi_{#1}\rangle}

\def\phicurk{|\phi_{cur}\rangle}
\def\phishk{|\phi_{sh}\rangle}

\def\xik{|\xi\rangle}

\def\xicurk{|\xi_{cur}\rangle}
\def\xishk{|\xi_{sh}\rangle}

\def\smzero{{\scriptscriptstyle (0)}}
\def\smone{{\scriptscriptstyle (1)}}
\def\smtwo{{\scriptscriptstyle (2)}}

\def\smzero{{\scriptscriptstyle (0)}}
\def\smone{{\scriptscriptstyle (1)}}
\def\smtwo{{\scriptscriptstyle (2)}}

\def\mubf{{\boldsymbol{\mu}}}

\def\smponetwo{{\scriptscriptstyle [1,2]}}

\def\Cwt{\widetilde{C}}

\def\ewt{\widetilde{e}}
\def\cwt{\widetilde{c}}

\def\rwt{\widetilde{r}}

\def\alpar{\alpha\partial}
\def\albpar{\bar\alpha\partial}

\def\eb{\bar{e}}
\def\mb{\bar{m}}

\def\msv{m}
\def\sigg{u}

\begin{document}

\preprint{FIAN-TD-2008-16; arXiv: 0805.3472 [hep-th]}

\title{Shadows, currents, and AdS fields}

\author{ R.R. Metsaev}

\email{metsaev@lpi.ru}

\affiliation{ Department of Theoretical Physics, P.N. Lebedev Physical
Institute, Leninsky prospect 53,  Moscow 119991, Russia}

\begin{abstract}
Conformal totally symmetric arbitrary spin currents and shadow fields in flat
space-time of dimension greater than or equal to four are studied. Gauge
invariant formulation for such currents and shadow fields is developed. Gauge
symmetries are realized by involving the Stueckelberg fields. Realization of
global conformal boost symmetries is obtained. Gauge invariant differential
constraints for currents and shadow fields are obtained. AdS/CFT
correspondence for currents and shadow fields and the respective normalizable
and non-normalizable solutions of massless totally symmetric arbitrary spin
AdS fields is studied. The bulk fields are considered in modified de Donder
gauge that leads to decoupled equations of motion. We demonstrate that
leftover on-shell gauge symmetries of bulk fields correspond to gauge
symmetries of boundary currents and shadow fields, while the modified de
Donder gauge conditions for bulk fields correspond to differential
constraints for boundary conformal currents and shadow fields. Breaking
conformal symmetries, we find interrelations between the gauge invariant
formulation of the currents and shadow fields and the gauge invariant
formulation of massive fields.
\end{abstract}


\maketitle

\section{Introduction}

In view of the aesthetic features of conformal field theory  an interest in
this theory was periodically renewed (see \cite{Fradkin:1985am} and
references therein). Conjectured duality \cite{Maldacena:1997re} of large $N$
conformal ${\cal N}=4$ supersymmetric Yang-Mills (SYM) theory and type IIB
superstring theory in $AdS_5 \times S^5$ has triggered intensive and in-depth
study of various aspects of conformal fields. In space-time of dimension
$d\geq 4$, conformal fields studied in this paper can be separated into two
groups: conformal currents and shadow fields. This is to say that field
having Lorentz algebra spin $s$ and conformal dimension $\Delta = s+d-2$,
is referred to as conformal current with canonical dimension%
\footnote{We note that conformal currents with $s=1$, $\Delta= d-1$ and
$s=2$, $\Delta = d$, correspond to conserved vector current and conserved
traceless rank-2 tensor field (energy-momentum tensor) respectively.
Conserved conformal currents can be built from massless scalar, spinor and
spin-1 fields (see e.g. \cite{Konstein:2000bi}). Discussion of higher-spin
conformal conserved charges bilinear in $4d$ massless fields of arbitrary
spins may be found in \cite{Gelfond:2006be}.},
while field having Lorentz algebra spin $s$ and dual conformal dimension
$\Delta = 2-s$ is referred to as shadow field%
\footnote{ It is the shadow fields that are used to discuss conformal
invariant equations of motion and Lagrangian formulations (see e.g.
\cite{Fradkin:1985am,Segal:2002gd,Erdmenger:1997gy,Boulanger:2001he}).
Discussion of equations for mixed-symmetry conformal fields with discrete
$\Delta$ may be found in \cite{Shaynkman:2004vu}.}.
In the framework of $AdS/CFT$ correspondence, the conformal currents and
shadow fields manifest themselves in two related ways at least. First, the
conformal currents appear as boundary values of {\it normalazible} solutions
of equations of motion for bulk fields of IIB supergravity in $AdS_5\times
S^5$ background, while the shadow fields appear as boundary values of {\it
non-normalazible} solutions of equations of motion for bulk fields of IIB
supergravity (see
e.g. \cite{Balasubramanian:1998sn}-\cite{Metsaev:2005ws}%
\footnote{In the earlier literature, discussion of shadow field dualities may
be found in \cite{Petkou:1994ad,Petkou:1996jc}.}).
Second, the conformal currents, which are dual to string theory states, can
be built in terms of fields of SYM theory. In view of these relations to IIB
suprgravity/superstring in $AdS_5\times S^5$ and SYM theory we think that
various alternative formulations of conformal currents and shadow fields will
be useful to understand string/gauge theory dualities better.

The purpose of this paper is to develop gauge invariant formulation for
conformal currents and shadow fields. In this paper, we discuss bosonic
arbitrary spin conformal currents and shadow fields in space-time of
dimension $d \geq 4$. Our approach to the conformal currents and shadow
fields can be summarized as follows.
\\
{\bf i}) Starting with field content of the standard formulation of currents
(and shadow fields), we introduce additional field degrees of freedom
(D.o.F), i.e., we extend space of fields entering the standard conformal
field theory. We note that these additional field D.o.F are similar to the
ones used in gauge invariant formulation of massive fields. Sometimes, such
additional field D.o.F are referred to as Stueckelberg fields.
\\
{\bf ii}) On the extended space of currents (and shadow fields), we introduce
new differential constraints, gauge transformations, and conformal algebra
transformations.
\\
{\bf iii}) The new differential constraints are invariant under the gauge
transformations and the conformal algebra transformations.
\\
{\bf iv})  The gauge symmetries and the new differential constraints make it
possible to match our approach and the standard one, i.e., by appropriate
gauge fixing of the Stueckelberg fields and by solving differential
constraints we obtain standard formulation of conformal currents and shadow
fields.

We apply our approach to the study of $AdS/CFT$ correspondence at the level
massless modes/currents, shadow fields matching. We shall demonstrate that
normalizable modes of massless $AdS$ fields are related to conformal
currents, while non-normalizable modes of massless $AdS$ fields are related
to shadow fields. In the earlier literature, such correspondence was studied
for scalar field in \cite{Balasubramanian:1998sn,Balasubramanian:2000pq} and
for massless arbitrary spin fields taken to be in light-cone gauge in
Ref.\cite{Metsaev:1999ui}. In the latter reference, we have also developed
light-cone formulation of $CFT$. Light-cone formulation of $CFT$ breaks
boundary Lorentz symmetries and therefore is not commonly used. It is
desirable therefore to develop $AdS/CFT$ correspondence for arbitrary spin
fields by maintaining boundary Lorentz symmetries%
\footnote{ One of popular gauges that respects boundary Lorentz symmetries is
the radial gauge. However, in contrast to our approach, the radial gauge does
not allow to treat normalizable and non-normalizable solutions of bulk
equations of motion on an equal footing.}.
This is that we do, among other things, in this paper. Our approach to the
study of $AdS/CFT$ correspondence can be summarized as follows.
\\
{\bf i}) We use modified Lorentz gauge (for spin-1 field) found in
Ref.\cite{Metsaev:1999ui} and modified de Donder gauge (for spin $s\geq 2$
fields) found in Ref.\cite{Metsaev:2008ks}. Remarkable property of these
gauges is that they lead to the simple {\it decoupled} bulk equations of
motion which can be solved in terms of Bessel function
and this simplifies considerably study of $AdS/CFT$ correspondence%
\footnote{ To our knowledge, our modified Lorenz gauge (for spin-1 field) and
modified de Donder gauges (for spin $s\geq 2$ fields) are unique
first-derivative gauges that lead to decoupled bulk equations of motion.
Another gauge that also leads to decoupled bulk equations of motion is the
light-cone gauge (see Ref.\cite{Metsaev:1999ui}). But, light-cone gauge
breaks boundary Lorentz symmetries.}.
\\
{\bf ii}) The number of boundary gauge conformal currents (or shadow fields)
involved in our gauge invariant approach coincides with the number of bulk
massless gauge $AdS$ fields involved in approach of
Ref.\cite{Fronsdal:1978vb}. Note however that, instead of approach in
Ref.\cite{Fronsdal:1978vb}, we use $CFT$ adapted formulation of arbitrary
spin $AdS$ field theory developed in \cite{Metsaev:2008ks}.
\\
{\bf iii}) The number of gauge transformation parameters involved in our
gauge invariant approach to currents (or shadow fields) coincides with the
number of gauge transformation parameters of bulk massless gauge $AdS$ fields
involved in the standard approach of Ref.\cite{Fronsdal:1978vb}.
\\
{\bf iv}) Our modified Lorentz gauge (for spin-1 field) and modified de
Donder gauge (for spin $s\geq 2$ fields) turn out to be related to the new
differential constraints we obtained in the framework of gauge invariant
approach to conformal currents (and shadow fields).
\\
{\bf v}) {\it Leftover on-shell} gauge symmetries of massless bulk $AdS$
fields are related to the gauge symmetries of boundary conformal currents (or
shadow fields).

The rest of the paper  is organized as follows.

In Sec. \ref{sec02},  we summarize the notation used in this paper and
briefly review the standard approach to conformal currents and shadow fields.

In Sections \ref{sec03} and \ref{sec04}, we  start with  the respective
examples of spin-1 conformal current and spin-1 shadow field. We illustrate
our gauge invariant approach to describing conformal current and shadow
field.

Sections \ref{sec05} and \ref{sec06} are devoted to spin-2 conformal current
and spin-2 shadow field respectively. We develop our gauge invariant approach
and demonstrate how our spin-2 current is related with the standard
energy-momentum tensor of $CFT$. We discuss also how our spin-2 shadow field
is related to the one appearing in the standard approach to $CFT$.

In Sections \ref{sec07} and \ref{sec08}, we develop gauge invariant approach
to arbitrary spin-$s$ conformal current and shadow field respectively. Fixing
Stueckelberg gauge symmetries and solving differential constraints for
current and shadow field we prove equivalence of our gauge invariant approach
and the standard approach to $CFT$.

In Sec. \ref{sec09}, we discuss two-point current-shadow field interaction
vertex.

Sec. \ref{sec10} is devoted to the study of $AdS/CFT$ correspondence for
massless low spin, $s=1,2$, bulk $AdS$ fields and boundary low spin, $s=1,2$,
currents and shadow fields.

Sec. \ref{sec11} is devoted to the study of $AdS/CFT$ correspondence for
massless arbitrary spin bulk $AdS$ field and boundary arbitrary spin current
and shadow field.

In Sec. \ref{sec12}, we discuss interrelations between our gauge invariant
approach to currents (and shadow fields) and gauge invariant (Stueckelberg)
approach to massive fields in flat space. In due course we discuss de
Donder-like gauge condition for arbitrary spin-$s$, $s \geq 2$, massive field
in the framework of Stueckelberg approach to massive field. The de
Donder-like gauge we find leads to surprisingly simple gauge-fixed action for
massive arbitrary spin field.

We collect various technical details in four appendices. In Appendix
\ref{appen01}, we discuss restrictions imposed on the two-point
current-shadow field interaction vertex by gauge symmetries and by dilatation
symmetries. In Appendix \ref{appen02}, we discuss restrictions imposed on
conformal boost transformations by gauge symmetries. In Appendix
\ref{appen03}, we review the modified Lorentz condition for spin-1 massless
$AdS$ field and modified de Donder gauge for massless spin-2 $AdS$ field,
while Appendix \ref{appen03new} is devoted to modified de Donder gauge for
fields propagating in conformal space. In Appendix \ref{appen04}, we present
some details of matching of the leftover gauge symmetries of bulk $AdS$
fields and the gauge symmetries of boundary currents and shadow fields.

\section{Preliminaries}
\label{sec02}

\subsection{Notation}

Our conventions are as follows. $x^a$ denotes coordinates in $d$-dimensional
flat space-time, while $\partial_a$ denotes derivatives with respect to
$x^a$, $\partial_a \equiv \partial /
\partial x^a$. Vector indices of the Lorentz algebra $so(d-1,1)$ take
the values $a,b,c,e=0,1,\ldots ,d-1$. We use mostly positive flat metric
tensor $\eta^{ab}$. To simplify our expressions we drop $\eta_{ab}$ in scalar
products, i.e., we use $X^aY^a \equiv \eta_{ab}X^a Y^b$.

We use a set of the creation operators $\alpha^a$, $\alpha^z$, and the
respective set of annihilation operators $\bar{\alpha}^a$, $\bar{\alpha}^z$,
\beq
\label{04092008-03} && [\bar{\alpha}^a,\alpha^b]=\eta^{ab}\,, \qquad
[\bar\alpha^z,\alpha^z]=1\,,
\\[5pt]
\label{04092008-04} && \bar\alpha^a |0\rangle = 0\,,\qquad  \bar\alpha^z
|0\rangle = 0\,,
\\[5pt]
\label{04092008-05} && \alpha^{a\dagger} = \bar\alpha^a\,, \qquad
\alpha^{z\dagger} = \bar\alpha^z \,.
\eeq
These operators will often be referred to as oscillators in what follows%
\footnote{ We use oscillator formulation
\cite{Lopatin:1987hz,Vasiliev:1987tk,Labastida:1987kw} to handle the many
indices appearing for tensor fields. It can also be reformulated as an
algebra acting on the symmetric-spinor bundle on the manifold $M$
\cite{Hallowell:2005np}.}.
The oscillators $\alpha^a$, $\bar\alpha^a$ and $\alpha^z$, $\bar\alpha^z$,
transform in the respective vector and scalar representations of the
$so(d-1,1)$ Lorentz algebra.

Throughout this paper we use operators constructed out of the derivatives and
the oscillators,
\beq
& \Box=\partial^a\partial^a\,,\quad \alpha\partial
=\alpha^a\partial^a\,,\quad \bar\alpha\partial =\bar\alpha^a\partial^a\,, &
\\[3pt]
& \alpha^2 = \alpha^a\alpha^a\,,\qquad \bar\alpha^2 =
\bar\alpha^a\bar\alpha^a\,, &
\\[3pt]
& N_\alpha \equiv \alpha^a \bar\alpha^a  \,,
\qquad
N_z \equiv \alpha^z \bar\alpha^z \,. &
\eeq

\subsection{Global conformal symmetries }

In $d$-dimensional flat space-time, the conformal algebra $so(d,2)$ consists
of translation generators $P^a$, dilatation generator $D$, conformal boost
generators $K^a$, and generators of the $so(d-1,1)$ Lorentz algebra $J^{ab}$.
We assume the following normalization for commutators of the conformal
algebra:
\beq
\label{ppkk}
&& {}[D,P^a]=-P^a\,, \hspace{0.5cm}  {}[P^a,J^{bc}]=\eta^{ab}P^c
-\eta^{ac}P^b, \ \ \ \
\\
&& [D,K^a]=K^a\,, \hspace{0.7cm} [K^a,J^{bc}]=\eta^{ab}K^c - \eta^{ac}K^b, \
\ \ \ \ \
\\[5pt]
\label{pkjj} && \hspace{1.5cm} {}[P^a,K^b]=\eta^{ab}D - J^{ab}\,,
\\
&& \hspace{1.5cm} [J^{ab},J^{ce}]=\eta^{bc}J^{ae}+3\hbox{ terms} \,.
\eeq

Let $\phik$ denotes conformal current (or shadow field) in flat space-time of
dimension $d\geq 4$. Under conformal algebra transformations the $\phik$
transforms as
\be \label{04092008-01} \delta_{\hat{G}} \phik  = \hat{G} \phik \,, \ee
where realization of the conformal algebra generators $\hat{G}$ in
terms of differential operators takes the form
\beq
\label{conalggenlis01} && P^a = \partial^a \,,
\\[3pt]
\label{conalggenlis02} && J^{ab} = x^a\partial^b -  x^b\partial^a + M^{ab}\,,
\\[3pt]
\label{conalggenlis03} && D = x\partial  + \Delta\,,
\\[3pt]
\label{conalggenlis04} && K^a = K_{\Delta,M}^a + R^a\,,
\eeq
and we use the notation
\beq
\label{kdelmdef01} && K_{\Delta,M}^a \equiv -\frac{1}{2}x^2\partial^a + x^a D
+ M^{ab}x^b\,,
\eeq
\be x\partial \equiv x^a \partial^a \,, \qquad x^2 = x^a x^a\,.\ee
In \rf{conalggenlis02}-\rf{conalggenlis04}, $\Delta$ is operator of conformal
dimension, $M^{ab}$ is spin operator of the Lorentz algebra,
\be  [M^{ab},M^{ce}]=\eta^{bc}M^{ae}+3\hbox{ terms} \,. \ee
The spin operator of the Lorentz algebra is well known for arbitrary spin
conformal currents and shadow fields. The spin operator of currents and
shadow fields studied in this paper takes the form
\be \label{mabdef0001} M^{ab} \equiv \alpha^a \bar\alpha^b -
\alpha^b\bar\alpha^a\,.\ee
$R^a$ is operator depending, in general, on derivatives with respect to
space-time coordinates%
\footnote{For conformal currents and shadow fields studied in this paper, the
operator $R^a$ does not depend on derivatives. Dependence on derivatives of
$R^a$ appears e.g., in ordinary-derivative approach to conformal fields
\cite{Metsaev:2007fq}.}
and not depending on space-time coordinates $x^a$,
\be [P^a,R^b]=0\,.\ee
In standard formulation of conformal currents and shadow fields, the operator
$R^a$ is equal to zero, while in gauge invariant approach that we develop in
this paper, the operator $R^a$ is non-trivial. This implies that, in the
framework of gauge invariant approach, complete description of the conformal
currents and shadow fields requires, among other things, finding the operator
$R^a$.

\subsection{ Standard approach to conformal currents and shadow fields}

We begin with brief review of the standard approach to conformal currents and
shadow fields. To keep our presentation as simple as possible we restrict our
attention to the case of arbitrary spin {\it totally symmetric} conformal
currents and shadow fields which have the appropriate canonical conformal
dimensions given below. In this section we recall main facts of conformal
field theory about these currents and shadow fields.

{\bf Conformal current with the canonical conformal dimension}. Consider
totally symmetric rank-$s$ tensor field $T^{a_1\ldots a_s}$ of the Lorentz
algebra $so(d-1,1)$. The field is referred to as spin-$s$ {\it conformal
current with canonical dimension} if $T^{a_1\ldots a_s}$ satisfies the
constraints

\be  \label{16052008-13} T^{aaa_3\ldots a_s}=0\,, \qquad
\partial^a T^{aa_2\ldots a_s}=0
\ee
and has the conformal dimension\!
\footnote{ The fact that expression in r.h.s. of \rf{candim} is the lowest
energy value of totally symmetric spin-$s$ massless fields propagating in
$AdS_{d+1}$ space was demonstrated in Ref.\cite{Metsaev:1994ys}.
Generalization of relation \rf{candim} to mixed-symmetry fields in $AdS$ may
be found in Ref.\cite{Metsaev:1995re}.}
\be \label{candim} \Delta = s + d - 2\,, \ee
which is referred to as {\it the canonical conformal dimension of spin-$s$
conformal current}. Taking into account that the operator $R^a$ of the
conformal current $T^{a_1\ldots a_s}$ is equal to zero, using the well-known
spin operator $M^{ab}$ of the totally symmetric traceless current
$T^{a_1\ldots a_s}$ and $\Delta$ in \rf{candim}, one can make sure that
constraints \rf{16052008-13} are invariant under conformal algebra
transformations \rf{04092008-01}.

{\bf Shadow field with the canonical conformal dimension}. Consider totally
symmetric rank-$s$ tensor field $\Phi^{a_1\ldots a_s}$ of the Lorentz algebra
$so(d-1,1)$. The field $\Phi^{a_1\ldots a_s}$ is referred to as shadow field
if it meets the following requirements:

{\bf i}) The field $\Phi^{a_1\ldots a_s}$ is traceless,
\be \label{16052008-10} \Phi^{aaa_3\ldots a_s}=0 \,. \ee

{\bf ii}) The field $\Phi^{a_1\ldots a_s}$ transforms under the conformal
algebra symmetries so that the following two point current-shadow field
interaction vertex
\be \label{16052008-12} \LL = \frac{1}{s!}\, \Phi^{a_1\ldots a_s}
T^{a_1\ldots a_s} \ee
is invariant (up to total derivative) under conformal algebra
transformations.

We now note that:
\\
{\bf i}) Taking into account conformal dimension of current \rf{candim} and
requiring vertex $\LL$ \rf{16052008-12} to be invariant under the dilatation
transformation we obtain conformal dimension of the spin-$s$ shadow field,
\be\label{candimsh} \Delta = 2 -s\,, \ee
which is referred to as {\it the canonical conformal dimension of spin-$s$
shadow field}. Taking into account that the operator $R^a$ of the conformal
current $T^{a_1\ldots a_s}$ is equal to zero and requiring vertex $\LL$
\rf{16052008-12} to be invariant under the conformal boost transformations we
find that the operator $R^a$ of the shadow field $\Phi^{a_1\ldots a_s}$ is
also equal to zero.
\\
{\bf ii})  Divergence-free constraint \rf{16052008-13} and requirement for
the vertex $\LL$ to be invariant imply that the shadow field is defined by
module of gauge transformation
\be \label{16052008-11} \delta \Phi^{a_1 \ldots a_1} = \Pi^{tr}
\partial^{(a_1} \xi^{a_2\ldots a_s)}\,, \ee
where $\xi^{a_1\ldots a_{s-1}}$ is traceless parameter of gauge
transformation and the projector $\Pi^{tr}$ is inserted to respect traceless
constraint \rf{16052008-10}.

\section{ Gauge invariant formulation of spin-1 conformal current}
\label{sec03}

To discuss gauge invariant formulation of spin-1 conformal current in flat
space of dimension $d\geq 4$ we use one vector field $\phi_{cur}^a$ and one
scalar field $\phi_{cur}$:
\beq \label{14052008-01}
\phi_{cur}^a\,,\qquad \phi_{cur}\,.
\eeq
The fields $\phi_{cur}^a$ and $\phi_{cur}$ transform in the respective vector
and scalar irreps of the Lorentz algebra $so(d-1,1)$. We note that fields
$\phi_{cur}^a$ and $\phi_{cur}$ \rf{14052008-01} have the conformal
dimensions
\be \label{14052008-02} \Delta_{\phi_{cur}^a} = d-1\,,\qquad
\qquad \Delta_{\phi_{cur}} = d-2\,.\ee

We now introduce the following differential constraint:%
\footnote{ Constraint \rf{14052008-10} can simply be obtained by adapting
standard procedure of introducing the Stueckelberg field for massive spin-1
field. Namely, representing standard conserved spin-1 current as $T_{cur}^a =
\phi_{cur}^a + \partial^a \phi_{cur}$ and using conservation law $\partial^a
T_{cur}^a=0$ we obtain \rf{14052008-10}. For spin $s>2$ fields, such
procedure involves complicated higher-derivative expressions and turns out to
be not convenient for developing gauge invariant approach to both massive and
conformal theories.}
\be \label{14052008-10} \partial^a \phi_{cur}^a + \Box \phi_{cur} =  0\,. \ee
It is easy to see that this constraint is invariant under gauge
transformations
\beq
\label{14052008-08} && \delta \phi_{cur}^a = \partial^a \xi_{cur}\,,
\\[3pt]
\label{14052008-09} && \delta \phi_{cur} =  - \xi_{cur} \,,
\eeq
where $\xi_{cur}$ is a gauge transformation parameter.

In order to obtain realization of conformal algebra symmetries we use the
oscillators. We collect fields \rf{14052008-01} into a ket-vector
$|\phi_{cur}\rangle$ defined by
\be |\phi_{cur}\rangle  = ( \phi_{cur}^a \alpha^a + \phi_{cur} \alpha^z )
|0\rangle\,. \ee
Realization of the spin operator $M^{ab}$ on $|\phi_{cur}\rangle$ is given in
\rf{mabdef0001}, while realization of the operator $\Delta$,
\be \Delta = d-1 - N_z\,,\ee
can be read from \rf{14052008-02}. We then find that a realization of the
operator $R^a$ on $|\phi_{cur}\rangle$ takes the form
\be
\label{Racur1} R^a = (2-d) \alpha^a \bar\alpha^z\,.
\ee
Using this, we make sure that constraint \rf{14052008-10} is invariant under
transformations of the conformal algebra \rf{04092008-01}. In terms of the
fields $\phi_{cur}^a$, $\phi_{cur}$, action of operator $R^a$ \rf{Racur1} can
be represented as
\beq
&& R^a\phi_{cur}^b = (2-d)\eta^{ab}\phi_{cur} \,,
\\
&& R^a \phi_{cur} = 0\,.
\eeq

From \rf{14052008-09}, we see that the scalar field $\phi_{cur}$ transforms
as Stueckelberg field, i.e., this field can be gauged away via Stueckelberg
gauge fixing, $\phi_{cur}=0$. If we gauge away the scalar field, then the
remaining vector field $\phi_{cur}^a$ becomes, according to constraint
\rf{14052008-10}, divergence-free. In other words, our constraint
\rf{14052008-10} taken to be in the gauge $\phi_{cur}=0$ leads to the
well-known divergence-free constraint of the standard approach%
\footnote{ As in standard approach to $CFT$, our currents can be considered
either as fundamental field degrees of freedom or as composite operators. At
the group theoretical level, we study in this paper, this distinction is
immaterial.}.

We note that our approach can be related with the standard one without gauge
fixing. Consider vector field
\be \label{19052008-30new}
T_{cur}^a  =  \phi_{cur}^a + \partial^a \phi_{cur} \,. \ee
It is easy to see that:
\\
{\bf i}) $T_{cur}^a$ is invariant under gauge transformations
\rf{14052008-08},\rf{14052008-09}.
\\
{\bf ii}) denoting the left hand side of \rf{14052008-10} by $C_{cur}$ we get
\be \label{19052008-30newnew} \partial^a T_{cur}^a = C_{cur}\,,\ee
i.e., constraint $C_{cur}=0$ \rf{14052008-10} amounts to
\be \partial^a T_{cur}^a = 0\,.\ee
To summarize, in our approach, the gauge invariant vector field $T_{cur}^a$
\rf{19052008-30new} is counterpart of the conserved current in standard
formulation of $CFT$.

\section{ Gauge invariant formulation of spin-1 shadow field}
\label{sec04}

To discuss gauge invariant formulation of spin-1 shadow field in space of
dimension $d\geq 4$ we use one vector field $\phi_{sh}^a$ and one scalar
field $\phi_{sh}$:
\beq \label{14052008-01sh}
\phi_{sh}^a\,,\qquad \phi_{sh}\,.
\eeq
The fields $\phi_{sh}^a$ and $\phi_{sh}$ transform in the respective vector
and scalar representations of the Lorentz algebra $so(d-1,1)$. We note that
these fields have the conformal dimensions
\be \label{14052008-02sh} \Delta_{\phi_{sh}^a} = 1\,,\qquad
\qquad \Delta_{\phi_{sh}} = 2\,.\ee

We now introduce the following differential constraint:
\be \label{15052008-02new} \partial^a \phi_{sh}^a + \phi_{sh} =  0\,. \ee
It is easy to see that this constraint is invariant under gauge
transformations
\beq
\label{14052008-08sh} && \delta \phi_{sh}^a = \partial^a \xi_{sh}\,,
\\[3pt]
\label{14052008-09sh} && \delta \phi_{sh} =  - \Box \xi_{sh} \,,
\eeq
where $\xi_{sh}$ is a gauge transformation parameter.

As before, to obtain realization of conformal algebra symmetries we use the
oscillators and introduce a ket-vector $|\phi_{sh}\rangle$ defined by
\be \label{04092008-02} |\phi_{sh}\rangle  = ( \phi_{sh}^a \alpha^a +
\phi_{sh} \alpha^z ) |0\rangle\,. \ee
Realization of the spin operator $M^{ab}$ on $|\phi_{sh}\rangle$ is given in
\rf{mabdef0001}, while realization of the operator $\Delta$,
\be \Delta = 1 + N_z\,, \ee
can be read from \rf{14052008-02sh}. We then find that a realization of the
operator $R^a$ on $|\phi_{sh}\rangle$ takes the form
\be
\label{Rash1} R^a = (d-2) \alpha^z \bar\alpha^a\,.
\ee
Using this, we check that constraint \rf{15052008-02new} is invariant under
transformations of the conformal algebra \rf{04092008-01}. In terms of the
fields $\phi_{sh}^a$, $\phi_{sh}$, action of operator $R^a$ \rf{Rash1} can be
represented as
\beq
&& R^a\phi_{sh}^b = 0 \,,
\\
&& R^a \phi_{sh} = (d-2)\phi_{sh}^a\,.
\eeq

Gauge transformation of the scalar field $\phi_{sh}$ \rf{14052008-09sh}
involves Dalambertian operator $\Box$, i.e., this transformation is not
realized as the standard Stueckelberg (Goldstone) gauge symmetry. Therefore
the scalar field appearing in the gauge invariant formulation of spin-1
shadow field cannot be referred to as Stueckelberg field. We note that our
field $\phi_{sh}^a$ can be identified with the shadow field $\Phi^a$ of the
standard approach to $CFT$.

As in the case of conformal current, we can introduce gauge invariant field
$T_{sh}^a$,
\be
\label{04092009-11} T_{sh}^a  =  \Box \phi_{sh}^a +  \partial^a \phi_{sh}\,.
\ee
One can check that:
\\
{\bf i}) $T_{sh}^a$ is invariant under gauge transformations
\rf{14052008-08sh},\rf{14052008-09sh}.
\\
{\bf ii}) differential constraint for gauge fields \rf{15052008-02new} leads
to divergence-free constraint for the field $T_{sh}^a$,
\be \label{16052008-01new} \partial^a T_{sh}^a = 0\,.\ee
However, constraint \rf{15052008-02new} is not equivalent to
\rf{16052008-01new}. Namely, if we denote the left hand side of
\rf{15052008-02new} by $C_{sh}$, then we get
\be  \label{16052008-01newnew} \partial^a T_{sh}^a = \Box C_{sh}\,.\ee
We see that constraint $C_{sh}=0$ \rf{15052008-02new} leads to constraint
\rf{16052008-01new}, while constraint \rf{16052008-01new} does not imply the
constraint $C_{sh}=0$, in general.

\section{ Gauge invariant formulation of spin-2 conformal current}
\label{sec05}

To discuss gauge invariant formulation of spin-2 conformal current in flat
space of dimension $d\geq 4$ we use one rank-2 tensor field
$\phi_{cur}^{ab}$, one vector field $\phi_{cur}^a$ and one scalar field
$\phi_{cur}$:
\beq \label{14052008-01gr}
\phi_{cur}^{ab}\,,\qquad \phi_{cur}^a\,,\qquad \phi_{cur}\,.
\eeq
The fields $\phi_{cur}^{ab}$, $\phi_{cur}^a$ and $\phi_{cur}$ transform in
the respective rank-2 tensor, vector and scalar representations of the
Lorentz algebra $so(d-1,1)$. Note that the field $\phi_{cur}^{ab}$ is not
traceless. We note that fields \rf{14052008-01gr} have the conformal
dimensions
\be \label{14052008-02gr} \Delta_{\phi_{cur}^{ab}} = d\,,\qquad
\Delta_{\phi_{cur}^a} = d-1\,,
\qquad \Delta_{\phi_{cur}} = d-2\,.\ee

We now introduce the following differential constraints:
\beq
\label{080423-01} && \partial^b \phi_{cur}^{ab} - \half \partial^a
\phi_{cur}^{bb} + \Box \phi_{cur}^a = 0 \,,
\\[5pt]
\label{080423-02} && \partial^a \phi_{cur}^a + \half \phi_{cur}^{aa} + \sigg
\Box \phi_{cur} = 0 \,,
\eeq
\be \label{080423-02(add)} \sigg \equiv \sqrt{2}
\Bigl(\frac{d-1}{d-2}\Bigr)^{1/2}\,. \ee
One can make sure that these constraints are invariant under gauge
transformations
\beq
\label{14052008-05} && \hspace{-0.5cm} \delta \phi_{cur}^{ab} =\partial^a
\xi_{cur}^b +
\partial^b \xi_{cur}^a + \frac{2}{d-2} \eta^{ab} \Box \xi_{cur}\,,
\\[5pt]
\label{14052008-06} && \hspace{-0.5cm}  \delta \phi_{cur}^a = \partial^a
\xi_{cur} - \xi_{cur}^a \,,
\\[5pt]
\label{14052008-07} && \hspace{-0.5cm}  \delta \phi_{cur} = - \sigg
\xi_{cur}\,,
\eeq
where $\xi_{cur}^a$, $\xi_{cur}$ are gauge transformation parameters.

In order to obtain realization of conformal algebra symmetries in an
easy--to--use form we use oscillators \rf{04092008-03} and collect fields
\rf{14052008-01gr} into a ket-vector $|\phi_{cur}\rangle$ defined by
\be |\phi_{cur}\rangle  = ( \half \phi_{cur}^{ab} \alpha^a \alpha^b +
\phi_{cur}^a \alpha^a \alpha^z + \frac{1}{\sqrt{2}} \phi_{cur} \alpha^z
\alpha^z ) |0\rangle\,. \ee
Realization of the spin operator $M^{ab}$ on $|\phi_{cur}\rangle$ is given in
\rf{mabdef0001}, while realization of the operator $\Delta$,
\be \Delta = d - N_z\,, \ee
can be read from \rf{14052008-02gr}. We then find that a realization of the
operator $R^a$ on $|\phi_{cur}\rangle$ takes the form
\beq \label{20052008-13}
R^a  & = & \bar{r} \Bigl( \Cwt^a +  \frac{2}{d(d -2)}\alpha^2
\bar{C}_\perp^a\Bigr)\,,
\\[5pt]
\label{20052008-13(02)} && \Cwt^a \equiv \alpha^a -
\frac{1}{d-2}\alpha^2\bar\alpha^a
\\[5pt]
\label{20052008-13(03)}&& \bar{C}_\perp^a \equiv \bar\alpha^a - \half
\alpha^a \bar\alpha^2 \,,
\\[5pt]
&& \bar{r} \equiv - \sqrt{(d - N_z)(d - 2 N_z)}\ \bar\alpha^z \,.
\eeq
Using this, we check that constraints \rf{080423-01},\rf{080423-02} are
invariant under transformations of the conformal algebra \rf{04092008-01}.

From \rf{14052008-06},\rf{14052008-07}, we see that the vector and scalar
fields $\phi_{cur}^a$, $\phi_{cur}$ transform as Stueckelberg fields, i.e.,
these fields can be gauged away via Stueckelberg gauge fixing,
$\phi_{cur}^a=0$, $\phi_{cur}=0$. If we gauge away these fields, then the
remaining rank-2 tensor field $\phi_{cur}^{ab}$ becomes, according to
constraints \rf{080423-01},\rf{080423-02}, divergence-free and traceless. In
other words, our constraints taken to be in the gauge $\phi_{cur}^a=0$,
$\phi_{cur}=0$ lead to the well-known divergence-free and tracelessness
constraints of the standard approach.

Our approach can be related with the standard one without gauge fixing, i.e.,
by maintaining gauge symmetries. To this end we construct the following
tensor field:
\beq \label{19052008-30}
T_{cur}^{ab} & = & \phi_{cur}^{ab} + \partial^a \phi_{cur}^b + \partial^b
\phi_{cur}^a
\nonumber \\[5pt]
& + & \frac{2}{\sigg} \partial^a \partial^b \phi_{cur} + \frac{2}{(d-2)\sigg}
\eta^{ab} \Box  \phi_{cur}\,. \ \ \ \ \ \ \ \ \ \
\eeq
One can make sure that:
\\
{\bf i}) $T_{cur}^{ab}$ is invariant under gauge transformations
\rf{14052008-05}-\rf{14052008-07}.
\\
{\bf ii}) Denoting the respective left hand sides of \rf{080423-01} and
\rf{080423-02} by $C_{cur}^a$ and $C_{cur}$ we get
\be \label{07092008-01} \partial^b T_{cur}^{ab}- \half \partial^a
T_{cur}^{bb} = C_{cur}^a\,, \qquad T_{cur}^{aa} = 2 C_{cur}\,, \ee
i.e., the constraints $C_{cur}^a=0$, $C_{cur}=0$ amount to
\be \partial^b T_{cur}^{ab} = 0 \,, \qquad \quad T_{cur}^{aa} =0\,.\ee
In our approach, the gauge invariant tensor field $T_{cur}^{ab}$
\rf{19052008-30} is counterpart of the energy-momentum tensor appearing in
standard formulation of $CFT$.

\section{ Gauge invariant formulation of spin-2 shadow field}
\label{sec06}

To discuss gauge invariant formulation of spin-2 shadow field in flat space
of dimension $d\geq 4$ we use one rank-2 tensor field $\phi_{sh}^{ab}$, one
vector field $\phi_{sh}^a$ and one scalar field $\phi_{sh}$:
\beq \label{14052008-01grsh}
\phi_{sh}^{ab}\,,\qquad\phi_{sh}^a\,,\qquad \phi_{sh}\,.
\eeq
The fields $\phi_{sh}^{ab}$, $\phi_{sh}^a$ and $\phi_{sh}$ transform in the
respective rank-2 tensor, vector and scalar representations of the Lorentz
algebra $so(d-1,1)$. We note that these fields have the conformal dimensions
\be \label{14052008-02grsh} \Delta_{\phi_{sh}^{ab}} = 0\,,\qquad
\Delta_{\phi_{sh}^a} = 1\,,
\qquad \Delta_{\phi_{sh}} = 2\,.\ee

We now introduce the following differential constraints:
\beq
\label{05152008-10new} && \partial^b \phi_{sh}^{ab} - \half \partial^a
\phi_{sh}^{bb} + \phi_{sh}^a = 0 \,,
\\[7pt]
\label{05152008-11new}&& \partial^a \phi_{sh}^a + \half \Box \phi_{sh}^{aa} +
\sigg \phi_{sh} = 0 \,,
\eeq
where $\sigg$ is given in \rf{080423-02(add)}. One can make sure that these
constraints are invariant under gauge transformations
\beq
\label{14052008-11} && \delta \phi_{sh}^{ab} =\partial^a \xi_{sh}^b +
\partial^b \xi_{sh}^a + \frac{2}{d-2} \eta^{ab} \xi_{sh}\,,
\\[7pt]
\label{14052008-12} && \delta \phi_{sh}^a = \partial^a \xi_{sh} - \Box
\xi_{sh}^a \,,
\\[7pt]
\label{14052008-13} && \delta \phi_{sh} = -\sigg \Box \xi_{sh}\,,
\eeq
where $\xi_{sh}^a$, $\xi_{sh}$ are gauge transformation parameters.

In order to obtain realization of conformal algebra symmetries we use the
oscillators and introduce a ket-vector $|\phi_{sh}\rangle$ defined by
\be |\phi_{sh}\rangle  = ( \half \phi_{sh}^{ab} \alpha^a \alpha^b +
\phi_{sh}^a \alpha^a \alpha^z + \frac{1}{\sqrt{2}}\phi_{sh} \alpha^z \alpha^z
) |0\rangle\,. \ee
Realization of the spin operator $M^{ab}$ on $|\phi_{sh}\rangle$ is given in
\rf{mabdef0001}, while realization of the operator $\Delta$,
\be
\Delta  =  N_z\,,
\ee
can be read from \rf{14052008-02grsh}. We then find that a realization of the
operator $R^a$ on $|\phi_{sh}\rangle$ takes the form
\beq
\label{20052008-14} R^a & = & r \Bigl( \bar\alpha^a - \frac{1}{d}\alpha^a
\bar\alpha^2\Bigr)\,,
\\[7pt]
&& r \equiv \alpha^z \sqrt{(d -  N_z)(d - 2 N_z)} \,.
\eeq
Using this, we check that constraints \rf{05152008-10new},\rf{05152008-11new}
are invariant under transformations of the conformal algebra
\rf{04092008-01}.

Gauge transformations of the scalar field $\phi_{sh}$ \rf{14052008-13} and
the vector field $\phi_{sh}^a$ \rf{14052008-12} involve Dalambertian operator
$\Box$. Therefore these transformations are not realized as the standard
Stueckelberg gauge symmetries, i.e., the scalar and vector fields cannot be
referred to as Stueckelberg fields. In contrast with the gauge invariant
approach to spin-2 current, the scalar and the vector fields appearing in the
gauge invariant approach to spin-2 shadow field are not Stueckelberg fields
and they cannot be gauged away via Stueckelberg gauge fixing. All that we can
do is to express these fields in terms of the rank-2 tensor field
$\phi_{sh}^{ab}$ by using constraints
\rf{05152008-10new},\rf{05152008-11new}. On the other hand, from
\rf{14052008-11}, we see that the trace of the rank-2 tensor field
$\phi_{sh}^{ab}$ transforms as Stueckelberg field, i.e., $\phi_{sh}^{aa}$ can
be gauged away via Stueckelberg gauge fixing, $\phi_{sh}^{aa}=0$. Imposing
the gauge $\phi_{sh}^{aa}=0$, we obtain traceless field $\phi_{sh}^{ab}$
which can be identified with the shadow field $\Phi^{ab}$ of the standard
approach to $CFT$.

As in the case of conformal current, we can introduce gauge invariant field
$T_{sh}^{ab}$,
\beq \label{06092008-12}
T_{sh}^{ab} & = & \Box^2 \phi_{sh}^{ab} + \Box(\partial^a \phi_{sh}^b +
\partial^b \phi_{sh}^a)
\nonumber\\[5pt]
&  + & \frac{2}{\sigg} \partial^a \partial^b \phi_{sh} + \frac{2}{(d-2)\sigg}
\eta^{ab} \Box  \phi_{sh}\,. \ \ \ \ \ \ \
\eeq
One can check that:
\\
{\bf i}) $T_{sh}^{ab}$ is invariant under gauge transformations
\rf{14052008-11}-\rf{14052008-13}.
\\
{\bf ii}) differential constraints for gauge fields
\rf{05152008-10new},\rf{05152008-11new} lead to divergence-free and
tracelessness constraints for the field $T_{sh}^{ab}$,
\be \label{16052008-01} \partial^b T_{sh}^{ab} = 0 \,,\qquad \qquad
T_{sh}^{aa} =0 \,.\ee
However, constraints \rf{16052008-01} are not equivalent to
\rf{05152008-10new},\rf{05152008-11new}. Namely, if we denote the respective
left hand sides of \rf{05152008-10new} and \rf{05152008-11new} by $C_{sh}^a$
and $C_{sh}$, then we obtain
\be \label{0409200-06} \partial^b T_{sh}^{ab} -\half \partial^a T_{sh}^{bb} =
\Box^2 C_{sh}^a \,, \qquad T_{sh}^{aa} = 2 \Box C_{sh} \,.\ee
From \rf{0409200-06}, we see that the constraints $C_{sh}^a=0$, $C_{sh}=0$
lead to constraints \rf{16052008-01}, while constraints \rf{16052008-01} do
not imply the constraints $C_{sh}^a=0$, $C_{sh}=0$, in general.

\section{ Gauge invariant formulation of
arbitrary spin conformal current} \label{sec07}

{\bf Field content}. To discuss gauge invariant formulation of arbitrary
spin-$s$ conformal current in flat space of dimension $d\geq 4$ we use the
following fields:
\be \label{phiset01}
\phi_{cur,\, s'}^{a_1\ldots a_{s'}}\,, \hspace{2cm} s'=0,1,\ldots,s;
\ee
where the subscript $s'$ denotes that the field $\phi_{cur,\, s'}^{a_1\ldots
a_{s'}}$ is rank-$s'$ tensor field of the Lorentz algebra $so(d-1,1)$.

We note that:\\
{\bf i}) In \rf{phiset01}, the fields $\phi_{cur,\,0}$ and $\phi_{cur,\,1}^a$
are the respective scalar and vector fields of the Lorentz algebra, while the
fields $\phi_{cur,\, s'}^{a_1\ldots a_{s'}}$, $s'>1$, are rank-$s'$ totally
symmetric tensor fields of the Lorentz algebra $so(d-1,1)$.
\\
{\bf ii}) The tensor fields $\phi_{cur,\,s'}^{a_1\ldots a_{s'}}$ with $s'\geq
4 $ satisfy the double-tracelessness constraint
\be \label{doutracon01} \phi_{cur,\, s'}^{aabba_5\ldots a_{s'}}=0\,, \qquad
s'=4,5,\ldots, s\,. \ee
\\
{\bf iii}) The fields $\phi_{cur, s'}^{a_1\ldots a_{s'}}$ have the following
conformal dimensions:
\be
\label{condimarbspi01} \Delta(\phi_{cur,\, s'}^{a_1\ldots a_{s'}}) = s' + d -
2\,.
\ee

In order to obtain the gauge invariant description in an easy--to--use form
we use the oscillators \rf{04092008-03} and introduce a ket-vector
$|\phi_{cur}\rangle$ defined by
\beq
&& \label{phikdef01}   |\phi_{cur}\rangle \equiv \sum_{s'=0}^s
\alpha_z^{s-s'}|\phi_{cur,\, s'}\rangle \,,
\\[5pt]
&&  \label{phikdef02} |\phi_{cur,\, s'}\rangle \equiv
\frac{\alpha^{a_1} \ldots \alpha^{a_{s'}}}{s'!\sqrt{(s-s')!}}
\, \phi_{cur,\, s'}^{a_1\ldots a_{s'}} |0\rangle\,.
\eeq
From \rf{phikdef01},\rf{phikdef02}, we see that the ket-vector
$|\phi_{cur}\rangle$ is degree-$s$ homogeneous polynomial in the oscillators
$\alpha^a$, $\alpha^z$, while the ket-vector $\phikst{s'}$ is degree-$s'$
homogeneous polynomial in the oscillators $\alpha^a$, i.e., these ket-vectors
satisfy the relations
\beq \label{10092008-05}
&& (N_\alpha + N_z - s)|\phi_{cur}\rangle = 0 \,,
\\[3pt]
&& (N_\alpha -s') |\phi_{cur,\, s'}\rangle = 0\,.
\eeq
In terms of the ket-vector $\phik$, double-tracelessness
constraint \rf{doutracon01} takes the form%
\footnote{ In this paper we adapt the formulation in terms of the double
tracelless gauge fields \cite{Fronsdal:1978vb}. Adaptation of approach in
Ref.\cite{Fronsdal:1978vb} to massive fields may be found in
\cite{Zinoviev:2001dt,Metsaev:2006zy}. Discussion of various formulations in
terms of unconstrained gauge fields may be found in
\cite{Francia:2002aa}-\cite{Buchbinder:2007ak}. For recent review, see
\cite{Fotopoulos:2008ka}. Discussion of other formulations which seem to be
most suitable for the theory of interacting fields may be found e.g. in
\cite{Alkalaev:2003qv,Skvortsov:2008vs}.}
\beq \label{10092008-02}
&& (\bar{\alpha}^2)^2 |\phi_{cur}\rangle  = 0 \,.
\eeq

{\bf Differential constraint}. We find the following differential constraint
for the conformal current:
\beq
\label{17052008-05} && \hspace{-1.7cm} \bar{C}_{cur}|\phi_{cur}\rangle  =  0
\,,
\\[5pt]
\label{17052008-07} &&  \hspace{-0.7cm}  \bar{C}_{cur}  =  \bar{C}_\perp   +
c_1 \bar\alpha^2 + c_2 \Box \Pi^\smponetwo\,,
\\[5pt]
\label{17052008-06} &&  \hspace{-0.7cm}  \bar{C}_\perp = \albpar - \half
\alpar \bar\alpha^2\,,
\\[5pt]
\label{17052008-08} &&  \hspace{-0.7cm}   \Pi^\smponetwo = 1 -\alpha^2
\frac{1}{2(2N_\alpha +d)}\bar\alpha^2\,,
\\[5pt]
\label{17052008-09} &&  \hspace{-0.7cm}  c_1 = \half \alpha^z \ewt_1\,,
\qquad
c_2 =  \ewt_1 \bar\alpha^z\,,
\\[5pt]
\label{masewtdef01} &&  \hspace{-0.7cm}  \ewt_1 =
\Bigl(\frac{2s+d-4-N_z}{2s+d-4-2N_z}\Bigr)^{1/2}\,.
\eeq
One can make sure that constraint \rf{17052008-05} is invariant under gauge
transformation and conformal algebra transformations which we discuss below.
Details of the derivation of constraint \rf{17052008-05} may be found in the
Appendix \ref{appen01}.

{\bf Gauge symmetries}. We now discuss gauge symmetries of the conformal
current. To this end we introduce the following gauge transformation
parameters:
\beq \label{epsilonset01}
&& \xi_{cur,\, s'}^{a_1\ldots a_{s'}}\,,\qquad\qquad s'=0,1,\ldots, s-1\,. \
\ \ \ \
\eeq
We note that\\
{\bf i}) In \rf{epsilonset01}, the gauge transformation parameters
$\xi_{cur,\,0}$ and $\xi_{cur,\,1}^a$ are the respective scalar and vector
fields of the Lorentz algebra, while the gauge transformation parameters
$\xi_{cur,\,s'}^{a_1\ldots a_{s'}}$, $s'>1$, are rank-$s'$ totally symmetric
tensor fields of the Lorentz algebra $so(d-1,1)$.
\\
{\bf ii}) The gauge transformation parameters $\xi_{cur,\,s'}^{a_1\ldots
a_{s'}}$ with $s'\geq 2 $ satisfy the tracelessness constraint
\be \label{epsdoutracon01} \xi_{cur,\, s'}^{aaa_3\ldots a_{s'}}=0\,, \qquad
s'= 2,3,\ldots, s-1\,. \ee
{\bf iii}) The gauge transformation parameters $\xi_{cur\, s'}^{a_1\ldots
a_{s'}}$ have the conformal dimensions
\be \label{epscondimarbspi01} \Delta(\xi_{cur,\,s'}^{a_1\ldots a_{s'}}) = s'
+ d-3\,.\ee

Now, as usually, we collect the gauge transformation parameters in ket-vector
$|\xi_{cur}\rangle$ defined by
\beq
&&  |\xi_{cur}\rangle \equiv \sum_{s'=0}^{s-1} \alpha_z^{s-1-s'}|\xi_{cur,\,
s'}\rangle \,,
\\[5pt]
&& |\xi_{cur,\,s'}\rangle \equiv
\frac{\alpha^{a_1} \ldots \alpha^{a_{s'}}}{s'!\sqrt{(s -1 - s')!}}
\, \xi_{cur,\, s'}^{a_1\ldots a_{s'}} |0\rangle. \
\eeq
The ket-vectors $|\xi_{cur}\rangle$, $|\xi_{cur,\,s'}\rangle$ satisfy the
algebraic constraints
\beq
&& (N_\alpha + N_z - s +1 ) |\xi_{cur}\rangle = 0 \,,
\\[7pt]
&& (N_\alpha -s')|\xi_{cur,\,s'}\rangle  = 0\,,
\eeq
which tell us that $|\xi_{cur}\rangle$ is a degree-$(s-1)$ homogeneous
polynomial in the oscillators $\alpha^a$, $\alpha^z$, while the ket-vector
$|\xi_{cur,\,s'}\rangle$ is degree-$s'$ homogeneous polynomial in the
oscillators $\alpha^a$.

In terms of the ket-vector $|\xi_{cur}\rangle$, tracelessness constraint
\rf{epsdoutracon01} takes the form
\be \label{08092008-08} \bar\alpha^2 |\xi_{cur}\rangle = 0 \,.\ee

Gauge transformation can entirely be written in terms of $|\phi_{cur}\rangle$
and $|\xi_{cur}\rangle$. This is to say that gauge transformation takes the
form
\beq \label{gautraarbspi01}
\delta |\phi_{cur}\rangle & = &  ( \alpar + b_1 + b_2 \alpha^2\Box )
|\xi_{cur} \rangle\,,
\\[10pt]
&& b_1 = - \alpha^z \ewt_1 \,,
\\[3pt]
&& b_2 =  \frac{1}{2s + d- 6 -2N_z} \ewt_1 \bar\alpha^z\,,
\eeq
where $\ewt_1$ is given in \rf{masewtdef01}. We note that constraint
\rf{17052008-05} is invariant under gauge transformation \rf{gautraarbspi01}.
Details of the derivation of gauge transformation \rf{gautraarbspi01} may be
found in the Appendix \ref{appen01}.

{\bf Realization of conformal algebra symmetries}. To complete the gauge
invariant formulation of the spin-$s$ conformal current we provide
realization of the conformal algebra symmetries on space of the ket-vector
$|\phi_{cur}\rangle$. All that is required is to fix the operators $M^{ab}$,
$\Delta$, and $R^a$ and insert then these operators into
\rf{conalggenlis01}-\rf{conalggenlis04}. Realization of the spin operator
$M^{ab}$ on ket-vector $\phicurk$ \rf{phikdef01} is given in \rf{mabdef0001},
while realization of the operator $\Delta$,
\beq \label{08092008-03}
&& \Delta  =  s+ d-2 - N_z\,,
\eeq
can be read from \rf{condimarbspi01}. In the gauge invariant formulation,
finding the operator $R^a$ provides the real difficulty. Representation of
the operator $R^a$ we find is given by
\beq \label{20052008-15}
&& \hspace{-0.3cm} R^a = \bar{r} \Bigl( \Cwt^a + \alpha^2
\frac{2}{(2N_\alpha+d -2)(2N_\alpha+d)} \bar{C}_\perp^a\Bigr)\,, \ \ \
\\[10pt]
\label{20052008-15(01)} && \hspace{-0.3cm} \Cwt^a \equiv \alpha^a - \alpha^2
\frac{1}{2N_\alpha + d-2}\bar\alpha^a\,,
\\[7pt]
\label{20052008-15(02)} && \hspace{-0.3cm} \bar{C}_\perp^a \equiv
\bar\alpha^a - \frac{1}{2} \alpha^a \bar\alpha^2 \,,
\\[7pt]
&& \hspace{-0.3cm} \bar{r} \equiv -
\Bigl((2s+d-4-N_z)(2s+d-4-2N_z)\Bigr)^{1/2} \bar\alpha^z\,.
\nonumber\\
&&
\eeq Details of the derivation of operator $R^a$ \rf{20052008-15} may be
found in the Appendix \ref{appen02}.

{\bf Equivalence of the gauge invariant and standard approaches}. We begin
with comment on the structure of gauge transformation \rf{gautraarbspi01}.
Making use of simplified notation for conformal currents, gauge
transformation parameters, derivatives, and flat metric tensor
\be
\phi_{cur,\, s'} \sim \phi_{cur,\, s'}^{a_1\ldots a_{s'}},\ \ \ \xi_{cur, s'}
\sim \xi_{cur,\,s'}^{a_1\ldots a_{s'}},\ \ \
\partial \sim \partial^a, \ \ \ \eta \sim \eta^{ab},\ee
gauge transformation \rf{gautraarbspi01} can schematically be represented as
\beq \label{gautraspi2def02} && \delta \phi_{cur,\,s'} \sim
\partial \xi_{cur,\, s'-1} + \xi_{cur,\,s'}  +
\eta \Box \xi_{cur,\, s'-2}  \,, \ \ \ \
\nonumber\\[3pt]
&& \hspace{3cm} s' = 2,3,\ldots,s\,,
\\[3pt]
\label{gautraspi2def03} && \delta \phi_{cur,\, 1} \sim \partial \xi_{cur,\,0}
+ \xi_{cur,\, 1} \,,
\\[3pt]
\label{gautraspi2def04} && \delta \phi_{cur,\, 0} \sim \xi_{cur,\, 0} \,,
\eeq
where we assume $\xi_{cur,\,s}\equiv 0$. From
\rf{gautraspi2def02}-\rf{gautraspi2def04}, we see that all gauge
transformations are realized as Stueckelberg (Goldstone) gauge
transformations. We now find currents that are realized as Stueckelberg
fields. To this end we note that the currents $\phi_{cur,\,s'}$ with $s'\geq
2 $ can be decomposed into traceless tensor fields as
\beq
&& \phi_{cur,\, s'} = \phi_{cur,\, s'}^{\rm T} \oplus \phi_{cur,\, s'-2}^{\rm
TT}\,,
\nonumber\\[3pt]
&& \hspace{2cm} s'= 2,3,\ldots,s\,,  \eeq
where $\phi_{cur,\,s'}^{\rm T}$ and $\phi_{cur,\,s'-2}^{\rm TT}$ stand for
the respective rank-$s'$ and rank-$(s'-2)$ traceless tensors of the Lorentz
algebra $so(d-1,1)$. From \rf{gautraspi2def02}-\rf{gautraspi2def04}, we see
that we can impose the gauge conditions
\beq
&& \label{18052008-01} \phi_{cur,\, 0}=0\,,\qquad \phi_{cur,\, 1}=0\,,\qquad
\phi_{cur,\, s'}^{\rm T} = 0\,,
\nonumber\\[3pt]
&& \hspace{3cm} s'= 2,3,\ldots,s-1\,. \eeq
Currents given in \rf{18052008-01} are Stueckelberg fields in our approach.

We now discuss restrictions imposed by differential constraint
\rf{17052008-05}. To this end we note that our gauge conditions
\rf{18052008-01} can be written in terms of the ket-vectors
$|\phi_{cur,\,s'}\rangle$ as%
\footnote{In terms of the ket-vector $|\phi_{cur}\rangle$, gauge conditions
\rf{18052008-01} can simply be represented as
$\bar\alpha^z\Pi^{\smponetwo}|\phi_{cur}\rangle=0$.}
\be \label{17052008-02} \Pi^\smponetwo |\phi_{cur,\,s'}\rangle = 0\,, \qquad
s'= 0,1,\ldots ,s-1\,,\ee
which, in turn, can be represented as
\be \label{17052008-03} |\phi_{cur,\,s'}\rangle = \alpha^2
\frac{1}{2(2N_\alpha +d)}\bar\alpha^2 |\phi_{cur,\,s'}\rangle,\ \ \
s'=0,1,\ldots,s-1.\ee
Making use of gauge conditions \rf{17052008-02} in differential constraint
\rf{17052008-05} leads to
\beq
&&\hspace{-0.5cm} \label{17052008-04} (\albpar - \half \alpar
\bar\alpha^2)|\phi_{cur,\,s'}\rangle\! + \! \half \ewt_{1,s-s'-1}
\bar\alpha^2 |\phi_{cur,\,s'+1}\rangle\! =\! 0,
\nonumber\\[3pt]
&& \hspace{3cm} s'= 0,1,\ldots ,s\,,
\eeq
where $\ewt_{1,n} \equiv \ewt_1|_{N_z=n}$. Using \rf{17052008-03},
Eqs.\rf{17052008-04} can be represented as
\beq \label{17052008-11}
&& \frac{2N_\alpha + d-4}{2N_\alpha + d-2} \Bigl( \alpar  - \alpha^2
\frac{1}{2N_\alpha + d}\albpar\Bigr)\bar\alpha^2|\phi_{cur,\,s'}\rangle
\nonumber\\[3pt]
&& + \half \ewt_{1,s-s'-1} \bar\alpha^2 |\phi_{cur,\,s'+1}\rangle=0\,,
\eeq
when $s'= 0,1,2,\ldots ,s-1$, while for $s'=s$, \rf{17052008-04} amounts to
\beq \label{17052008-11new}
(\albpar - \half \alpar \bar\alpha^2)|\phi_{cur,\,s}\rangle=0 \,. \eeq
Taking into account \rf{17052008-11} and gauge conditions
$|\phi_{cur,\,0}\rangle = 0$, $|\phi_{cur,\,1}\rangle=0$ we obtain
\be  \label{17052008-12} \bar\alpha^2|\phi_{cur,\,s'}\rangle=0\,,\qquad
s'=0,1,\ldots, s\,.\ee
Relations \rf{17052008-03} and \rf{17052008-12} imply
\be  \label{17052008-13} |\phi_{cur,\,s'}\rangle=0\,,\qquad s'=0,1,\ldots,
s-1\,.\ee
Thus, we are left with the one spin-$s$ traceless current
$|\phi_{cur,\,s}\rangle$ which turns out to be divergence-free because of
\rf{17052008-11new},
\be \albpar|\phi_{cur,\,s}\rangle = 0\,.\ee
This implies that our gauge invariant approach is equivalent to the standard
one.

\section{ Gauge invariant formulation of
arbitrary spin shadow field } \label{sec08}

{\bf Field content}. To discuss gauge invariant formulation of arbitrary
spin-$s$ shadow field in flat space of dimension $d\geq 4$ we use the
following fields:
\be \label{phiset01sh}
\phi_{sh,\, s'}^{a_1\ldots a_{s'}}\,, \hspace{2cm} s'=0,1,\ldots, s;
\ee
where the subscript $s'$ denotes that the field $\phi_{sh,\,s'}^{a_1\ldots
a_{s'}}$ is rank-$s'$ tensor field of the Lorentz algebra $so(d-1,1)$.

We note that:\\
{\bf i}) In \rf{phiset01sh}, the fields $\phi_{sh,\,0}$ and $\phi_{sh,\,1}^a$
are the respective scalar and vector fields of the Lorentz algebra, while the
fields $\phi_{sh,\, s'}^{a_1\ldots a_{s'}}$, $s'>1$, are rank-$s'$ totally
symmetric tensor fields of the Lorentz algebra $so(d-1,1)$.
\\
{\bf ii}) The tensor fields $\phi_{sh,\,s'}^{a_1\ldots a_{s'}}$ with $s'\geq
4 $ satisfy the double-tracelessness constraint
\be \label{doutracon01sh} \phi_{sh,\, s'}^{aabba_5\ldots a_{s'}}=0\,, \qquad
s' = 4,5,\ldots ,s \,. \ee
\\
{\bf iii}) The fields $\phi_{sh, s'}^{a_1\ldots a_{s'}}$ have the following
conformal dimensions:
\be
\label{condimarbspi01sh} \Delta(\phi_{sh,\, s'}^{a_1\ldots a_{s'}}) = 2-s'\,.
\ee

In order to obtain the gauge invariant description in an easy--to--use form
we use the oscillators and introduce a ket-vector $|\phi_{sh}\rangle$ defined
by
\beq
&&  \label{phikdef01sh} |\phi_{sh}\rangle \equiv \sum_{s'=0}^s
\alpha_z^{s-s'}|\phi_{sh,\, s'}\rangle \,,
\\[5pt]
&& \label{phikdef02sh} |\phi_{sh,\, s'}\rangle \equiv
\frac{\alpha^{a_1} \ldots \alpha^{a_{s'}}}{s'!\sqrt{(s-s')!}}
\, \phi_{sh,\, s'}^{a_1\ldots a_{s'}} |0\rangle\,.
\eeq
From \rf{phikdef01sh},\rf{phikdef02sh}, we see that the ket-vectors
$|\phi_{sh}\rangle$, $|\phi_{sh,\,s'}\rangle$ satisfy the algebraic
constraints
\beq \label{09092008-04}
&& (N_\alpha + N_z - s)|\phi_{sh}\rangle = 0 \,,
\\[7pt]
&& (N_\alpha -s')|\phi_{sh,\, s'}\rangle = 0\,.
\eeq
These constraints tell us that $|\phi_{sh}\rangle$ is a degree-$s$
homogeneous polynomial in the oscillators $\alpha^a$, $\alpha^z$, while the
ket-vector $\phikst{sh,\,s'}$ is degree-$s'$ homogeneous polynomial in the
oscillators $\alpha^a$. In terms of the ket-vector $|\phi_{sh}\rangle$,
double-tracelessness constraint \rf{doutracon01sh} takes the form
\beq \label{09092008-11}
&& (\bar{\alpha}^2)^2 |\phi_{sh}\rangle  = 0 \,.
\eeq

{\bf Differential constraint}. We find the following differential constraint
for the shadow field:
\beq
\label{17052008-05sh} && \hspace{-1cm} \bar{C}_{sh}|\phi_{sh}\rangle  =  0
\,,
\\[7pt]
\label{17052008-07sh} && \bar{C}_{sh}  =  \bar{C}_\perp   +  c_1 \bar\alpha^2
\Box + c_2\Pi^\smponetwo\,,
\\[7pt]
\label{17052008-06sh} && \bar{C}_\perp = \albpar - \half \alpar
\bar\alpha^2\,,
\\[7pt]
\label{17052008-08sh} &&  \Pi^\smponetwo = 1 -\alpha^2 \frac{1}{2(2N_\alpha
+d)}\bar\alpha^2\,,
\\[7pt]
\label{17052008-09sh} && c_1 = \half \alpha^z \ewt_1\,,
\qquad
c_2 =  \ewt_1 \bar\alpha^z\,,
\\[7pt]
\label{masewtdef01sh} && \ewt_1 =
\Bigl(\frac{2s+d-4-N_z}{2s+d-4-2N_z}\Bigr)^{1/2}\,.
\eeq
One can make sure that constraint \rf{17052008-05sh} is invariant under gauge
transformation and conformal algebra transformations which we discuss below.
Details of the derivation of constraint \rf{17052008-05sh} may be found in
the Appendix \ref{appen01}.

{\bf Gauge symmetries of shadow field}. We now discuss gauge symmetries of
the shadow field. To this end we introduce the following gauge transformation
parameters:
\beq \label{epsilonset01sh}
&& \xi_{sh,\, s'}^{a_1\ldots a_{s'}}\,,\qquad\qquad s'=0,1,\ldots, s-1\,.
\eeq
We note that\\
{\bf i}) In \rf{epsilonset01sh}, the gauge transformation parameters
$\xi_{sh,\,0}$ and $\xi_{sh,\,1}^a$ are the respective scalar and vector
fields of the Lorentz algebra, while the gauge transformation parameters
$\xi_{sh,\,s'}^{a_1\ldots a_{s'}}$, $s'>1$, are rank-$s'$ totally symmetric
tensor fields of the Lorentz algebra $so(d-1,1)$.
\\
{\bf ii}) The gauge transformation parameters $\xi_{sh,\,s'}^{a_1\ldots
a_{s'}}$ with $s'\geq 2 $ satisfy the tracelessness constraint
\be \label{epsdoutracon01sh} \xi_{sh,\, s'}^{aaa_3\ldots a_{s'}}=0\,, \qquad
s' =2,3,\ldots, s-1\,. \ee
{\bf iii}) The gauge transformation parameters $\xi_{sh,\, s'}^{a_1\ldots
a_{s'}}$ have the conformal dimensions
\be \label{epscondimarbspi01sh} \Delta(\xi_{sh,\,s'}^{a_1\ldots a_{s'}}) =
1-s'\,.\ee

Now, as usually, we collect gauge transformation parameters in ket-vector
$|\xi_{sh}\rangle$ defined by
\beq
&& |\xi_{sh}\rangle \equiv \sum_{s'=0}^{s-1} \alpha_z^{s-1-s'}|\xi_{sh,\,
s'}\rangle \,,
\\[5pt]
&&  |\xi_{sh,\,s'}\rangle \equiv
\frac{\alpha^{a_1} \ldots \alpha^{a_{s'}}}{s'!\sqrt{(s -1 - s')!}}
\, \xi_{sh,\, s'}^{a_1\ldots a_{s'}} |0\rangle\,.  \eeq
The ket-vectors $|\xi_{sh}\rangle$, $|\xi_{sh,\,s'}\rangle$ satisfy the
algebraic constraints
\beq
&& (N_\alpha + N_z - s +1 ) |\xi_{sh}\rangle = 0 \,,
\\[7pt]
&&( N_\alpha -s') |\xi_{sh,\,s'}\rangle  = 0  \,,
\eeq
which tell us that $|\xi_{sh}\rangle$ is a degree-$(s-1)$ homogeneous
polynomial in the oscillators $\alpha^a$, $\alpha^z$, while the ket-vector
$|\xi_{sh,\,s'}\rangle$ is degree-$s'$ homogeneous polynomial in the
oscillators $\alpha^a$. In terms of the ket-vector $|\xi_{sh}\rangle$,
tracelessness constraint \rf{epsdoutracon01sh} takes the form
\be \label{08092008-09} \bar\alpha^2 |\xi_{sh}\rangle = 0 \,.\ee

Gauge transformation can entirely be written in terms of $|\phi_{sh}\rangle$
and $|\xi_{sh}\rangle$. This is to say that gauge transformation takes the
form
\beq \label{gautraarbspi01sh}
\delta |\phi_{sh}\rangle & = &  ( \alpar + b_1 \Box  + b_2 \alpha^2 )
|\xi_{sh}\rangle \,,
\\[15pt]
\label{13092008-06} && b_1 = - \alpha^z \ewt_1 \,,
\\[5pt]
\label{13092008-07} && b_2 =  \frac{1}{2s + d- 6 -2N_z} \ewt_1
\bar\alpha^z\,,
\eeq
where $\ewt_1$ is given in \rf{masewtdef01sh}. We note that constraint
\rf{17052008-05sh} is invariant under gauge transformation
\rf{gautraarbspi01sh}. Details of the derivation of gauge transformation
\rf{gautraarbspi01sh} may be found in the Appendix \ref{appen01}.

{\bf Realization of conformal algebra symmetries}. To complete gauge
invariant formulation of spin-$s$ shadow field we should provide realization
of the conformal algebra symmetries on the space of the ket-vector
$|\phi_{sh}\rangle$, i.e. we should find operators $M^{ab}$, $\Delta$ and
$R^a$ to insert them into \rf{conalggenlis01}-\rf{conalggenlis04}.
Realization of the spin operator $M^{ab}$ on ket-vector $\phishk$
\rf{phikdef01sh} is given in \rf{mabdef0001}, while realization of the
operator $\Delta$,
\beq \label{08092008-04}
&& \Delta  =  2- s + N_z\,,
\eeq
can be read from \rf{condimarbspi01sh}. Representation of the operator $R^a$
we find is given by
\beq \label{20052008-16}
&& \hspace{-1.2cm} R^a  =  r \Bigl( \bar\alpha^a - \alpha^a
\frac{1}{2N_\alpha+d } \bar\alpha^2\Bigr)\,,
\\[7pt]
&& \hspace{-1.2cm} r \equiv \alpha^z
\Bigl((2s+d-4-N_z)(2s+d-4-2N_z)\Bigr)^{1/2}.
\eeq
Details of the derivation of operator $R^a$ \rf{20052008-16} may be found in
the Appendix \ref{appen02}.

{\bf Equivalence of the gauge invariant and standard approaches}. We begin
with comment on the structure of gauge transformation \rf{gautraarbspi01sh}
and identification of Stueckelberg shadow fields in the gauge invariant
approach. Making use of simplified notation for shadow fields, gauge
transformation parameters, derivatives, and flat metric tensor
\be
\phi_{sh,\, s'} \sim \phi_{sh,\, s'}^{a_1\ldots a_{s'}},\quad \xi_{sh,\, s'}
\sim \xi_{sh,\,s'}^{a_1\ldots a_{s'}},\ \ \
\partial \sim \partial^a,\ \ \ \eta \sim \eta^{ab},\ee
gauge transformation \rf{gautraarbspi01sh} can schematically be represented
as
\beq \label{gautraspi2def02sh} && \delta \phi_{sh,\,s'} \sim
\partial \xi_{sh,\, s'-1} + \Box \xi_{sh,\,s'}  +
\eta \xi_{sh,\, s'-2}  \,,
\nonumber\\[3pt]
&& \hspace{3cm} s' = 2,3,\ldots, s\,,
\\[3pt]
\label{gautraspi2def03sh} && \delta \phi_{sh,\, 1} \sim \partial \xi_{sh,\,0}
+ \Box \xi_{sh,\, 1} \,,
\\[3pt]
\label{gautraspi2def04sh} && \delta \phi_{sh,\, 0} \sim \Box \xi_{sh,\, 0}
\,,
\eeq
where we assume $\xi_{sh,\,s}\equiv 0$. We now find shadow fields that are
realized as Stueckelberg fields. To this end we note that the fields
$\phi_{sh,\,s'}$ with $s' \geq 2$ can be decomposed into traceless tensor
fields as
\be \phi_{sh,\, s'} = \phi_{sh,\, s'}^{\rm T} \oplus \phi_{sh,\, s'-2}^{\rm
TT}\,, \qquad s'= 2,3,\ldots, s\,,  \ee
where $\phi_{sh,\, s'}^{\rm T}$ and $\phi_{sh,\, s'-2}^{\rm TT}$ stand for
the respective rank-$s'$ and rank-$(s'-2)$ traceless tensors of the Lorentz
algebra $so(d-1,1)$. From \rf{gautraspi2def02sh}-\rf{gautraspi2def04sh}, we
see that, in contrast to conformal currents, the following shadow fields
\be \label{18052008-01sh} \phi_{sh,\, 0}\,,\qquad \phi_{sh,\, 1}\,,\qquad
\phi_{sh,\, s'}^{\rm T}\,, \qquad s'= 2,3,\ldots, s\,, \ee
{\it cannot be gauged away} via Stueckelberg gauge fixing. From
\rf{gautraspi2def02sh}, it is easy to see that by using gauge symmetries
related to the gauge transformation parameters
\be \xi_{sh,\,s'} \,,\qquad s'=0,1,\ldots, s-2\,,\ee
we can impose the following gauge conditions
\be \label{17052008-01shnew} \phi_{sh,\, s'-2}^{\rm TT} = 0 \,, \qquad s' =
2,3,\ldots, s\,, \ee
and we note that fields in \rf{17052008-01shnew} are the Stueckelberg fields
in the framework of gauge invariant approach.

We now discuss restrictions imposed by differential constraint
\rf{17052008-05sh}. To this end we note that our gauge conditions
\rf{17052008-01shnew} can be written in terms of the ket-vectors
$|\phi_{sh,\,s'}\rangle$ as
\be \label{17052008-02sh} \bar\alpha^2|\phi_{sh,\,s'}\rangle = 0\,, \qquad s'
= 2,3,\ldots, s\,.\ee
Making use of gauge conditions \rf{17052008-02sh} in \rf{17052008-05sh} leads
to

\beq
&& \label{17052008-04sh} \albpar|\phi_{sh,\,s'}\rangle +
(s-s'+1)\ewt_{1,s-s'} |\phi_{sh,\,s'-1}\rangle = 0 \,,
\nonumber\\[3pt]
&& \hspace{3cm} s' = 0,1,\ldots, s\,.
\eeq
Relations \rf{17052008-04sh} imply that the fields $|\phi_{sh,\,s'}\rangle$,
$s'=0,1,\ldots,s-1$, can be expressed in terms of the one field
$|\phi_{sh,\,s}\rangle$ subject to the tracelessness constraint (see
\rf{17052008-02sh} when $s'=s$). Thus, we are left with the one spin-$s$
traceless shadow field $|\phi_{sh,\,s}\rangle$ and one surviving gauge
symmetry generated by the gauge transformation parameter $\xi_{sh,\,s-1}$.
This implies that our gauge invariant approach is equivalent to the standard
one.

\section{ Two point current-shadow field interaction vertex}
\label{sec09}

We now discuss two-point current-shadow field interaction vertex. In the
gauge invariant approach, interaction vertex is determined by requiring the
vertex to be invariant under both gauge transformations of currents and
shadow fields. Also, the interaction vertex should be invariant under
conformal algebra transformations.

{\bf Spin-1}. We begin with spin-1 fields. Let us consider the following
vertex:
\be  \label{20052008-12}  \LL = \phi_{cur}^a \phi_{sh}^a +
\phi_{cur}\phi_{sh}\,. \ee
Under gauge transformations of the current \rf{14052008-08},\rf{14052008-09},
the variation of vertex \rf{20052008-12} takes the form (up to total
derivative)
\be  \delta_{ \xi_{cur}} \LL = - \xi_{cur} (\partial^a \phi_{sh}^a +
\phi_{sh})\,. \ee
From this expression, we see that the vertex $\LL$ is invariant under gauge
transformations of the current provided the shadow field satisfies
differential constraint \rf{15052008-02new}. We then find that under gauge
transformations of the shadow field \rf{14052008-08sh},\rf{14052008-09sh} the
variation of the vertex $\LL$ takes the form (up to total derivative)
\be \delta_{ \xi_{sh}} \LL =  -\xi_{sh} (\partial^a \phi_{cur}^a + \Box
\phi_{cur})\,, \ee
i.e., the vertex $\LL$ is invariant under gauge transformations of the shadow
field provided the current satisfies differential constraint
\rf{14052008-10}.

Making use of the representation for generators of the conformal algebra
obtained in the Sections \ref{sec03},\ref{sec04} we check that vertex $\LL$
\rf{20052008-12} is also invariant under the conformal algebra
transformations.

{\bf Spin-2}. We proceed with spin-2 fields. One can make sure that the
following vertex
\be \label{20052008-10} \LL = \half \phi_{cur}^{ab}\phi_{sh}^{ab} -
\frac{1}{4} \phi_{cur}^{aa}\phi_{sh}^{bb} + \phi_{cur}^a \phi_{sh}^a +
\phi_{cur} \phi_{sh} \ee
is invariant under gauge transformations of the spin-2 shadow field
\rf{14052008-11}-\rf{14052008-13} provided the spin-2 current satisfies
differential constraints \rf{080423-01},\rf{080423-02}. Vertex
\rf{20052008-10} is also invariant under gauge transformations of the spin-2
current \rf{14052008-05}-\rf{14052008-07} provided the spin-2 shadow field
satisfies differential constraints \rf{05152008-10new},\rf{05152008-11new}.
Using the representation for generators of the conformal algebra obtained in
the Sections \ref{sec05},\ref{sec06} we check that vertex $\LL$
\rf{20052008-10} is invariant under the conformal algebra transformations.

{\bf Arbitrary spin current and shadow field}. For the case of arbitrary spin
current and shadow field the gauge invariant vertex takes the form

\be \label{20052008-11} \LL = \langle \phi_{cur}| (1 - \frac{1}{4}\alpha^2
\bar\alpha^2) |\phi_{sh}\rangle \,.\ee
This vertex is invariant under gauge transformation of the shadow field
\rf{gautraarbspi01sh} provided the current satisfies differential constraint
\rf{17052008-05}. Vertex \rf{20052008-11} is also invariant under gauge
transformation of the current \rf{gautraarbspi01} provided the shadow field
satisfies differential constraint \rf{17052008-05sh}. Using the
representation for generators of the conformal algebra obtained in the
Sections \ref{sec07},\ref{sec08} we check that $\LL$ \rf{20052008-11} is also
invariant under the conformal algebra transformations. Details of the
derivation of the vertex $\LL$ may be found in the Appendix \ref{appen01}.

\section{ AdS/CFT correspondence }
\label{sec10}

We now apply our results to the study of $AdS/CFT$ correspondence for bulk
massless fields and boundary conformal currents and shadow fields. We
demonstrate that normalizable solutions of bulk equations of motion are
related to conformal currents, while non-normalizable solutions of bulk
equations of motion are related to shadow fields. As is well known
investigation of $AdS/CFT$ correspondence for massless fields requires
analysis of some subtleties related to the fact that global transformations
of bulk massless fields are defined up to local gauge transformations. In our
approach, these complications are easily controllable because of the
following reasons:
\\
{\bf i}) We use the modified Lorentz gauge for spin-1 field and the modified
de Donder gauge for spin $s\geq 2$ fields. These gauges lead to the {\it
decoupled} bulk equations of motion for arbitrary spin $AdS$ fields and this
considerably simplifies the study of $AdS/CFT$ correspondence. We note that
the most convenient way to dial with the modified gauges is to use the
Poincar\'e parametrization of $AdS_{d+1}$ space,
\be \label{lineelem01} ds^2 = \frac{1}{z^2}(dx^a dx^a + dz\, dz)\,. \ee
{\bf ii}) The modified gauges are invariant under the leftover on-shell gauge
symmetries of bulk $AdS$ fields. Note however that, in our approach, we have
gauge
symmetries not only at $AdS$ side, but also at the boundary $CFT$%
\footnote{ Note that in the standard approach to $CFT$ only the shadow fields
are transformed under gauge transformations, while in our gauge invariant
approach both the currents and shadow fields are transformed under gauge
transformations. Thus, our approach allows us to study the currents and
shadow fields on an equal footing.}.
It turns out that these gauge symmetries are also related via $AdS/CFT$
correspondence. Namely, the leftover on-shell gauge symmetries of bulk $AdS$
fields are related with the gauge symmetries of currents and shadow fields we
obtained in the framework of our gauge invariant approach to $CFT$ in the
Sections \ref{sec03}-\ref{sec08}.
\\
{\bf iii}) In AdS space and at the boundary, we have the same number of gauge
fields and the same number of gauge transformation parameters. Also, our
$AdS$ fields, currents, and shadow fields satisfy the same algebraic
constraints.

\subsection{ AdS/CFT correspondence for spin-1 fields}

As a warm up let us consider spin-1 Maxwell field. In $AdS_{d+1}$ space, the
massless spin-1 field is described by fields $\phi^a(x,z)$ and $\phi(x,z)$
which are the respective vector and scalar fields of the $so(d-1,1)$ algebra.
In the modified Lorentz gauge,%
\footnote{Discussion of $AdS/CFT$ correspondence for spin-1 Maxwell field by
using radial gauge may be found in \cite{wit}.}
found in Ref.\cite{Metsaev:1999ui},
\be \label{15052008-01} \partial^a \phi^a + (\partial_z - \frac{d-3}{2z})\phi
= 0 \,,\ee
we obtain the decoupled equations of motion (for details, see Appendix
\ref{appen03}),
\beq
\label{15052008-02} && (\Box + \partial_z^2 - \frac{1}{z^2}(\nu_1^2-
\frac{1}{4}))\phi^a=0\,,
\\[7pt]
\label{15052008-03} && (\Box + \partial_z^2 -  \frac{1}{z^2}(\nu_0^2-
\frac{1}{4}) )\phi=0\,,
\\[10pt]
&& \label{15052008-04}  \nu_1 = \frac{d-2}{2}\,,\qquad  \nu_0 =
\frac{d-4}{2}\,.
\eeq
Gauge condition \rf{15052008-01} and equations of motion
\rf{15052008-02},\rf{15052008-03} are invariant under the leftover on-shell
gauge transformations
\beq
\label{15052008-05} && \delta \phi^a  = \partial^a \xi \,,
\\[5pt]
\label{15052008-06} && \delta \phi  = (\partial_z  + \frac{d-3}{2z}) \xi \,,
\eeq
where the gauge transformation parameter $\xi$ satisfies the equation of
motion

\be \label{15052008-07} (\Box + \partial_z^2 -\frac{1}{z^2}(\nu_1^2
-\frac{1}{4}))\xi = 0\,. \ee
It is easy to see that the normalizable solution of equations
\rf{15052008-02},\rf{15052008-03} takes the form
\beq
\label{15052008-08} && \phi_{norm}^a(x,z) = U_{\nu_1} \phi_{cur}^a(x)\,,
\\[5pt]
\label{15052008-09} && \phi_{norm}(x,z) = U_{\nu_0} (- \phi_{cur}(x))\,,
\eeq
while the non-normalizable solution is given by%
\footnote{To keep discussion from becoming unwieldy here and below we
restrict our attention to odd $d$. In this case, solutions given in
\rf{15052008-08},\rf{15052008-09} and \rf{15052008-10},\rf{15052008-11} are
independent.}
\beq
\label{15052008-10} && \phi_{non-norm}^a(x,z) = U_{-\nu_1} \phi_{sh}^a(x)\,,
\\[5pt]
\label{15052008-11} && \phi_{non-norm}(x,z) = U_{-\nu_0} \phi_{sh}(x)\,,
\eeq
where we introduce operator $U_\nu$ defined by
\be \label{15052008-12}  U_\nu \equiv \sqrt{qz} J_\nu (qz)q^{-\nu
-\half}\,,\qquad q^2 \equiv \Box\,,\ee
and $J_\nu$ stands for the Bessel function. Taking into account well-known
properties of the Bessel function we find that the asymptotic behavior of the
normalizabe solution is given by
\beq
\label{15052008-13} && \phi_{norm}^a(x,z) \ \ \stackrel{z\rightarrow
0}{\longrightarrow} \ \ z^{\nu_1 + \half} \phi_{cur}^a(x)\,,
\\[5pt]
\label{15052008-14} && \phi_{norm}(x,z) \ \ \stackrel{z\rightarrow
0}{\longrightarrow} \ \  z^{\nu_0 + \half}  \phi_{cur}(x)\,,
\eeq
while the asymptotic behavior of the non-normalizable solution takes the form
\beq
\label{15052008-15} && \phi_{non-norm}^a(x,z) \ \ \stackrel{z\rightarrow
0}{\longrightarrow} \ \ z^{-\nu_1 + \half} \phi_{sh}^a(x)\,,
\\[5pt]
\label{15052008-16} && \phi_{non-norm}(x,z) \ \ \stackrel{z\rightarrow
0}{\longrightarrow} \ \ z^{-\nu_0 + \half}  \phi_{sh}(x)\,.
\eeq
In \rf{15052008-13}-\rf{15052008-16}, we drop overall factors that do not
depend on $z$ and $\Box$. From \rf{15052008-13}-\rf{15052008-16}, we see that
$\phi_{cur}^a$, $\phi_{cur}$ are indeed boundary values of the normalizable
solution, while $\phi_{sh}^a$, $\phi_{sh}$ are boundary values of the
non-normalizable solution.

In the r.h.s. \rf{15052008-08},\rf{15052008-09} and
\rf{15052008-10},\rf{15052008-11} we use the respective notation
$\phi_{cur}^a$, $\phi_{cur}$ and $\phi_{sh}^a$, $\phi_{sh}$ since we are
going to demonstrate that these boundary values are indeed the conformal
currents and shadow fields entering our gauge invariant formulation in the
Sections \ref{sec03},\ref{sec04}. Namely one can prove the following
statements:

{\bf i}) {\it Leftover on-shell} gauge transformations
\rf{15052008-05},\rf{15052008-06} of normalizable solution
\rf{15052008-08},\rf{15052008-09} lead to gauge transformations
\rf{14052008-08},\rf{14052008-09} of the conformal currents $\phi_{cur}^a$,
$\phi_{cur}$, while {\it leftover on-shell} gauge transformations
\rf{15052008-05},\rf{15052008-06} of non-normalizable solution
\rf{15052008-10},\rf{15052008-11} lead to gauge transformations
\rf{14052008-08sh},\rf{14052008-09sh}  of the shadow fields $\phi_{sh}^a$,
$\phi_{sh}$.

{\bf ii}) For normalizable solution \rf{15052008-08},\rf{15052008-09},
modified Lorentz gauge condition  \rf{15052008-01}  leads to differential
constraint \rf{14052008-10} of the conformal currents $\phi_{cur}^a$,
$\phi_{cur}$, while, for non-normalizable solution
\rf{15052008-10},\rf{15052008-11}, modified Lorentz gauge condition
\rf{15052008-01}  leads to differential constraint \rf{15052008-02new} of the
shadow fields $\phi_{sh}^a$, $\phi_{sh}$.

{\bf iii}) Global $so(d,2)$ symmetries of the normalizable (non-normalizable)
massless spin-1 modes in $AdS_{d+1}$ become global $so(d,2)$ conformal
symmetries of the conformal spin-1 current (shadow field).%
\footnote{ In this Section, to avoid the repetition, we do not demonstrate
matching of the global $so(d,2)$-symmetries. Matching of the global
$so(d,2)$-symmetries for arbitrary spin fields is studied in the Section
\ref{sec11}.}

These statements can easily be proved by using the following relations for
the operator $U_\nu$:
\beq
\label{15052008-17} && (\partial_z +\frac{\nu -\half}{z})U_\nu = U_{\nu-1}\,,
\\[7pt]
\label{15052008-18} && (\partial_z - \frac{\nu + \half}{z})U_\nu =
U_{\nu+1}(-\Box) \,,
\\[7pt]
\label{15052008-19} && (\partial_z + \frac{\nu -\half}{z})U_{-\nu} =
U_{-\nu+1}(-\Box )\,,
\\[7pt]
\label{15052008-20} && (\partial_z -\frac{\nu +\half}{z})U_{-\nu} =
U_{-\nu-1}\,,
\eeq
which, in turn, can be obtained by using the following well-known identities
for the Bessel function:
\be
\label{15052008-21}  (\partial_z + \frac{\nu}{z}) J_{\nu }(z) =
J_{\nu-1}(z)\,,
\quad
( \partial_z - \frac{\nu}{z}) J_{\nu }(z) = - J_{\nu + 1}(z)\,.
\ee
As an illustration we demonstrate how constraint for the conformal current
\rf{14052008-10} can be obtained from modified Lorentz gauge condition
\rf{15052008-01}. To this end, adapting relation \rf{15052008-18} for
$\nu=\nu_0$ \rf{15052008-04}, we obtain
\be
\label{15052008-23} (\partial_z - \frac{d-3}{2z})U_{\nu_0} = U_{\nu_1}
(-\Box) \,.
\ee
Plugging normalizable solutions $\phi_{norm}^a$ \rf{15052008-08},
$\phi_{norm}$ \rf{15052008-09} in modified Lorentz gauge condition
\rf{15052008-01} and using \rf{15052008-23} we obtain the relation
\be
\label{15052008-24} \partial^a \phi_{norm}^a  + (\partial_z -
\frac{d-3}{2z})\phi_{norm}
=
U_{\nu_1}( \partial^a \phi_{cur}^a  + \Box \phi_{cur})\,,
\ee
i.e., our modified Lorentz gauge condition \rf{15052008-01} leads indeed to
differential constraint for the conformal current \rf{14052008-10}.

As second illustration, we demonstrate how gauge transformations of the
conformal current \rf{14052008-08},\rf{14052008-09} can be obtained from
leftover on-shell gauge transformations of massless $AdS$ field
\rf{15052008-05},\rf{15052008-06}. To this end we note that the respective
normalizable and non-normalizable solutions of equation for gauge
transformation parameter \rf{15052008-07} take the form
\beq
\label{15052008-25} && \xi_{norm}(x,z) = U_{\nu_1} \xi_{cur}(x) \,,
\\[5pt]
\label{15052008-26} && \xi_{non-norm}(x,z) = U_{-\nu_1} \xi_{sh}(x)\,. \eeq
Plugging \rf{15052008-08},\rf{15052008-25} in \rf{15052008-05} we see that
\rf{15052008-05} leads indeed to \rf{14052008-08}. To match the remaining
gauge transformations \rf{15052008-06} and \rf{14052008-09} we adapt
relation \rf{15052008-17} with $\nu=\nu_1$ to obtain
\be
\label{15052008-27} (\partial_z + \frac{d-3}{2z})U_{\nu_1} = U_{\nu_0}\,.
\ee
Plugging \rf{15052008-25} in \rf{15052008-06} and using \rf{15052008-27} we
obtain
\be \label{15052008-28} \delta \phi_{norm} = U_{\nu_0} \xi_{cur}\,. \ee
Taking into account \rf{15052008-09} we see that gauge transformations
\rf{15052008-06} and \rf{14052008-09} match.

In similar way, one can match: {\bf i}) the leftover on-shell gauge
transformations of the non-normalizable massless $AdS$ modes and the gauge
transformations of the shadow field; {\bf ii}) the modified Lorentz gauge
condition for the non-normalizable solution and the differential constraint
for the shadow field.

Gauge invariant fields $T_{cur}^a$, $T_{sh}^a$ given in \rf{19052008-30new},
\rf{04092009-11} can also be obtained via $AdS/CFT$ correspondence. We
consider a field strength $W^a$ constructed out of the massless fields
$\phi^a$, $\phi$,
\be \label{06092008-18}
W^a  =  (\partial_z + \frac{d-3}{2z}) \phi^a - \partial^a \phi\,,
\ee
and note that:
\\
{\bf i}) $W^a$ is invariant under gauge transformations
\rf{15052008-05},\rf{15052008-06}.
\\
{\bf ii}) plugging normalizable and non-normalizable solutions
\rf{15052008-08}-\rf{15052008-11} in \rf{06092008-18} and using
\rf{19052008-30new}, \rf{04092009-11} we obtain the respective relations
\beq
\label{06092008-19} && W_{norm}^a = U_{\nu_0} T_{cur}^a\,,
\\[3pt]
\label{06092008-20} && W_{non-norm}^a = U_{-\nu_0} (-T_{sh}^a)\,,
\eeq
i.e., for the normalizable solution, bulk field $W^a$ \rf{06092008-18}
corresponds to boundary gauge invariant field $T_{cur}^a$
\rf{19052008-30new}, while, for the non-normalizable solution, bulk field
$W^a$ \rf{06092008-18} corresponds to boundary gauge invariant field
$T_{sh}^a$ \rf{04092009-11}.
\\
{\bf iii}) denoting the left hand side of \rf{15052008-01} by $C_{mod}$, we
get
\be \label{06092008-21}
\partial^a W^a = (\partial_z + \frac{d-3}{2z}) C_{mod}\,.
\ee
We then check that plugging the normalizable solution in \rf{06092008-21} and
using \rf{06092008-19} gives \rf{19052008-30newnew}, while plugging the
non-normalizable solution in \rf{06092008-21} and using \rf{06092008-20}
gives \rf{16052008-01newnew}.

\subsection{ AdS/CFT correspondence for spin-2 fields}

We now proceed with the discussion of $AdS/CFT$ correspondence for bulk
massless spin-2 $AdS$ field and boundary spin-2 conformal current and shadow
field. To this end we use the modified de Donder gauge condition for the
massless spin-2 $AdS$ field \cite{Metsaev:2008ks}
\footnote{ Discussion of $AdS/CFT$ correspondence for massless spin-2 field
taken to be in radial gauge may be found in \cite{Liu:1998bu,aru}.}.
In Ref.\cite{Metsaev:2008ks}, we found that the suitable modification of the
standard de Donder gauge condition leads to the decoupled equations of motion
for the massless spin-2 $AdS$ field. We begin therefore with the presentation
of our result in Ref.\cite{Metsaev:2008ks}. Some useful details may be found
in the Appendix \ref{appen03}.

In $AdS_{d+1}$ space, massless spin-2 field is described by fields
$\phi^{ab}(x,z)$, $\phi^a(x,z)$, $\phi(x,z)$. The field $\phi^{ab}$ is rank-2
tensor field of the $so(d-1,1)$ algebra, while $\phi^a$ and $\phi$ are the
respective vector and scalar fields of the $so(d-1,1)$ algebra. Gauge
condition, which we refer to as modified de Donder gauge condition, is
defined to be
\beq
\label{15052008-29} && \hspace{-0.7cm} \partial^b \phi^{ab} -\half \partial^a
\phi^{bb} + (\partial_z - \frac{d-1}{2z})\phi^a =0\,,
\\[5pt]
\label{15052008-30} && \hspace{-0.7cm} \partial^a \phi^a -\half (\partial_z
+\frac{d-1}{2z})\phi^{aa} + \sigg (\partial_z -\frac{d-3}{2z})\phi=0\,,
\nonumber\\[-3pt]
\eeq
where $\sigg$ is given in \rf{080423-02(add)}. Remarkable property of this
gauge condition is that it leads to the decoupled equations of motion for the
fields $\phi^{ab}$, $\phi^a$, $\phi$,
\beq
\label{15052008-31}&& (\Box + \partial_z^2 - \frac{1}{z^2}(\nu_2^2
-\frac{1}{4}))\phi^{ab}=0\,,
\\[7pt]
\label{15052008-32}&& (\Box + \partial_z^2 - \frac{1}{z^2}(\nu_1^2
-\frac{1}{4}))\phi^a=0\,,
\\[7pt]
\label{15052008-33}&& (\Box + \partial_z^2 - \frac{1}{z^2}(\nu_0^2
-\frac{1}{4}))\phi=0\,,
\eeq
\be \label{15052008-34} \nu_2 =\frac{d}{2}\,,\qquad  \nu_1
=\frac{d-2}{2}\,,\qquad  \nu_0 =\frac{d-4}{2}\,.\ee

These equations and gauge condition \rf{15052008-29}, \rf{15052008-30} are
invariant under the leftover on-shell gauge transformations,
\beq
\label{15052008-35} && \hspace{-0.3cm} \delta \phi^{ab} = \partial^a \xi^b
+\partial^b \xi^a + \frac{2}{d-2}(\partial_z -\frac{d-1}{2z})\eta^{ab} \xi\,,
\qquad\quad
\\[5pt]
\label{15052008-36} && \hspace{-0.3cm} \delta \phi^a = \partial^a \xi +
(\partial_z +\frac{d-1}{2z}) \xi^a\,,
\\[5pt]
\label{15052008-37} && \hspace{-0.3cm} \delta \phi = \sigg (\partial_z
+\frac{d-3}{2z}) \xi\,,
\eeq
where the gauge transformation parameters $\xi^a$ and $\xi$ satisfy the
respective equations of motion,
\beq
\label{15052008-38} && (\Box + \partial_z^2 - \frac{1}{z^2}(\nu_2^2
-\frac{1}{4}))\xi^a=0\,, \ \ \ \ \
\\[7pt]
\label{15052008-39} && (\Box + \partial_z^2 - \frac{1}{z^2}(\nu_1^2
-\frac{1}{4}))\xi=0\,.
\eeq
Thus, we see that our modified de Donder gauge leads to the decoupled
equations of motion for both the gauge fields and gauge transformation
parameters. This streamlines investigation of $AdS/CFT$ correspondence.

First of all we note that the normalizable solution of equations of motion
\rf{15052008-31}-\rf{15052008-33} is given by
\beq
\label{15052008-40} && \phi_{norm}^{ab}(x,z) = U_{\nu_2}
\phi_{cur}^{ab}(x)\,,
\\[5pt]
\label{15052008-41}&& \phi_{norm}^a(x,z) = U_{\nu_1} (-\phi_{cur}^a(x))\,,
\\[5pt]
\label{15052008-42}&& \phi_{norm}(x,z) = U_{\nu_0} \phi_{cur}(x)\,,
\eeq
while the non-normalizable solution takes the form
\beq
\label{15052008-43}&& \phi_{non-norm}^{ab}(x,z) = U_{-\nu_2}
\phi_{sh}^{ab}(x)\,,
\\[5pt]
\label{15052008-44}&& \phi_{non-norm}^a(x,z) = U_{-\nu_1} \phi_{sh}^a(x)\,,
\\[5pt]
\label{15052008-45}&& \phi_{non-norm}(x,z) = U_{-\nu_0} \phi_{sh}(x)\,,
\eeq
where the operator $U_\nu$ is defined in \rf{15052008-12}. From these
relations, we find the asymptotic behavior of the normalizable solution
\beq
\label{15052008-46}&& \phi_{norm}^{ab}(x,z) \ \ \stackrel{z\rightarrow
0}{\longrightarrow} \ \ z^{\nu_2 + \half} \phi_{cur}^{ab}(x)\,,
\\[5pt]
\label{15052008-47}&& \phi_{norm}^a(x,z) \ \ \stackrel{z\rightarrow
0}{\longrightarrow} \ \
  z^{\nu_1 + \half} \phi_{cur}^a(x)\,,
\\[5pt]
\label{15052008-48}&& \phi_{norm}(x,z) \ \ \stackrel{z\rightarrow
0}{\longrightarrow} \ \ z^{\nu_0 + \half}  \phi_{cur}(x)\,,
\eeq
while the asymptotic behavior of the non-normalizable solution takes the form
\beq
\label{15052008-49}&& \phi_{non-norm}^{ab}(x,z) \ \ \stackrel{z\rightarrow
0}{\longrightarrow} \ \ z^{-\nu_2 + \half} \phi_{sh}^{ab}(x)\,,
\\[5pt]
\label{15052008-50}&& \phi_{non-norm}^a(x,z) \ \ \stackrel{z\rightarrow
0}{\longrightarrow} \ \ z^{-\nu_1 + \half} \phi_{sh}^a(x)\,,
\\[5pt]
\label{15052008-51}&& \phi_{non-norm}(x,z) \ \ \stackrel{z\rightarrow
0}{\longrightarrow} \ \ z^{-\nu_0 + \half}  \phi_{sh}(x)\,.
\eeq
From \rf{15052008-46}-\rf{15052008-51}, we see that the fields
$\phi_{cur}^{ab}$, $\phi_{cur}^a$, $\phi_{cur}$ are indeed boundary values of
the normalizable solution, while $\phi_{sh}^{ab}$, $\phi_{sh}^a$, $\phi_{sh}$
are boundary values of the non-normalizable solution.

In the r.h.s. \rf{15052008-40}-\rf{15052008-42} and
\rf{15052008-43}-\rf{15052008-45}, we use the respective notation
$\phi_{cur}^{ab}$, $\phi_{cur}^a$, $\phi_{cur}$ and $\phi_{sh}^{ab}$,
$\phi_{sh}^a$, $\phi_{sh}$ because these boundary values turn out to be the
spin-2 conformal currents and shadow fields entering our gauge invariant
formulation in the Sections \ref{sec05},\ref{sec06}. Namely, one can prove
the following statements:
\\
{\bf i}) {\it Leftover on-shell} gauge transformations
\rf{15052008-35}-\rf{15052008-37} of normalizable solution
\rf{15052008-40}-\rf{15052008-42} lead to gauge transformations
\rf{14052008-05}-\rf{14052008-07} of the conformal currents
$\phi_{cur}^{ab}$, $\phi_{cur}^a$, $\phi_{cur}$, while {\it leftover
on-shell} gauge transformations \rf{15052008-35}-\rf{15052008-37} of
non-normalizable solution \rf{15052008-43}-\rf{15052008-45} lead to gauge
transformations \rf{14052008-11}-\rf{14052008-13} of the shadow fields
$\phi_{sh}^{ab}$, $\phi_{sh}^a$, $\phi_{sh}$.
\\
{\bf ii}) For normalizable solution \rf{15052008-40}-\rf{15052008-42},
modified de Donder gauge condition \rf{15052008-29},\rf{15052008-30} leads to
differential constraints \rf{080423-01},\rf{080423-02} of the conformal
currents $\phi_{cur}^{ab}$, $\phi_{cur}^a$, $\phi_{cur}$ while, for
non-normalizable solution \rf{15052008-43}-\rf{15052008-45}, modified de
Donder gauge condition \rf{15052008-29},\rf{15052008-30} leads to
differential constraints \rf{05152008-10new},\rf{05152008-11new} of the
shadow fields $\phi_{sh}^{ab}$, $\phi_{sh}^a$, $\phi_{sh}$.
\\
{\bf iii}) Global $so(d,2)$ bulk symmetries of the normalizable
(non-normalizable) massless spin-2 modes in $AdS_{d+1}$  become global
$so(d,2)$ boundary conformal symmetries of the spin-2 current (shadow field).

These statements can easily be proved in the same way as in the case of
massless spin-1 field. To do that one needs to use relations for the operator
$U_\nu$ given in \rf{15052008-17}-\rf{15052008-20}. Also, one needs to take
into account the following normalizable solution of equations of motion for
the gauge transformation parameters in \rf{15052008-38},\rf{15052008-39}:
\beq
&& \xi_{norm}^a (x,z) = U_{\nu_2} \xi_{cur}^a(x)\,,
\\[5pt]
&& \xi_{norm}(x,z) = U_{\nu_1} (-\xi_{cur}(x))\,,
\eeq
and the appropriate non-normalizable solution given by
\beq
&& \xi_{non-norm}^a (x,z) = U_{\nu_2} \xi_{sh}^a(x)\,,
\\[5pt]
&& \xi_{non-norm}(x,z) = U_{\nu_1} \xi_{sh}(x)\,.
\eeq

We note that gauge invariant fields $T_{cur}^{ab}$, $T_{sh}^{ab}$ given in
\rf{19052008-30} \rf{06092008-12} can also be obtained via $AdS/CFT$
correspondence. To this end we consider a field strength $W^{ab}$ constructed
out of the massless fields $\phi^{ab}$, $\phi^a$, $\phi$,
\beq \label{06092008-11}
W^{ab} & = & (\partial_z + \frac{d-3}{2z}) (\partial_z + \frac{d-1}{2z})
\phi^{ab}
\nonumber\\[3pt]
&  - &  (\partial_z + \frac{d-3}{2z}) (\partial^a\phi^b +\partial^b\phi^a)
\nonumber\\[3pt]
& + &  \frac{2}{\sigg}\partial^a \partial^b \phi +
\frac{2}{(d-2)\sigg}\eta^{ab}\Box  \phi \,,\qquad
\eeq
where $\sigg$ is given in \rf{080423-02(add)}. We note that:
\\
{\bf i}) $W^{ab}$ is invariant under on-shell gauge transformations
\rf{15052008-35}-\rf{15052008-37}.
\\
{\bf ii}) plugging normalizable and non-normalizable solutions
\rf{15052008-40}-\rf{15052008-45} in \rf{06092008-11} and using
\rf{19052008-30}, \rf{06092008-12} we obtain the respective relations
\beq
\label{06092008-13} && W_{norm}^{ab} = U_{\nu_0} T_{cur}^{ab}\,,
\\[3pt]
\label{06092008-14} && W_{non-norm}^{ab} = U_{-\nu_0}(- T_{sh}^{ab})\,,
\eeq
i.e., we see that, for the normalizable solution, bulk tensor field $W^{ab}$
\rf{06092008-11} corresponds to boundary gauge invariant field $T_{cur}^{ab}$
\rf{19052008-30}, while, for the non-normalizable solution, bulk tensor field
$W^{ab}$ \rf{06092008-11} corresponds to boundary gauge invariant field
$T_{sh}^{ab}$ \rf{06092008-12}.
\\
{\bf iii}) denoting the respective left hand sides of \rf{15052008-29} and
\rf{15052008-30} by $C_{mod}^a$ and $C_{mod}$ we get
\beq \label{06092008-15}
&& \hspace{-1cm}  \partial^b W^{ab} - \half \partial^a W^{bb} = (\partial_z +
\frac{d-3}{2z}) (\partial_z + \frac{d-1}{2z})C_{mod}^a\,,
\nonumber\\
&&
\\
\label{06092008-16} && \hspace{-1cm}  W^{aa} = -2 (\partial_z +
\frac{d-3}{2z}) C_{mod}\,.
\eeq
We then check that plugging the normalizable solution in
\rf{06092008-15},\rf{06092008-16} and using \rf{06092008-13} gives
\rf{07092008-01}, while plugging the non-normalizable solution in
\rf{06092008-15},\rf{06092008-16} and using \rf{06092008-14} gives
\rf{0409200-06}.

\section{ AdS/CFT correspondence for arbitrary spin fields}\label{sec11}

We proceed with the discussion of $AdS/CFT$ correspondence for bulk massless
arbitrary spin-$s$ $AdS$ field and boundary spin-$s$ conformal current and
shadow field. To discuss the correspondence we use the {\it modified} de
Donder gauge condition for bulk massless arbitrary spin field
\footnote{ In light-cone gauge, $AdS/CFT$ correspondence for arbitrary spin
massless fields was studied in Ref.\cite{Metsaev:1999ui}. In radial gauge,
$AdS/CFT$ correspondence for arbitrary spin massless fields was considered in
Ref. \cite{Germani:2004jf}.}.
In Ref.\cite{Metsaev:2008ks} we found that some modification of the standard
de Donder gauge condition%
\footnote{ Recent interesting applications of the {\it standard} de Donder
gauge to the various problems of higher-spin fields may be found in
Refs.\cite{Guttenberg:2008qe,Manvelyan:2008ks}.}
leads to the {\it decoupled} equations of motion for arbitrary spin $AdS$ field%
\footnote{ We believe that our modified de Donder gauge will also be useful
for better understanding of various aspects of AdS/QCD correspondence which
are discussed e.g. in \cite{Andreev:2002aw}-\cite{Abidin:2008ku}.}.
We begin therefore with the presentation of our result in
Ref.\cite{Metsaev:2008ks}. In $AdS_{d+1}$ space, massless spin-$s$ field is
described by the following scalar, vector, and totally symmetric tensor
fields of the Lorentz algebra $so(d-1,1)$:
\be \label{18052008-02}
\phi_{s'}^{a_1\ldots a_{s'}}\,, \hspace{1cm} s'=0,1,\ldots,s.
\ee
The fields $\phi_{s'}^{a_1\ldots a_{s'}}$ with $s'>3$ are double-traceless,
\be \label{18052008-03} \phi_{s'}^{aabba_5\ldots a_{s'}}=0\,, \hspace{1cm}
s'=4,5,\ldots,s. \ee

In order to obtain the gauge invariant description in an easy--to--use form
we use the oscillators and introduce a ket-vector $|\phi\rangle$ defined by
\beq
\label{18052008-04} && \hspace{-1cm}  |\phi\rangle \equiv \sum_{s'=0}^s
\alpha_z^{s-s'}|\phi_{s'}\rangle \,,
\\[5pt]
\label{18052008-05} && \hspace{-1cm} |\phi_{s'}\rangle \equiv
\frac{\alpha^{a_1} \ldots \alpha^{a_{s'}}}{s'!\sqrt{(s - s')!}}
\, \phi_{ s'}^{a_1\ldots a_{s'}} |0\rangle\,.
\eeq
From \rf{18052008-04},\rf{18052008-05}, we see that the ket-vector
$|\phi\rangle$ is degree-$s$ homogeneous polynomial in the oscillators
$\alpha^a$, $\alpha^z$, while the ket-vector $|\phi_{s'}\rangle$ is
degree-$s'$ homogeneous polynomial in the oscillators $\alpha^a$, i.e., these
ket-vectors satisfy the relations
\beq
\label{18052008-06} && (N_\alpha + N_z - s)|\phi\rangle = 0 \,,
\\[5pt]
\label{18052008-07} && (N_\alpha - s')|\phi_{s'}\rangle = 0 \,.
\eeq
In terms of the ket-vector $|\phi\rangle$, double-tracelessness constraint
\rf{18052008-03} takes the form
\beq
&& (\bar{\alpha}^2)^2 |\phi\rangle  = 0 \,.
\eeq

Gauge condition, which we refer to as modified de Donder gauge condition, is
defined to be

\beq \label{080405-01}
&& \hspace{-0.7cm} \bar{C}_{mod}^{} \phik  =  0 \,,
\\[5pt]
\label{080405-01add} && \bar{C}_{mod}^{} \equiv  \albpar - \half \alpar
\bar\alpha^2 + \half e_1 \bar\alpha^2 - \eb_1\Pi^\smponetwo \,,
\\[5pt]
\label{18052008-08} && \Pi^\smponetwo \equiv 1 -\alpha^2\frac{1}{2(2N_\alpha
+d)}\bar\alpha^2\,,
\\[5pt]
\label{18052008-09} && e_1 = e_{1,1} \Bigl( \partial_z + \frac{2s + d -5
-2N_z}{2z}\Bigr)\,,
\\[6pt]
\label{18052008-10} && \eb_1 = \Bigl(\partial_z - \frac{2s + d -5
-2N_z}{2z}\Bigr) \eb_{1,1}\,,
\\[6pt]
\label{18052008-11} && e_{1,1} =  - \alpha^z \ewt_1
\qquad
\eb_{1,1} =  - \ewt_1 \bar\alpha^z\,,
\\[6pt]
\label{18052008-12new} && \ewt_1 =
\Bigl(\frac{2s+d-4-N_z}{2s+d-4-2N_z}\Bigr)^{1/2}\,.
\eeq
In this gauge, we obtain the decoupled equations of motion for massless
arbitrary spin-$s$ $AdS$ field $\phik$ ,
\beq
\label{18052008-13} && \Bigl(\Box + \partial_z^2 - \frac{1}{z^2} (\nu^2
-\frac{1}{4} ) \Bigr)\phik = 0\,,
\\[5pt]
\label{18052008-14} &&\qquad  \nu \equiv s + \frac{d-4}{2} - N_z\,.
\eeq
Gauge condition \rf{080405-01}  and equations \rf{18052008-13} are
invariant under the leftover on-shell gauge transformation
\be
\label{18052008-15} \delta \phik =  ( \alpar - e_1  -
\frac{\alpha^2}{2s+d-6-2N_z}\eb_1 ) \xik \,,
\ee
where $e_1$, $\eb_1$ are given in \rf{18052008-09},\rf{18052008-10} and gauge
transformation ket-vector $\xik$ satisfies the equations of motion
\beq \label{13092008-03}
\label{18052008-18} && \Bigl(\Box + \partial_z^2 - \frac{1}{z^2} (\nu^2
-\frac{1}{4} ) \Bigr)\xik = 0\,,
\eeq
with $\nu$ given \rf{18052008-14}. In terms of $so(d-1,1)$ algebra tensor
fields, the ket-vector $\xik$ is represented as
\beq
\label{08092008-06} && |\xi\rangle \equiv \sum_{s'=0}^{s-1}
\alpha_z^{s-1-s'}|\xi_{s'}\rangle \,,
\\[5pt]
\label{08092008-07} && |\xi_{s'}\rangle \equiv
\frac{\alpha^{a_1} \ldots \alpha^{a_{s'}}}{s'!\sqrt{(s -1 - s')!}}
\, \xi_{s'}^{a_1\ldots a_{s'}} |0\rangle\,, \ \ \ \ \ \ \ \
\eeq
and satisfies the standard tracelessness constraint
\be \label{08092008-05} \bar\alpha^2|\xi\rangle =0 \,.\ee

We note that the gauge invariant description of the conformal currents (or
shadow fields) given in the Sections \ref{sec07},\ref{sec08} and the
description of $AdS$ fields given in this Section turn out to be very
convenient for the studying $AdS/CFT$ correspondence because of the following
reasons:
\\
{\bf i}) the number of gauge fields involved in the gauge invariant
description of the spin-$s$ conformal current (or shadow field) in
$d$-dimensional space is equal to the number of gauge fields involved in the
gauge invariant description of the massless spin-$s$ field in $AdS_{d+1}$
(see \rf{phiset01},\rf{phiset01sh} and \rf{18052008-02}). Note also that the
conformal current, shadow field, and $AdS$ field satisfy the same
double-tracelessness constraint (see \rf{doutracon01},\rf{doutracon01sh} and
\rf{18052008-03})
\\
{\bf ii}) the number of gauge transformation parameters involved in the gauge
invariant description of the spin-$s$ conformal current (or shadow field) in
$d$-dimensional space is equal to the number of gauge transformation
parameters involved in the gauge invariant description of the massless
spin-$s$ field in $AdS_{d+1}$ (see \rf{epsilonset01},\rf{epsilonset01sh} and
\rf{08092008-06}, \rf{08092008-07}). Also, all these gauge transformation
parameters satisfy the same tracelessness constraint (see
\rf{08092008-08},\rf{08092008-09} and \rf{08092008-05})
\\
{\bf iii}) in the Poincar\'e parametrization of $AdS_{d+1}$ space, the
$d$-dimensional Poincar\'e symmetries of $AdS_{d+1}$ field theory are
manifest. In the conformal current/shadow field theory, the $d$-dimensional
Poincar\'e symmetries are also manifest, i.e. manifest Poincar\'e symmetries
of $AdS$ field theory and $CFT$ match.

We now discuss solutions of equations of motion in \rf{18052008-13}. It is
easy to see that the respective normalizable and non-normalizable solutions
of equations \rf{18052008-13} take the form
\beq
\label{18052008-19}  &&  |\phi_{norm}(x,z)\rangle = U_\nu (-)^{N_z}
|\phi_{cur}(x)\rangle\,,
\\[5pt]
\label{18052008-20}  &&  |\phi_{non-norm}(x,z)\rangle = U_{-\nu}
|\phi_{sh}(x)\rangle\,,
\eeq
where the operator $U_\nu$ is defined in \rf{15052008-12}. From these
relations, we find the asymptotic behavior of our solutions
\beq
&& |\phi_{norm}(x,z)\rangle \qquad \ \stackrel{z\rightarrow
0}{\longrightarrow} \ \ \ z^{\nu + \half} |\phi_{cur}(x)\rangle\,,
\\[10pt]
&& |\phi_{non-norm}(x,z)\rangle \ \ \stackrel{z\rightarrow
0}{\longrightarrow} \ \ \ z^{-\nu + \half} |\phi_{sh}(x)\rangle\,. \qquad
\eeq

Now we are ready to formulate our statements:

{\bf i}) {\it Leftover on-shell} gauge transformation \rf{18052008-15} of
normalizable solution \rf{18052008-19} leads to gauge transformation
\rf{gautraarbspi01} of the current $\phicurk$, while {\it leftover on-shell}
gauge transformation \rf{18052008-15} of non-normalizable solution
\rf{18052008-20} leads to gauge transformation \rf{gautraarbspi01sh} of the
shadow field $\phishk$.
\\
{\bf ii}) For normalizable solution \rf{18052008-19}, modified de Donder
gauge condition \rf{080405-01} leads to differential constraint
\rf{17052008-05} of the current $\phicurk$, while, for non-normalizable
solution \rf{18052008-20}, modified de Donder gauge condition \rf{080405-01}
leads to differential constraint \rf{17052008-05sh} of the shadow field
$\phishk$
\footnote{ We expect that use of standard de Donder gauge condition leads to
an isomorphic realization of conformal symmetries. At present time, it is
difficult to check this statement explicitly because standard de Donder gauge
condition leads to coupled equations. Analysis of these equations is
complicated and their solution is not known in closed form so far (see e.g.
Ref.\cite{Manvelyan:2008ks}).}.

{\bf iii}) Global $so(d,2)$ bulk symmetries of the normalizable
(non-normalizable) massless spin-$s$ modes in $AdS_{d+1}$  become global
$so(d,2)$ boundary conformal symmetries of the spin-$s$ current (shadow
field).

We note that all these statements can straightforwardly be proved by using
the following relations for the operator $U_\nu$:
\beq
&& \label{20052008-20} {\bf e}_1 U_\nu = U_\nu \alpha^z \,,
\\[5pt]
 \label{20052008-21}&& {\bf \eb}_1 U_\nu = U_\nu (-\Box \bar\alpha^z) \,,
\\[5pt]
\label{20052008-22}&& {\bf e}_1 U_{-\nu} = U_{-\nu}(-\Box \alpha^z )\,,
\\[5pt]
 \label{20052008-23}&& {\bf \eb}_1 U_{-\nu} = U_{-\nu}\, \bar\alpha^z \,,
\\[5pt]
 \label{20052008-24}&& {\bf e}_1 (z U_{\nu+1}) = z U_{\nu+1} \alpha^z \,,
\\[5pt]
\label{20052008-25}&& {\bf \eb}_1 (z U_{\nu+1})  = 2 U_\nu\bar\alpha^z  -
z\Box U_{\nu+1}\bar\alpha^z \,,
\\[5pt]
\label{20052008-26} && {\bf e}_1 (z U_{-\nu+1})  = 2 U_{-\nu} \alpha^z  - z
\Box U_{-\nu+1} \alpha^z \,,
\\[5pt]
\label{20052008-27} && {\bf \eb}_1 (z U_{-\nu+1}) = z U_{-\nu+1} \bar\alpha^z
\,,
\eeq
where $\nu$ is given in \rf{18052008-14} and we use the notation
\beq
&& {\bf e}_1 \equiv \alpha^z (\partial_z + \frac{\nu -\half}{z})\,,
\\[3pt]
&& {\bf \eb}_1 \equiv  (\partial_z - \frac{\nu -\half}{z})\bar\alpha^z\,.
\eeq

Also, one needs to take into account the following normalizable and
non-normalizable solutions of equations of motion for the gauge
transformation parameters in \rf{13092008-03},
\beq
\label{13092008-01}  &&  |\xi_{norm}(x,z)\rangle = U_\nu (-)^{N_z}
|\xi_{cur}(x)\rangle\,,
\\[3pt]
\label{13092008-02}  &&  |\xi_{non-norm}(x,z)\rangle = U_{-\nu}
|\xi_{sh}(x)\rangle\,.
\eeq

As an illustration we demonstrate how the gauge transformation of the shadow
field can be obtained from the leftover on-shell gauge transformation of the
massless non-normalizable $AdS$ modes.  To this end, we note that, one the
one hand, gauge transformation of $|\phi_{non-norm}\rangle$ takes the form
(see \rf{18052008-20})
\be
\label{13092008-04}  \delta |\phi_{non-norm}(x,z)\rangle = U_{-\nu} \delta
|\phi_{sh}(x)\rangle\,.
\ee
On the other hand, plugging \rf{13092008-02} in \rf{18052008-15} and using
\rf{20052008-22},\rf{20052008-23} we obtain the relations
\beq
\label{13092008-05} && \delta |\phi_{non-norm}(x,z)\rangle
\\
&&  =  ( \alpar - e_1  - \frac{\alpha^2}{2s+d-6-2N_z}\eb_1
)U_{-\nu}|\xi_{sh}\rangle
\nonumber\\[5pt]
&& = U_{-\nu}( \alpar + b_1 \Box  + b_2 \alpha^2 ) |\xi_{sh}\rangle \,,
\nonumber
\eeq
where the $b_1$-, $b_2$-operators entering gauge transformation
\rf{gautraarbspi01sh} of the shadow field are given in
\rf{13092008-06},\rf{13092008-07}. Comparing of \rf{13092008-04} and
\rf{13092008-05} we see that leftover on-shell gauge transformation
\rf{18052008-15} of the massless non-normalizable $AdS$ modes
\rf{18052008-20} leads indeed to gauge transformation \rf{gautraarbspi01sh}
of the shadow field.

In a similar way, using \rf{20052008-20},\rf{20052008-21} we learn that the
leftover on-shell gauge transformation of the massless normalizable $AdS$
modes leads to the gauge transformation of the current.

\subsection{ Matching of bulk and boundary global symmetries}

We finish our study  of $AdS/CFT$ correspondence with the comparison of bulk
and boundary global symmetries. On the one hand, global symmetries of
conformal currents and shadow fields are described by the conformal algebra
$so(d,2)$. On the other hand, relativistic symmetries of the $AdS_{d+1}$
field dynamics are also described by the $so(d,2)$ algebra. For application
to the study of $AdS/CFT$ correspondence, it is convenient to realize the
bulk $so(d,2)$ algebra symmetries by using nomenclature of the conformal
algebra. This is to say that to discuss the bulk $so(d,2)$-symmetries we use
basis of the $so(d,2)$ algebra which consists of translation generators
$P^a$, conformal boost generators $K^a$, dilatation generator $D$, and
generators of the $so(d-1,1)$ algebra, $J^{ab}$. In this basis, the $so(d,2)$
algebra transformations of the massless spin-$s$ $AdS_{d+1}$ field $\phik$
take the form $\delta_{\hat{G}}\phik  = \hat{G} \phik$, where realization of
the $so(d,2)$ algebra generators $\hat{G}$ in terms of differential operators
is given by
\beq
\label{conalggenlis01ads} && P^a = \partial^a \,,
\\[3pt]
\label{conalggenlis02ads} && J^{ab} = x^a\partial^b -  x^b\partial^a +
M^{ab}\,,
\\[3pt]
\label{conalggenlis03ads} && D = x\partial + \Delta\,, \qquad \Delta =
z\partial_z + \frac{d-1}{2}\,, \qquad\quad
\\[3pt]
\label{conalggenlis04ads} && K^a = K_{\Delta,M}^a + R^a\,,
\eeq
\beq
\label{14092008-06} && R^a  =  R_\smzero^a + R_\smone^a\,,
\\[3pt]
\label{14092008-07} && R_\smzero^a = - z \Cwt^a \eb_{1,1} + ze_{1,1}
\bar\alpha^a\,,
\\[3pt]
\label{14092008-08} && R_\smone^a =  -\half z^2 \partial^a\,,
\eeq
and the operators $M^{ab}$ and $\Cwt^a$ are given in \rf{mabdef0001}  and
\rf{20052008-15(01)} respectively, while $K_{\Delta,M}^a$ and $e_{1,1}$ are
given in \rf{kdelmdef01} and \rf{18052008-11} respectively.

We note that representation for generators given in
\rf{conalggenlis01ads}-\rf{conalggenlis04ads} is valid for gauge invariant
theory of $AdS$ field. This to say that our modified Lorentz and de Donder
gauges respect the Poicar\'e and dilatation symmetries, but break
$K^a$-symmetries. In other words, expressions for generators $P^a$, $J^{ab}$
and $D$ given in \rf{conalggenlis01ads}-\rf{conalggenlis03ads} are still
valid for the gauge-fixed $AdS$ fields, while expression for the generator
$K^a$ \rf{conalggenlis04ads} should be modified to restore conformal boost
symmetries for the gauge-fixed $AdS$ fields. Therefore let us first to
demonstrate matching of the Poincar\'e and dilatation symmetries. What is
required is to demonstrate matching of the $so(d,2)$ algebra generators for
bulk $AdS$ fields given in \rf{conalggenlis01ads}-\rf{conalggenlis03ads} and
ones for boundary currents (or shadow fields) given in
\rf{conalggenlis01}-\rf{conalggenlis03}. As for generators of the Poincar\'e
algebra, $P^a$, $J^{ab}$, they already coincide on both sides (see formulas
\rf{conalggenlis01}, \rf{conalggenlis02} and the respective formulas
\rf{conalggenlis01ads},\rf{conalggenlis02ads}). Next, consider the dilatation
generator $D$. Here we need explicit form of solution to bulk theory
equations of motion given in \rf{18052008-19},\rf{18052008-20}. Using the
notation $D_{_{AdS}}$ and $D_{_{CFT}}$ to indicate the respective
realizations of the dilatation generator $D$ on the bulk fields
\rf{conalggenlis03ads} and the conformal currents and shadow fields
\rf{conalggenlis03} we obtain the relations
\beq
&& D_{_{AdS}} |\phi_{norm}\rangle =  U_\nu D_{_{CFT}} |\phi_{cur}\rangle\,,
\\[3pt]
&& D_{_{AdS}} |\phi_{non-norm}\rangle =  U_{-\nu} D_{_{CFT}}
|\phi_{sh}\rangle\,,
\eeq
where the expressions for $D_{_{CFT}}$ corresponding to $\phicurk$ and
$\phishk$ can be obtained from \rf{conalggenlis03} and the respective
conformal dimension operators $\Delta$ given in \rf{08092008-03} and
\rf{08092008-04}. Thus, the generators $D_{_{AdS}}$ and $D_{_{CFT}}$ also
match.

We now turn to matching of the conformal boost $K^a$-symmetries. Technically,
this is the most difficult point of the analysis because matching of the
$K^a$-symmetries requires analysis of some subtleties of our gauge fixing for
$AdS$ field. We now discuss these subtleties.

As we have already said our modified Lorentz and de Donder gauges break the
$K^a$-symmetries. This implies that generator $K^a$ given in
\rf{conalggenlis04ads} should be modified to restore the conformal boost
symmetries of the gauge-fixed $AdS$ field theory. In order to restore these
broken $K^a$-symmetries we should, following standard procedure, add
compensating gauge transformations to maintain the conformal boost
$K^a$-symmetries. Thus, in order to find improved $K^a$-transformations of
the gauge-fixed $AdS$ field $\phik$ we start with the generic global
$K^a$-transformations \rf{conalggenlis04ads} supplemented by the appropriate
compensating gauge transformation
\be \label{20052008-34}
K^a_{impr} \phik = K^a \phik + \delta_{\xi^{K^a} }\phik\,, \ee
where the gauge transformation $\delta_{\xi^{K^a} }\phik$ is obtained from
\rf{18052008-15} by substituting $\xik \rightarrow |\xi^{K^a}\rangle$. The
compensating gauge transformation parameter $|\xi^{K^a}\rangle$ can as
usually be found by requiring improved transformation \rf{20052008-34} to
maintain the gauge condition \rf{080405-01},
\be \label{20052008-30} \bar{C}_{mod} K_{impr}^a  \phik = 0 \,, \ee
where the operator $\bar{C}_{mod}$ is given in \rf{080405-01add}. Plugging
\rf{20052008-34} in \rf{20052008-30}, we find that Eq.\rf{20052008-30} leads
to the equation
\be \label{20052008-31}
\Bigl(\Box + \partial_z^2 - \frac{1}{z^2}(\nu^2
-\frac{1}{4})\Bigr)|\xi^{K^a}\rangle  - 2 \bar{C}_\perp^a \phik = 0\,,
\ee
where $\nu$ is given in \rf{18052008-14} and $\bar{C}_\perp^a$ is defined in
\rf{20052008-15(02)}. Thus, we obtain the non-homogeneous second-order
differential equation for the compensating gauge transformation parameter
$|\xi^{K^a}\rangle$. Plugging normalizable solution \rf{18052008-19} and
non-normalizable solution \rf{18052008-20} in \rf{20052008-31} we find the
respective solutions to the compensating gauge transformation parameters,
\beq
\label{20052008-32} && |\xi_{norm}^{K^a}(x,z)\rangle = z U_{\nu+1}
\bar{C}_\perp^a (-)^{N_z}|\phi_{cur}\rangle\,,
\\[7pt]
\label{20052008-33} && |\xi_{non-norm}^{K^a}(x,z)\rangle = z U_{-\nu+1}
\bar{C}_\perp^a |\phi_{sh}\rangle\,.
\eeq
Making use of solutions \rf{20052008-32}, \rf{20052008-33} in
\rf{20052008-34}, we obtain the improved $K^a$-transformations. We then make
sure that the improved $K^a$-transformations of the
normalizable/non-normalizable bulk $AdS$ modes lead to the conformal boost
transformations for the current/shadow fields obtained in the Section
\ref{sec07}/\ref{sec08} . This can easily be proved by using relations for
the operator $U_\nu$ given in \rf{20052008-20}-\rf{20052008-27}. Details may
be found in the Appendix \ref{appen04}.

The results presented here should have interesting generalizations to
mixed-symmetry fields. In the case of mixed-symmetry fields we could, in
principle, redo our analysis by using equations of Ref.\cite{Metsaev:1995re}
given in Lorentz/de Donder gauge conditions. However, as in the case of
totally symmetric fields, these gauge conditions lead to coupled equations.
Analysis of these coupled equations is complicated and their solution is not
known in closed form so far. On the other hand, promising gauge invariant
approach to mixed-symmetry $AdS$ fields was recently developed in
Ref.\cite{Alkalaev:2003qv}. It would be interesting to generalize our
modified de Donder gauge to the mixed-symmetry fields by using this approach.
This will it make possible to extend our analysis to the case of
mixed-symmetry fields.

\section{ Interrelations between gauge invariant approaches to currents,
shadow fields and massive fields in flat space}\label{sec12}

The gauge invariant description of conformal currents and shadow fields
involves Stueckelberg fields. As is well known, the gauge invariant
description of massive field is also formulated by using Stueckelberg fields.
It is worth mentioning that the number of Stueckelberg fields in the gauge
invariant approach to the spin-$s$ current coincides with the number of
Stueckelberg fields in the gauge invariant approach to the spin-$s$ massive
field. Moreover, there are other interesting interrelations between the gauge
invariant approaches to conformal currents, shadow fields and massive fields.
These interrelations are realized by breaking the conformal symmetries and
can be summarized as follows.
\\
{\bf i})  the gauge transformations of the massive fields can be obtained
from the ones of the conformal currents (or shadow fields) by making the
replacement
\be \label{19052008-05} \Box \rightarrow m^2 \ee
in the gauge transformations of the conformal currents (or shadow fields) and
by making the appropriate re-scaling of the conformal currents (or shadow
fields).
\\
{\bf ii}) Lorentz-like gauge for the massive spin-1 field and de Donder-like
gauge for the massive spin $s\geq 2$ fields can be obtained by making
replacement \rf{19052008-05} in the differential constraints of the conformal
currents (or shadow fields) and by making the appropriate re-scaling of the
conformal currents (or shadow fields).

We note that it is substitution \rf{19052008-05} that breaks the conformal
symmetries. Substitution \rf{19052008-05} is similar to the one used in the
procedure of the standard dimensional reduction from {\it massless} field in
$d+1$ dimensional flat space to {\it massive} field in $d$-dimensional flat
space. Note however that, in our approach, we break the conformal symmetries
of $d$-dimensional space down to the $d$-dimensional Poincar\'e symmetries,
while the standard procedure of dimensional reduction breaks the
$d+1$-dimensional Poincar\'e symmetries down to the $d$-dimensional
Poincar\'e symmetries.

We now demonstrate the interrelations for various spin fields in turn. In due
course we present our de Donder-like gauge for massive spin-$s$, $s > 2$,
fields. To our knowledge this gauge has not been discussed in the earlier
literature.

{\bf Interrelations for spin-1 fields}. In the gauge invariant approach,
massive spin-1 field is described by gauge fields $\phi_m^a$, $\phi_m^{}$
with Lagrangian
\be \label{19052008-01} \LL = -\frac{1}{4}F^{ab}F^{ab} - \half (m
\phi_{\msv}^a +
\partial^a \phi_{\msv})^2\,, \ee
$F^{ab}=\partial^a \phi_{\msv}^b - \partial^b \phi_{\msv}^a$, which is
invariant under the gauge transformations
\be  \label{19052008-02} \delta \phi_{\msv}^a = \partial^a\xi_{\msv}\,,\qquad
\delta \phi_{\msv} = - m \xi_{\msv}\,. \ee
It easy to see that gauge transformations \rf{19052008-02} can be obtained by
substituting
\be  \label{19052008-03} \phi_{cur}^a \rightarrow \phi_{\msv}^a\,, \qquad
\phi_{cur}\rightarrow \frac{1}{m} \phi_{\msv}\,,\qquad \xi_{cur}^{}
\rightarrow \xi_{\msv}^{} \ee
in gauge transformations  of the spin-1 current
\rf{14052008-08},\rf{14052008-09}. Also, it is easy to see that gauge
transformations \rf{19052008-02} can be obtained by substituting
\rf{19052008-05} and
\be  \label{19052008-04} \phi_{sh}^a \rightarrow \phi_{\msv}^a\,,\qquad
\phi_{sh} \rightarrow m \phi_{\msv},\qquad \xi_{sh} \rightarrow  \xi_{\msv}
\ee
in gauge transformations of the spin-1 shadow field \rf{14052008-08sh},
\rf{14052008-09sh}.

We now consider the interrelations between gauge condition for the massive
field spin-1 field and the differential constraints for the current and
shadow field. Let us consider the following well-known Lorentz-like gauge
condition for the massive spin-1 gauge fields and the corresponding
gauge-fixed equations
\beq
&& \label{19052008-06} \partial^a\phi_{\msv}^a + m \phi_{\msv} = 0 \,,
\\
&& \label{19052008-06new} (\Box -  m^2) \phi_{\msv}^a = 0 \,,\eeq
which are invariant under leftover on-shell gauge transformations
\rf{19052008-02} if the gauge transformation parameter satisfies the equation
\be \label{gaufixeqmotspi01} (\Box - m^2)\xi_{\msv} = 0\,.\ee
Note that gauge-fixed equations \rf{19052008-06new} can be obtained from the
appropriate gauge-fixed Lagrangian. Namely, denoting the left hand side of
\rf{19052008-06} by $C_{\msv}$ we obtain the well-known gauge-fixed
Lagrangian
\beq
&& \LL_{total} \equiv \LL - \half C_\msv C_\msv\,,
\\[3pt]
&& \LL_{total} = \frac{1}{2} \phi_{\msv}^a(\Box-m^2)\phi_{\msv}^a\,,
\eeq
which leads to equations \rf{19052008-06new}.

We now note that Lorentz-like gauge condition for massive gauge fields
\rf{19052008-06} can be obtained from differential constraint for the
conformal current \rf{14052008-10} (or shadow field \rf{15052008-02new}) by
making substitutions \rf{19052008-05} \rf{19052008-03}, \rf{19052008-04}.

{\bf Interrelations for spin-2 fields}. In the gauge invariant approach,
massive spin-2 field is described by gauge fields $\phi_\msv^{ab}$,
$\phi_\msv^a$, $\phi_\msv^{}$ with Lagrangian \cite{Zinoviev:2001dt}
\beq \label{19052008-10}
\LL & = & \frac{1}{4} \phi_\msv^{ab} (E_{_{EH}}\phi_{\msv})^{ab} + \half
\phi_\msv^a (E_{_{Max}} \phi_\msv)^a + \half \phi_\msv^{}\Box\phi_\msv^{}
\nonumber\\[3pt]
& + & m \phi_\msv^a (\partial^b \phi_\msv^{ba}  - \partial^a \phi_\msv^{bb} -
\sigg\partial^a \phi_\msv^{})
\nonumber\\[3pt]
& - & \frac{m^2}{4} \phi_\msv^{ab} \phi_\msv^{ab} +  \frac{m^2}{4}
\phi_\msv^{aa} \phi_\msv^{bb}
\nonumber\\[3pt]
& + & \frac{\sigg m^2}{2}\phi_m^{}\phi_m^{aa} + \frac{dm^2}{2(d-2)}\phi_m^2
\,,
\eeq
where the respective second-derivative Einstein-Hilbert and Maxwell operators
$E_{_{EH}}$, $E_{_{Max}}$ are given by
\beq
&& (E_{_{EH}}\phi)^{ab} = \Box \phi^{ab} -\partial^a\partial^c\phi^{cb} -
\partial^b\partial^c\phi^{ca}  + \partial^a \partial^b \phi^{cc}
\nonumber\\
&& \hspace{1.52cm} + \ \eta^{ab}(\partial^c\partial^e\phi^{ce} - \Box
\phi^{cc}) \,,
\\[5pt]
&& (E_{_{Max}}\phi)^a = \Box \phi^a  -\partial^a \partial^b\phi^b \,,
\eeq
and $\sigg$ is defined in \rf{080423-02(add)}. Lagrangian \rf{19052008-10} is
invariant under the gauge transformations
\beq
\label{19052008-11} && \hspace{-0.7cm} \delta \phi_{\msv}^{ab} =\partial^a
\xi_{\msv}^b +
\partial^b \xi_{\msv}^a
 + \frac{2m }{d-2} \eta^{ab} \xi_{\msv}\,,
\\[5pt]
\label{19052008-12} && \hspace{-0.7cm} \delta \phi_{\msv}^a = \partial^a
\xi_{\msv} - m \xi_{\msv}^a \,,
\\[5pt]
\label{19052008-13} && \hspace{-0.7cm} \delta \phi_{\msv} = -\sigg m
\xi_\msv\,.
\eeq
It is easy to see that these transformations can be obtained by making
substitutions \rf{19052008-05} and
\beq
& \label{19052008-14} \phi_{cur}^{ab} \rightarrow \phi_{\msv}^{ab}\,, \quad
\phi_{cur}^a \rightarrow \frac{1}{m} \phi_{\msv}^a \,,\quad \phi_{cur}
\rightarrow \frac{1}{m^2} \phi_{\msv} \,,           & \qquad \ \
\\[3pt]
& \xi_{cur}^a \rightarrow \xi_{\msv}^a \,, \quad \xi_{cur} \rightarrow
\frac{1}{m} \xi_{\msv} \,,               &
\eeq
\beq
&  \label{19052008-15} \phi_{sh}^{ab} \rightarrow  \phi_{\msv}^{ab} \,,\quad
\phi_{sh}^a \rightarrow m \phi_{\msv}^a \,,\quad \phi_{sh} \rightarrow m^2
\phi_{\msv} \,,       & \qquad \ \
\\[3pt]
& \xi_{sh}^a \rightarrow  \xi_{\msv}^a \,,\qquad  \xi_{sh} \rightarrow m
\xi_{\msv} \,, &
\eeq
in gauge transformations of the current \rf{14052008-05}-\rf{14052008-07} (or
shadow field \rf{14052008-11}-\rf{14052008-13}).

Now let us consider interrelations between gauge conditions for the massive
gauge fields and the differential constraints for the current and shadow
field. We find the following de Donder-like gauge condition for the massive
gauge fields:
\beq
\label{19052008-16} && \hspace{-1cm} \partial^b \phi_{\msv}^{ab} - \half
\partial^a \phi_{\msv}^{bb} + m \phi_{\msv}^a = 0 \,,
\\[5pt]
\label{19052008-17} && \hspace{-1cm} \partial^a \phi_{\msv}^a + \half m
\phi_{\msv}^{aa} + \sigg m \phi_{\msv}^{} = 0 \,.
\eeq
The surprise is that gauge condition \rf{19052008-16},\rf{19052008-17} leads
to the {\it decoupled} equations of motion for the massive gauge fields,
\beq
& \label{19052008-19} (\Box - m^2)\phi_{\msv}^{ab} = 0, \qquad  (\Box -
m^2)\phi_{\msv}^a = 0,
\nonumber\\[3pt]
&(\Box - m^2)\phi_{\msv}^{} = 0\,. &
\eeq
The gauge condition and equations of motion are invariant under leftover
on-shell gauge transformations \rf{19052008-11}-\rf{19052008-13}, where the
gauge transformation parameters satisfy the equations
\be  \label{19052008-20} (\Box - m^2)\xi_{\msv}^a = 0\,,\qquad  (\Box -
m^2)\xi_{\msv}^{} = 0\,.\ee

Note that gauge-fixed equations \rf{19052008-19} can be obtained from the
appropriate gauge-fixed Lagrangian. Namely, if we denote the respective left
hand sides of \rf{19052008-16} and \rf{19052008-17} by $C_{\msv}^a$ and
$C_{\msv}$, and define the gauge-fixed Lagrangian as
\be \LL_{total} = \LL - \half C_\msv^a C_\msv^a - \half C_\msv C_\msv\,, \ee
then we get the surprisingly simple gauge-fixed Lagrangian:
\beq \label{gaufixlag01}
\LL_{total} & = & \frac{1}{4} \phi_{\msv}^{ab}(\Box-m^2)\phi_{\msv}^{ab} -
\frac{1}{8} \phi_{\msv}^{aa} (\Box-m^2) \phi_{\msv}^{bb}
\nonumber\\
&  + &\half \phi_{\msv}^a (\Box-m^2) \phi_{\msv}^a + \half \phi_{\msv}
(\Box-m^2) \phi_{\msv}^{}, \qquad \ \ \ \ \  \eeq
which leads to equations \rf{19052008-19}. To our knowledge, for $d>4$, gauge
condition \rf{19052008-16}, \rf{19052008-17} and Lagrangian \rf{gaufixlag01}
have not been discussed in the earlier literature.

We now note that de Donder-like gauge condition for the massive gauge fields
\rf{19052008-16},\rf{19052008-17} can simply be obtained by making
substitutions \rf{19052008-05},\rf{19052008-14},\rf{19052008-15} in
differential constraints for the current \rf{080423-01},\rf{080423-02} (or
shadow field \rf{05152008-10new},\rf{05152008-11new}).

Also, we note that gauge invariant field $T_{cur}^{ab}$ \rf{19052008-30} (or
$T_{sh}^{ab}$ \rf{06092008-12}) can be related with the Pauli-Fierz field
entering spin-2 massive field theory. Thus, in the gauge invariant approach,
the Pauli-Fierz field has the following representation in terms of
the massive gauge fields:%
\footnote{ For $d=4$, formula \rf{19052008-31} was given in
Ref.\cite{Schwinger:1970xc}.}
\beq \label{19052008-31}
\Phi_{\scriptscriptstyle{PF}}^{ab} & = & \phi_{\msv}^{ab} + \frac{1}{m}
(\partial^a \phi_{\msv}^b +
\partial^b \phi_{\msv}^a)
\\[3pt]
& + & \frac{2}{\sigg m^2} \partial^a \partial^b \phi_{\msv} +
\frac{2}{(d-2)\sigg} \eta^{ab} \phi_{\msv}^{}\,,
\nonumber
\eeq
where $\sigg$ is given in \rf{080423-02(add)}. One can make sure that:
\\
{\bf i}) $\Phi_{\scriptscriptstyle{PF}}^{ab}$ is invariant under gauge
transformations \rf{19052008-11}-\rf{19052008-13};
\\
{\bf ii}) inserting the field $\Phi_{\scriptscriptstyle{PF}}^{ab}$ into the
Pauli-Fierz Lagrangian for the massive spin-2 field
\be \LL = \frac{1}{4} \Phi_{\scriptscriptstyle{PF}}^{ab}
(E_{_{EH}}\Phi_{\scriptscriptstyle{PF}})^{ab} -
\frac{m^2}{4}(\Phi_{\scriptscriptstyle{PF}}^{ab}\Phi_{\scriptscriptstyle{PF}}^{ab}
- \Phi_{\scriptscriptstyle{PF}}^{aa} \Phi_{\scriptscriptstyle{PF}}^{bb})\,,
\qquad \ \ee
gives gauge invariant Lagrangian \rf{19052008-10};
\\
{\bf iii}) $\Phi_{\scriptscriptstyle{PF}}^{ab}$ given in \rf{19052008-31} can
simply be obtained by making substitutions \rf{19052008-05},\rf{19052008-14}
in field $T_{cur}^{ab}$ \rf{19052008-30} (or by making substitutions
\rf{19052008-05},\rf{19052008-15} in field $T_{sh}^{ab}$ \rf{06092008-12}).

{\bf Interrelations for arbitrary spin fields}. We begin with presentation of
gauge invariant Lagrangian for massive spin-$s$ field in $d$-dimensional flat
space. Gauge fields entering the gauge invariant Lagrangian can be collected
in a ket-vector
\beq
\label{phimsvdef01} && \hspace{-0.7cm}  |\phi_{\msv}\rangle \equiv
\sum_{s'=0}^s \alpha_z^{s-s'}|\phi_{\msv,\, s'}\rangle \,,
\\[5pt]
&& \hspace{-0.7cm} |\phi_{\msv,\,s'}\rangle \equiv
\frac{\alpha^{a_1} \ldots \alpha^{a_{s'}}}{s'!\sqrt{(s - s')!}}
\, \phi_{\msv,\,s'}^{a_1\ldots a_{s'}} |0\rangle\,. \hspace{1.2cm}
\eeq

In terms of ket-vector \rf{phimsvdef01}, Lagrangian of the massive gauge
fields takes the form%
\footnote{ In terms of the tensor fields $\phi_{s'}^{a_1\ldots a_{s'}}$,
Lagrangian \rf{masspi2lag01} was found in \cite{Zinoviev:2001dt}.}
\be \label{masspi2lag01} \LL = \frac{1}{2} \langle \phi_{\msv}| E |
\phi_{\msv}\rangle\,, \ee
where operator $E$ is given by
\beq
&& \label{masFrosecordope01} E  =  E_\smtwo + E_\smone + E_\smzero\,,
\\[3pt]
&& E_\smtwo \equiv \Box - \alpha\partial\bar\alpha\partial +
\frac{1}{2}(\alpha\partial)^2\bar\alpha^2 + \frac{1}{2} \alpha^2
(\bar\alpha\partial)^2
\nonumber\\[3pt]
&& \qquad \ \ \ \ - \frac{1}{2}\alpha^2 \Box \bar\alpha^2
-\frac{1}{4}\alpha^2\alpha\partial\,\bar\alpha\partial\bar\alpha^2\,,
\\[3pt]
&& E_\smone \equiv  \eb_{1m} \AA  + e_{1m} \bar\AA \,,
\\[3pt]
&& E_\smzero \equiv m_1 + \alpha^2\bar\alpha^2m_2 + \mb_3 \alpha^2 + m_3
\bar\alpha^2\,,
\eeq

\beq
&& \AA \equiv  \alpar - \alpha^2 \albpar + \frac{1}{4}\alpha^2\alpar
\bar\alpha^2 \,,
\\[3pt]
&& \bar\AA \equiv \albpar -  \alpar\bar\alpha^2 + \frac{1}{4}\alpha^2 \albpar
\bar\alpha^2 \,,
\\[3pt]
&& e_{1m} = m \alpha^z \ewt_1 \,,
\qquad
\eb_{1m} =  -m \ewt_1 \bar\alpha^z  \,,
\\[5pt]
&& m_1 = \frac{2s+d-2-N_z}{2s+d-2-2N_z}(N_z-1) m^2 \,,
\\[5pt]
&& m_2 = \frac{2(2s+d-2) + (2s+d - 7)N_z - N_z^2}{4(2s+d-2-2N_z)} m^2 \,,
\nonumber\\
&&
\\
&& m_3 =  \half  e_{1m} e_{1m}\,,
\qquad
\mb_3  = \half \eb_{1m} \eb_{1m}\,.
\eeq
The Lagrangian is invariant under the gauge transformation
\be \label{masgautraarbspi01}
\delta |\phi_{\msv}\rangle  =  ( \alpar - e_{1m} -\frac{\alpha^2}{2s + d- 6
-2N_z}\eb_{1m} ) |\xi_{\msv}\rangle\,,
\ee
where the ket-vector of gauge transformation parameter $|\xi_{\msv}\rangle$
is represented in terms of $so(d-1,1)$ algebra tensor fields as
\beq
&& \hspace{-0.9cm} |\xi_{\msv}\rangle \equiv \sum_{s'=0}^{s-1}
\alpha_z^{s-1-s'}|\xi_{\msv,\,s'}\rangle \,,
\\[5pt]
&&\hspace{-0.9cm}  |\xi_{\msv,\,s'}\rangle \equiv
\frac{\alpha^{a_1} \ldots \alpha^{a_{s'}}}{s'!\sqrt{(s -1 - s')!}}
\xi_{\msv,\,s'}^{a_1\ldots a_{s'}} |0\rangle.
\eeq
The ket-vectors $|\phi_{\msv}\rangle$ and $|\xi_{\msv}\rangle$ satisfy the
respective double-tracelessness and tracelessness constraints
\be (\bar\alpha^2)^2|\phi_{\msv}\rangle =0 \,,\qquad
\bar\alpha^2|\xi_{\msv}\rangle =0 \,.\ee

Now let us consider the interrelations between the gauge invariant approaches
to massive field, conformal current, and shadow field. We begin with the
comparison of the gauge transformations.

Is is easy to see that gauge transformation \rf{masgautraarbspi01} can simply
be obtained by making substitutions \rf{19052008-05} and
\beq \label{19052008-21} & |\phi_{cur}\rangle \rightarrow m^{-N_z}
|\phi_{\msv}\rangle \,,\qquad |\xi_{cur}\rangle \rightarrow m^{-N_z}
|\xi_{\msv}^{}\rangle \,, & \qquad \ \ \
\\[5pt]
& \label{19052008-22} |\phi_{sh}\rangle \rightarrow m^{N_z}
|\phi_{\msv}\rangle \,,\qquad |\xi_{sh}\rangle \rightarrow m^{N_z}
|\xi_{\msv}^{}\rangle \,, & \eeq
in gauge transformation of the conformal current \rf{gautraarbspi01} (or
shadow field \rf{gautraarbspi01sh}).

We now proceed with the comparison of de Donder-like gauge for the massive
gauge fields and the differential constraints for the currents and shadow
fields. We find the following de Donder-like gauge condition for the massive
arbitrary spin-$s$ field
\beq
\label{19052008-23} && \bar{C}_m |\phi_\msv^{}\rangle = 0\,,
\\
\label{10052008-01} && \bar{C}_m \equiv  \albpar - \half \alpar \bar\alpha^2
+ \half e_{1m} \bar\alpha^2 - \eb_{1m}\Pi^\smponetwo\,. \qquad \ \ \
\eeq
We note that gauge condition \rf{19052008-23} leads to the decoupled
gauge-fixed equations of motion for the massive gauge fields
\be \label{19052008-27} (\Box - m^2)|\phi_{\msv}^{}\rangle =0\,.\ee
These gauge-fixed equations of motion and gauge condition \rf{19052008-23}
are invariant under leftover on-shell gauge transformations
\rf{masgautraarbspi01} if the gauge transformation parameter satisfies the
equation
\be  \label{19052008-28} (\Box - m^2)|\xi_{\msv}^{}\rangle = 0\,.\ee

Note that gauge-fixed equations \rf{19052008-27} can be obtained from the
appropriate gauge-fixed Lagrangian. Namely, if we define the gauge-fixed
Lagrangian as
\be \LL_{total} = \LL +  \half \langle \phi_m^{} |  C_\msv \bar{C}_\msv
|\phi_\msv^{}\rangle\,, \ee
where $\bar{C}_m$ is given in \rf{10052008-01}, while $C_m$ is defined by
\be \label{10052008-02} C_m \equiv \alpar - \half \alpha^2 \albpar + \half
\eb_{1m} \alpha^2 - e_{1m} \Pi^\smponetwo \,,
\ee
then we get the surprisingly simple gauge-fixed Lagrangian:
\be \label{gaufixlag02}
\LL_{total} =  \half \langle
\phi_{\msv}^{}|(1-\frac{1}{4}\alpha^2\bar\alpha^2) (\Box-m^2)
|\phi_{\msv}^{}\rangle\,, \ee
which leads to equations \rf{19052008-27}. To our knowledge, de Donder-like
gauge condition \rf{19052008-23} and gauge-fixed Lagrangian \rf{gaufixlag02}
have not been discussed in the earlier literature.

We now note that de Donder-like gauge for the massive gauge fields
\rf{19052008-23} can simply be obtained by making substitutions
\rf{19052008-05}, \rf{19052008-21} and \rf{19052008-22} in differential
constraints for the currents \rf{17052008-05} (or shadow fields
\rf{17052008-05sh}).

To summarize, we have obtained the gauge transformations and de Donder-like
gauges of the massive fields from the gauge transformations and the
differential constraints of the conformal currents (or shadow fields). It is
clear that we can formally inverse our substitutions, i.e., we can obtain the
gauge transformations and the differential constraints of the conformal
currents (or shadow fields) from the gauge transformations and de Donder-like
gauge of the massive gauge fields by using formally the inverse substitution,
i.e., first, by making the appropriate re-scaling of the massive gauge fields
and then making the substitution $m^2\rightarrow \Box$. By now, in the
literature, there are various approaches to gauge invariant formulations of
massive fields. Obviously, use of the just mentioned interrelations between
conformal currents (shadow fields) and massive fields might be helpful for
straightforward generalization of those approaches to the case of conformal
currents and shadow fields.

\begin{acknowledgments}
This work was supported by the RFBR Grant No.08-02-01118, RFBR Grant for
Leading Scientific Schools, Grant No. 1615.2008.2, by the Dynasty Foundation
and by the Alexander von Humboldt Foundation Grant PHYS0167.
\end{acknowledgments}

\appendix
\section{ Restrictions imposed by gauge invariance and by dilatation symmetry}
\label{appen01}

Under the dilatation transformations, the currents and shadows transform as
$\delta_D \phicurk = D_{cur}\phicurk$, $\delta_D \phishk = D_{sh}\phishk$,
where $D_{cur}$, $D_{sh}$ are given by
\beq
\label{03092008-02} && D_{cur} = x\partial + \Delta_{cur}\,, \qquad
\Delta_{cur} = \Delta_{0\,cur} - N_z\,, \qquad
\\
\label{03092008-03} && D_{sh} = x\partial + \Delta_{sh}\,, \qquad \ \ \
\Delta_{sh} =\Delta_{0\,sh} + N_z\,,
\eeq
where $\Delta_{0\,cur}$, $\Delta_{0\,sh}$ are constants. We now demonstrate
that the two-point current-shadow field interaction vertex
\be  \label{03092008-04} \LL =\langle \phi_{cur}| \mubf \phishk \,, \ee
is invariant under the dilatation transformations provided $\mubf$ takes the
following form:
\beq
\label{03092008-05} & \mubf = 1 + g_1\alpha^2 \bar\alpha^2 + g_2 \Box
\bar\alpha^2 \,, &
\\[3pt]
\label{03092008-06} & g_2 = \alpha^z\alpha^z \tilde{g}_2\,,& \eeq
where $g_1$, $\tilde{g}_2$ depend only on $N_z$. To this end we start with
the general expression for $\mubf$,
\beq  \label{03092008-07} && \mubf = 1 + g_1\alpha^2 \bar\alpha^2 + g_2'
\bar\alpha^2 + g_3'\alpha^2\,,
\\[3pt]
\label{03092008-08} && g_2' = \alpha^z\alpha^z g_2''\,,  \qquad g_3' = g_3''
\bar\alpha^z\bar\alpha^z \,, \eeq
where $g_1$, $g_2''$, $g_3''$ depend only on $N_z$ and $\Box$. Requiring
vertex $\LL$ \rf{03092008-04} to be invariant under the dilatation
transformation, $\delta_D\LL = 0$ (up to total derivative), gives the
equation
\be  \label{03092008-09} D_{cur}^\dagger \mubf + \mubf D_{sh} =0\,, \ee
which amounts to the following equations:
\beq
\label{03092008-01} && [x\partial + N_z ,\mubf]=0\,,
\\[3pt]
\label{03092008-10} && \Delta_{0\,cur} + \Delta_{0\,sh} = d\,.  \eeq
It is easily seen that solution to equation \rf{03092008-01} is given by
\be  \label{03092008-11} g_2'' = \Box \tilde{g}_2\,, \qquad g_3''=0\,,\ee
where $\tilde{g}_2$ depends only on $N_z$. Plugging this solution in
\rf{03092008-07} we see that $\mubf$ takes the form given in
\rf{03092008-05},\rf{03092008-06}.

We now find the restrictions imposed on the gauge transformations of
$\phicurk$ and $\phishk$ by the dilatation symmetry.  We are going to
demonstrate that the dilatation symmetry leads to the following gauge
transformations of the currents and shadows:
\beq
\label{03092008-14} && \delta \phicurk =  G_{cur}\xicurk\,,
\\
\label{03092008-15} &&\delta \phishk = G_{sh} \xishk \,,
\\[7pt]
\label{03092008-16}&& \qquad G_{cur} = \alpar + b_{1\,cur} +
b_{2\,cur}\alpha^2 \Box \,,
\\[3pt]
\label{03092008-17} && \qquad G_{sh} = \alpar + b_{1\,sh} \Box +
b_{2\,sh}\alpha^2 \,,
\\[7pt]
\label{08092008-01} && \qquad b_{1\,cur}= \alpha^z \tilde{b}_{1\,cur}\,,\quad
b_{2\,cur}= \tilde{b}_{2\,cur} \bar\alpha^z\,, \qquad
\\
\label{08092008-02} && \qquad  b_{1\,sh}= \alpha^z \tilde{b}_{1\,sh}\,,\quad
b_{2\,sh}= \tilde{b}_{2\,sh} \bar\alpha^z\,,
\eeq
where $\tilde{b}_{1\,cur}$, $\tilde{b}_{2\,cur}$, $\tilde{b}_{1\,sh}$,
$\tilde{b}_{2\,sh}$ depend only on $N_z$. To this end we note, that under the
dilatation transformations the gauge transformation parameters $\xicurk$ and
$\xishk$ transform as $\delta_D \xicurk = D_{\xi_{cur}}\xicurk$, $\delta_D
\xishk = D_{\xi_{sh}}\xishk$, where $D_{\xi_{cur}}$, $ D_{\xi_{sh}}$ are
given by
\be
\label{03092008-12}  D_{\xi_{cur}} = D_{cur} -1 \,, \qquad
D_{\xi_{sh}} = D_{sh}-1\,,
\ee
and $D_{cur}$, $D_{sh}$ are defined in \rf{03092008-02},\rf{03092008-03}. To
avoid the repetition we restrict our attention to the gauge transformation of
the current. We note that the general form of gauge transformation operator
$G_{cur}$ \rf{03092008-14} is given by
\be
\label{03092008-18} G_{cur} = \alpar + b_1' + b_2'\alpha^2 \,,
\ee
where $b_1'$, $b_2'$ depend on $\alpha^z$, $\bar\alpha^z$, and $\Box$.
Requiring the gauge symmetry to respect the dilatation transformation gives
the equation
\be \label{03092008-19}   D_{cur} G_{cur}  = G_{cur} D_{\xi_{cur}}\,.\ee
Plugging $G_{cur}$ \rf{03092008-18} in \rf{03092008-19} and using
\rf{03092008-02},\rf{03092008-12} we see that Eq.\rf{03092008-19} leads to
the following solution for $b_1'$, $b_2'$:
\be b_1' = \alpha^z \tilde{b}_{1\,cur} \qquad b_2' = \Box \tilde{b}_{2\,cur}
\bar\alpha^z \,,\ee
where $\tilde{b}_{1\,cur}$, $\tilde{b}_{2\,cur}$ depend only on $N_z$, i.e.,
we arrive at $G_{cur}$ given in \rf{03092008-16},\rf{08092008-01}.

In a quite similar way, one can obtain representation for $G_{sh}$ given in
\rf{03092008-17}, \rf{08092008-02}.

We now demonstrate that requiring the vertex $\LL$ to be invariant under
gauge transformations \rf{03092008-14},\rf{03092008-15} leads to the
following results:
\\
{\bf i}) the operators $\bar{C}_{cur}$, $\bar{C}_{sh}$ take form:
\beq
\bar{C}_{cur} & = &  \albpar- \half \alpar \bar\alpha^2
+ c_{1\,cur} \bar\alpha^2 + c_{2\,cur} \Box \Pi^\smponetwo\,, \qquad \ \
\\
\label{04092008-10} \bar{C}_{sh} & = &  \albpar- \half \alpar \bar\alpha^2
+ c_{1\,sh} \Box\bar\alpha^2 + c_{2\,sh} \Pi^\smponetwo\,;
\eeq
\\
{\bf ii}) the $c$-operators and $b$-operators are related as
\beq
\label{04092008-11x} && c_{1\,cur}  = - \half b_{1\,cur}\,,
\\[3pt]
\label{04092008-11xx} && c_{2\,sh} = (2s + d-6-2N_z) b_{2\,cur}\,,
\\[3pt]
\label{04092008-11} && c_{1\,cur} =  \half b_{2\,sh}^\dagger (2s+d
-6-2N_z)\,,
\\
\label{04092008-12} && c_{2\,cur}= - b_{1\,sh}^\dagger\,,
\\[5pt]
\label{08092008-10} && c_{1\,sh}  = - \half b_{1\,sh}\,,
\\[3pt]
\label{08092008-11} && c_{2\,sh} = (2s + d-6-2N_z) b_{2\,sh}\,.
\\[3pt]
\label{04092008-13} && c_{1\,sh} =  \half b_{2\,cur}^\dagger (2s+d
-6-2N_z)\,,
\\
\label{04092008-14} && c_{2\,sh}= - b_{1\,cur}^\dagger\,,
\eeq
i.e., the $c$-operators are represented similarly to the $b$-operators (see
\rf{08092008-01},\rf{08092008-02})
\beq
\label{08092008-12} && c_{1\,cur}= \alpha^z \cwt_{1\,cur}\,,\quad c_{2\,cur}=
\cwt_{2\,cur} \bar\alpha^z\,,
\\
\label{08092008-13} && c_{1\,sh}= \alpha^z \cwt_{1\,sh}\,,\qquad c_{2\,sh}=
\cwt_{2\,sh} \bar\alpha^z\,,
\eeq
where $\cwt$-operators depend only on $N_z$.
\\
{\bf iii}) the $\cwt$-operators satisfy the relations:
\be \label{09092008-01} \cwt_{1\,cur}\cwt_{2\,cur} = \half \ewt_1{}^2 \,,
\qquad \cwt_{1\,sh}\cwt_{2\,sh} = \half \ewt_1{}^2 \,,\ee
where $\ewt_1$ is defined in \rf{masewtdef01}.
\\
{\bf iv}) $g_1$ and $g_2$ are determined to be
\be \label{04092008-08} g_1 = -\frac{1}{4}\,, \qquad g_2 = 0 \,.\ee

Before to prove these results we note that the methods for finding the
operators $\bar{C}_{cur}$ and $\bar{C}_{sh}$ are quite similar. Therefore to
avoid the repetition we present details of the derivations of the operator
$\bar{C}_{sh}$.

To find the restrictions imposed on $\bar{C}_{sh}$ by requiring that $\LL$ be
invariant under the gauge transformation of $\phicurk$ we note the relation
(up to total derivative)
\beq
&& -\langle G_{cur} \xi_{cur}|\mubf \phishk
\\
&& =  \langle\xi_{cur}|\Bigl( \albpar + 2g_1 \alpar \bar\alpha^2 + g_2 \Box
\albpar \bar\alpha^2 -  b_{1\,cur}^\dagger
\nonumber \\
&& - (b_{1\,cur}^\dagger g_2 + b_{2\,cur}^\dagger +2b_{2\,cur}^\dagger
g_1(2N_\alpha+d))\Box\bar\alpha^2\Bigr)\phishk\,,
\nonumber
\eeq
which implies that the requirement of invariance of $\LL$ under the gauge
transformation of $\phicurk$,
\be \langle G_{cur} \xi_{cur}|\mubf \phishk = 0\,, \ee
leads to the constraint
\be\label{04092008-06} \bar{C}_{sh}\phishk = 0\,, \ee
with the following $\bar{C}_{sh}$:
\beq
\label{04092008-09} \bar{C}_{sh} & = & \Pi^\smponetwo ( \albpar + 2g_1 \alpar
\bar\alpha^2) + g_2 \Box \albpar \bar\alpha^2
\nonumber\\[5pt]
& - &  (b_{1\,cur}^\dagger g_2 + b_{2\,cur}^\dagger + 2 b_{2\,cur}^\dagger
g_1(2N_\alpha+d))\Box\bar\alpha^2 \qquad \
\nonumber\\[3pt]
& - &  b_{1\,cur}^\dagger\Pi^\smponetwo\,.
\eeq
We now find the restrictions on $\bar{C}_{sh}$ which are obtained by
requiring that constraint \rf{04092008-06} be invariant under the gauge
transformation of $\phishk$, i.e. we consider the equation
\be \label{04092008-07} \bar{C}_{sh} G_{sh} \xishk =0\,, \ee
where $G_{sh}$ and $\bar{C}_{sh}$ are given in \rf{03092008-17} and
\rf{04092008-09} respectively. Before studying all restrictions on
$\bar{C}_{sh}$ which are obtainable from \rf{04092008-07} we note that the
requirement for cancellation of $\alpar\albpar$- and $(\albpar)^2$-terms in
\rf{04092008-07} leads to $g_1$, $g_2$ given in \rf{04092008-08}. Plugging
these $g_1$, $g_2$ in \rf{04092008-09}, we obtain \rf{04092008-10} with
$c_{1\,sh}$, $c_{2\,sh}$ given in \rf{04092008-13},\rf{04092008-14}. Now we
are ready to find all restrictions on $\bar{C}_{sh}$ which are obtainable
from \rf{04092008-07}. Thus, using \rf{04092008-10} we represent the left
hand side of \rf{04092008-07} as
\beq
\label{10092008-01} \bar{C}_{sh} G_{sh}\xishk & =& ( \Box  X_1
+ \Box \albpar X_2
+  C X_3)\xishk\,,\qquad \ \
\eeq
\beq
C &\equiv & \alpar - \alpha^2 \frac{1}{2N_\alpha +d} \albpar \,,
\\[3pt]
X_1 &  \equiv &  1 + c_{2\,sh} b_{1\,sh}
\nonumber\\[3pt]
& + & 2(2s + d - 2 - 2N_z ) c_{1\,sh} b_{2\,sh}\,,
\\[5pt]
X_2 &  \equiv &  b_{1\,sh} +  2c_{1\,sh}\,,
\\[5pt]
X_3 &  \equiv &  c_{2\,sh} - (2s  + d - 6 - 2N_z )b_{2\,sh}\,.
\eeq
From \rf{10092008-01}, we see that Eq.\rf{04092008-07} amounts to the
equations $X_i\xishk=0$, $i=1,2,3$. Solution to equations $X_2\xishk = 0$,
$X_3\xishk=0$ is given by \rf{08092008-10},\rf{08092008-11}. Making use of
\rf{08092008-10},\rf{08092008-11} in Eq. $X_1\xishk =  0$ gives the equation
\be \label{08092008-14} \Bigl(c_{2\,sh} c_{1\,sh} -
\frac{2s+d-2-2N_z}{2s+d-4-2N_z} c_{1\,sh} c_{2\,sh} - \half\Bigr)\xishk= 0
\,.\ee
Using representation for the $c$-operators given in \rf{08092008-13} we find
that Eq.\rf{08092008-14} allows us to determine the quantity
$\cwt_{1\,sh}\cwt_{2\,sh}$ uniquely. The result is given in \rf{09092008-01}.

We finish the discussion in this Appendix by making remark on the similarity
transformation of the currents and shadows. As we have demonstrated,
requiring the differential constraints for the currents and shadows to be
invariant under the gauge transformations gives unique solution for the
products $\cwt_{1\,cur}\cwt_{2\,cur}$, $\cwt_{1\,sh}\cwt_{2\,sh}$
\rf{09092008-01}. From \rf{04092008-11x},\rf{04092008-14} and
\rf{04092008-12},\rf{08092008-10}, it is seen that the $\cwt$-operators are
related as
\be  \label{10092008-15} \cwt_{1,cur} = \half \cwt_{2,sh}\,,\qquad
\cwt_{1,sh} = \half \cwt_{2\,cur} \,. \ee
It turns out that there are no additional restrictions on the
$\cwt$-operators. This implies, there is arbitrariness in the choice of the
$\cwt$-operators. We note that this arbitrariness is related with the
similarity transformation of the currents and shadows,
\be \label{10092008-14} \phicurk \rightarrow U\phicurk \,,\qquad \phishk
\rightarrow U^{-1} \phishk \,,\ee
where $U$ is an arbitrary function of $N_z$ with the restriction that $U$ is
not equal to zero for allowed eigenvalues of $N_z$ equal to $0,1,\ldots,s$.
It is seen that transformation \rf{10092008-14} leaves the vertex $\LL$
invariant, but changes the $\cwt$-operators. Using this transformation one of
the $\cwt$-operators can be made arbitrary function of $N_z$ with the
restriction that this function is not equal to zero for allowed eigenvalues
of $N_z$ equal to $0,1,\ldots,s$. The remaining $\cwt$-operators are then
determined uniquely by relations \rf{09092008-01} and \rf{10092008-15}. In
this paper we use the following choice of the $\cwt$-operators:
\beq \label{11092008-02}
&& \cwt_{1\,cur} = \half \ewt_1\,, \qquad \cwt_{2\,cur} = \ewt_1 \,,
\\
\label{11092008-02xxx} && \cwt_{1\,sh} = \half \ewt_1\,, \qquad \ \ \
\cwt_{2\,sh} = \ewt_1 \,. \eeq
This choice turns out to be convenient for the study of $AdS/CFT$
correspondence.

\section{ Restrictions imposed by conformal boost symmetries}
\label{appen02}

In this Appendix, we use the notation $R_{cur}^a$ and $R_{sh}^a$ to indicate
the respective realizations of operator $R^a$ on the current $\phicurk$ and
the shadow field $\phishk$. Because the methods for finding the operators
$R_{cur}^a$ and $R_{sh}^a$ are quite similar we present details of the
derivation of the operator $R_{sh}^a$ and outline procedure of the derivation
of the operator $R_{cur}^a$.

We find the operator $R_{sh}^a$ by requiring that:
\\
{\bf i}) the differential constraint for $\phishk$ be invariant under
conformal boost transformations;
\\
{\bf ii}) the operator $R_{sh}^a$ be independent of the derivatives
$\partial^a$.

Before to analyze restrictions imposed on $R_{sh}^a$ by the conformal boost
symmetries we find general expression for the operator $R_{sh}^a$ that
respects the dilatation symmetry and algebraic constraint \rf{09092008-04}.
Requiring that the operator $R^a$ respects algebraic constraint
\rf{09092008-04} and the commutation relation $[D,K^a]=K^a$ gives
\beq
\label{09092008-05} && [N_\alpha + N_z ,R_{sh}^a] = 0
\\[3pt]
\label{09092008-06} && [\Delta_{sh}, R_{sh}^a] = R_{sh}^a\,.
\eeq
To derive \rf{09092008-06} we take into account that the operator $R_{sh}^a$
is independent of the space coordinates $x^a$ because of commutator
\rf{pkjj}. Also, we use our assumption that $R_{sh}^a$ is independent of the
derivatives $\partial^a$. Taking into account expression for $\Delta_{sh}$ in
\rf{03092008-03} it is easy to see that \rf{09092008-05},\rf{09092008-06}
amount to the commutators
\be
\label{09092008-07}  [N_\alpha, R_{sh}^a] = -R_{sh}^a\,,
\qquad
[N_z, R_{sh}^a] = R_{sh}^a\,.
\ee
General solution to \rf{09092008-07} is obvious:
\beq \label{09092008-10}
& R_{sh}^a = r_{0,1,sh}\bar\alpha^a + r_{0,2,sh} \alpha^a\bar\alpha^2 +
r_{0,3,sh} \alpha^2\bar\alpha^a\bar\alpha^2\,, & \qquad
\\[3pt]
& r_{0,k,sh} = \alpha^z \rwt_{0,k,sh}\,, \qquad  k=1,2,3, &
\eeq
where the operators $\rwt_{0,k,sh}$ depend only on $N_z$. Note that to derive
\rf{09092008-10} we take into account constraint \rf{09092008-11} which tells
us that the contribution of $(\bar\alpha^2)^n$-terms to $R_{sh}^a$ is
irrelevant when $n\geq 2$.

We now consider restrictions imposed on $R_{sh}^a$ by the conformal boost
symmetries. Consider the differential constraint for the shadow filed
$\phishk$,
\be \label{09092008-12} \bar{C}_{sh} \phishk = 0\,, \ee
where $\bar{C}_{sh}$ is given in \rf{04092008-10}. Requiring this constraint
to be invariant under the conformal boost transformations gives the equations
\be \label{09092008-03}\bar{C}_{sh} K^a \phishk = 0\,, \ee
where the conformal boost operator $K^a$ takes the form given in
\rf{conalggenlis04}, $K^a = K_{\Delta_{sh},M}^a +R_{sh}^a$. To analyze
equations \rf{09092008-03} we note the following helpful formulas:
\be \label{09092008-14}
[\bar{C}_{sh},K_{\Delta_{sh},M}^a] = x^a \bar{C}_{sh} + \bar{C}_{sh\smzero}^a
+ \bar{C}_{sh\smone}^a\,,
\ee

\beq
\label{09092008-16} && \hspace{-0.7cm} \bar{C}_{sh\smzero}^a \equiv
(\Delta_{sh} - N_\alpha - d +1) \bar{C}_\perp^a
\nonumber\\
&&\qquad \quad - \half (2N_\alpha + d -4) C^a \bar\alpha^2\,,
\\[5pt]
\label{09092008-17} && \hspace{-0.7cm}\bar{C}_{sh\smone}^a \equiv
(2\Delta_{sh} -d)c_1 \bar\alpha^2
\partial^a + 2c_1 M^{ab}\partial^b \bar\alpha^2\,,
\\[5pt]
\label{10092008-03} && \qquad C^a\equiv \alpha^a - \alpha^2
\frac{1}{2N_\alpha +d}\bar\alpha^a\,,
\eeq
\beq
\label{09092008-13} && \hspace{-1cm}\bar{C}_{sh} R^a \phishk =  \Ybf^a
\phishk\,,
\\[9pt]
\label{09092008-19} \Ybf^a & \equiv & Y_1 C^a \bar\alpha^2
+  Y_2 \Box \bar\alpha^a \bar\alpha^2
+ Y_3  \bar{C}_\perp^a
\nonumber \\[5pt]
& + & Y_4 \partial^a \bar\alpha^2
+  Y_5 M^{ab}\partial^b \bar\alpha^2
+  Y_6 C\bar\alpha^a \bar\alpha^2 \,,
\eeq
\beq
Y_1 &\equiv & \half c_{2\,sh} r_{0,1,sh} + c_{2\,sh} r_{0,2,sh}
\nonumber\\[5pt]
& - & \frac{2N_\alpha + d-4}{2(2N_\alpha +d-2)} r_{0,1,sh} c_{2\,sh}\,,
\\[5pt]
Y_2 &\equiv &  [c_{1\,sh},r_{0,1,sh}] + 2c_{1\,sh} r_{0,2,sh}
\nonumber\\[5pt]
& + & 2(2N_\alpha +d) c_{1\,sh} r_{0,3,sh}\,,
\\[5pt]
Y_3 & \equiv &  [c_{2\,sh}, r_{0,1,sh}]\,,
\\[5pt]
Y_4 & \equiv & \half r_{0,1,sh} + r_{0,2,sh}\,,
\\[5pt]
Y_5 & \equiv  &r_{0,2,sh}\,,
\\[5pt]
Y_6 & \equiv & -  r_{0,3,sh}(2N_\alpha + d-4)\,.
\eeq
Also, we note that to derive \rf{09092008-13} we use constraint
\rf{09092008-12}. Using \rf{09092008-12},\rf{09092008-14},\rf{09092008-13} it
is easy to see that equations \rf{09092008-03} lead to the equations,
\be \label{09092008-15}
(\Ybf^a + \bar{C}_{sh\smzero}^a + \bar{C}_{sh\smone}^a )\phishk = 0\,.
\ee
Taking into account \rf{09092008-16}-\rf{09092008-19} we see that equations
\rf{09092008-15} amount to the following equations:
\beq
\label{080423-30} && (Y_1 - \half (2N_\alpha + d-4)) C^a \bar\alpha^2 \phishk
= 0\,,
\\[5pt]
\label{080423-34} && Y_2 \bar\alpha^a \bar\alpha^2 \phishk = 0 \,,
\\[5pt]
\label{080423-31} &&  ( Y_3 + \Delta_{sh} - N_\alpha  - d +1 )
\bar{C}_\perp^a \phishk= 0\,,
\\[5pt]
\label{080423-33} && (Y_4 + (2\Delta_{sh}  -d)c_1 )  \bar\alpha^2 \phishk  =
0\,,
\\[5pt]
\label{080423-32} &&  (Y_5 + 2c_1 ) M^{ab}\bar\alpha^2 \phishk= 0\,,
\\[5pt]
\label{080423-34new} && Y_6 C\bar\alpha^a \bar\alpha^2 \phishk= 0 \,.
\eeq
Analysis of  Eqs.\rf{080423-30}-\rf{080423-34new} is straightforward. From
\rf{080423-32}, \rf{080423-34new}, we obtain
\beq
&& \label{080423-35} r_{0,2,sh} = -2c_{1\,sh}\,,
\\
&& \label{080423-35new} r_{0,3,sh} = 0\,. \eeq
From \rf{080423-33},\rf{080423-35}, we find
\be \label{080423-36} r_{0,1,sh} = 2(d+2 -2\Delta_{sh}) c_{1\,sh}\,.\ee
Using \rf{080423-35}-\rf{080423-36} we find that Eq.\rf{080423-34} is
satisfied automatically. Using \rf{080423-36} we represent Eq.\rf{080423-31}
as
\beq \label{080423-42}
&& \hspace{-1cm} \Bigl( 2(d-2\Delta_{sh}) c_{2\,sh} c_{1\,sh} -  2(d + 2 -
2\Delta_{sh}) c_{1\,sh} c_{2\,sh}
\nonumber\\[3pt]
&& + \Delta_{sh} + N_z  -s-d+2\Bigr)\bar{C}_\perp^a\phishk = 0 \,.
\eeq
Using solution for the $c$-operators given in
\rf{08092008-13},\rf{09092008-01}, we find that Eq.\rf{080423-42} is solved
by
\be \label{09092008-02} \Delta_{0\,sh} = 2-s \,.\ee
Finally, using \rf{080423-35},\rf{080423-36},\rf{09092008-02} and solution
for the $c$-operators given in \rf{08092008-13},\rf{09092008-01} we check
that Eq.\rf{080423-30} is satisfied automatically.

To summarize, taking into account solution for the $r$-operators given in
\rf{080423-35}-\rf{080423-36},\rf{09092008-02} and using \rf{09092008-04} we
cast the operator $R_{sh}^a$ into the following form:
\beq \label{10092008-06}
&& R_{sh}^a  =  r_{0,1,sh} \Bigl( \bar\alpha^a - \alpha^a
\frac{1}{2N_\alpha+d } \bar\alpha^2\Bigr)\,,
\\[5pt]
\label{10092008-06xxx} && \qquad r_{0,1,sh} = 2c_{1\,sh}(2s + d - 4
-2N_z)\,.
\eeq
Inserting $\cwt_{1\,sh}$ \rf{11092008-02xxx} in \rf{10092008-06xxx},
\rf{10092008-06} gives $R_{sh}^a$ \rf{20052008-16}.

In a similar way, we can find the operator $R_{cur}^a$. Requiring that the
operator $R_{cur}^a$ respects algebraic constraints \rf{10092008-05},
\rf{10092008-02} and the commutation relation $[D,K^a]=K^a$ gives
\beq \label{10092008-04}
&& \hspace{-1cm} R_{cur}^a = r_{0,1,cur}\Cwt^a + r_{0,2,cur} \alpha^2
\bar{C}_\perp^a + r_{0,3,cur}\alpha^2 C^a \bar\alpha^2\,,
\nonumber\\
&&
\\[3pt]
&& r_{0,k,cur} =  \rwt_{0,k,cur}\bar\alpha^z\,, \qquad  k=1,2,3,
\eeq
where the operators $\rwt_{0,k,cur}$ depend only on $N_z$. The operators
$\Cwt^a$, $\bar{C}_\perp^a$, $C^a$ are defined in
\rf{20052008-15(01)},\rf{20052008-15(02)},\rf{10092008-03} respectively.
Requiring the constraint $\bar{C}_{cur}\phicurk=0$ to be invariant under the
conformal boost symmetries leads to the following solution for the
$r$-operators:
\beq
\label{080309-24}&& r_{0,1,cur} = -(2s+d-4-2N_z)c_{2\,cur}\,,
\\[3pt]
\label{080309-25} && r_{0,2,cur} = -\frac{2}{2s+d-6-2N_z} c_{2\,cur}\,,
\\[3pt]
\label{080309-25ex} && r_{0,3,cur} = 0\,.
\eeq
Inserting these $r$-operators in \rf{10092008-04} and using  \rf{10092008-05}
we cast the operator $R_{cur}^a$ into the following form:
\be \label{10092008-07} R_{cur}^a  =  r_{0,1,cur} \Bigl( \Cwt^a + \alpha^2
\frac{2}{(2N_\alpha+d -2)(2N_\alpha+d)} \bar{C}_\perp^a\Bigr)\,.
\ee
With the choice of the $\cwt_{2\,cur}$-operator made in \rf{11092008-02}, the
operator $R_{cur}^a$ \rf{10092008-07} takes the form given in
\rf{20052008-15}.

Alternatively, the operator $R_{cur}^a$ can be evaluated by using $R_{sh}^a$
\rf{10092008-06} and requiring the vertex $\LL$,
\be \label{11092008-01}
\LL   =  \langle\phi_{cur}| \mubf  \phishk\,,
\qquad  \mubf \equiv 1 - \frac{1}{4}\alpha^2\bar\alpha^2\,,
\ee
to be invariant under the conformal boost transformations. To this end let us
use the notation $K_{cur}^a$ and $K_{sh}^a$ to indicate the respective
realizations of the operator $K^a$ on the current $\phicurk$ and the shadow
field $\phishk$. Requiring vertex $\LL$ \rf{11092008-01} to be invariant
under the conformal boost transformations gives the relation (up to total
derivative)
\be
\langle\phi_{cur}| \mubf  K_{sh}^a\phishk = - \langle K_{cur}^a \phi_{cur}
|\mubf \phishk\,.
\ee
Taking into account that the operators $K_{\Delta_{cur},M}^a$,
$K_{\Delta_{sh},M}^a$ satisfy the relation (up to total derivative)
\beq
\langle\phi_{cur}| \mubf  K_{\Delta_{sh},M}^a\phishk = - \langle
K_{\Delta_{cur},M}^a \phi_{cur} |\mubf \phishk\,,\qquad
\eeq
we conclude that the operators $R_{cur}^a$ and $R_{sh}^a$ should satisfy the
relation
\beq \label{10092008-08}
\langle\phi_{cur}| \mubf  R_{sh}^a\phishk = - \langle R_{cur}^a \phi_{cur}
|\mubf \phishk\,.
\eeq
Using \rf{10092008-06} and \rf{10092008-15} we make sure that relation
\rf{10092008-08} leads to $R_{cur}^a$ given in \rf{10092008-07}. This
provides additional check to our calculations.

\section{ Modified Lorentz and de Donder gauge conditions}
\label{appen03}

In this Appendix, we explain some details of the derivation of the modified
Lorentz and de Donder gauge conditions.

{\bf Spin-1 field}. We use field $\Phi^A$ carrying flat Lorentz algebra
$so(d,1)$ vector indices $A,B=0,1,\ldots, d-1,d$. The field $\Phi^A$ is
related with field carrying the base manifold indices $\Phi^\mu$,
$\mu=0,1,\ldots,d$, in a standard way $\Phi^A= e_\mu^A\Phi^\mu$, where
$e_\mu^A$ is vielbein of $AdS_{d+1}$ space. For the Poincar\'e
parametrization of $AdS_{d+1}$ space \rf{lineelem01}, vielbein $e^A=e^A_\mu
dx^\mu$ and Lorentz connection, $de^A+\omega^{AB}\wedge e^B=0$, are given by
\be\label{eomcho01} e_\mu^A=\frac{1}{z}\delta^A_\mu\,,\qquad
\omega^{AB}_\mu=\frac{1}{z}(\delta^A_z\delta^B_\mu
-\delta^B_z\delta^A_\mu)\,, \ee
where $\delta_\mu^A$ is Kronecker delta symbol. We use a covariant derivative
with the flat indices $\DD^A$,
\be \DD_A \equiv e_A^\mu \DD_\mu\,,\qquad \DD^A = \eta^{AB}\DD_B\,,\ee
where $e_A^\mu$ is inverse of $AdS$ vielbein, $e_\mu^A e_B^\mu = \delta_B^A$
and $\eta^{AB}$ is flat metric tensor. With choice made in \rf{eomcho01}, the
covariant derivative takes the form
\beq
&& \DD^A \Phi^B = \hat{\partial}^A \Phi^B + \omega^{ABC}\Phi^C\,,
\\
&& \hat{\partial}^A = z\partial^A \,,\qquad  \omega^{ABC} =
\eta^{AC}\delta_z^B - \eta^{AB}\delta_z^C\,,\eeq
where we adapt the following conventions for the derivatives and coordinates:
$\partial^A=\eta^{AB}\partial_B$, $\partial_A=\partial/\partial x^A $, $x^A
\equiv \delta_\mu^A x^\mu$, $x^A=x^a,x^d$, $x^d\equiv z$.

With these conventions, equations of motion of massless spin-1 field $AdS$
field $\Phi^A=\Phi^a,\Phi^z$,
\be \label{06092008-01} \DD^A F^{AB} = 0\,,\qquad
F^{AB}=\DD^A\Phi^B-\DD^B\Phi^A\,, \ee
can be represented as
\beq
&&  \label{06092008-02} (\hat{\partial}^2-d\hat{\partial}_z + d- 1)\Phi^A -
\hat{\partial}^A (\DD\Phi + 2 \Phi^z)
\nonumber\\[3pt]
&& + 2\delta_z^A \DD\Phi + (d+1) \delta_z^A \Phi^z=0\,, \eeq
where $\hat{\partial}^2 \equiv \hat{\partial}^A \hat{\partial}^A$,
$\DD\Phi\equiv \DD^A\Phi^A$. Our modified Lorentz gauge condition is defined
by the relation, \cite{Metsaev:1999ui},
\be  \label{06092008-03} \DD^A\Phi^A + 2\Phi^z =0\,,  \ee
which, in the Poincar\'e coordinates, can be represented as
\be  \label{06092008-04} \hat{\partial}^A \Phi^A + (2-d)\Phi^z = 0\,.\ee
Using \rf{06092008-03} in gauge invariant equations of motion
\rf{06092008-02} leads to the decoupled gauge-fixed equations of motion
\be  \label{06092008-05} (\hat{\partial}^2-d\hat{\partial}_z + d - 1 )\Phi^A
+(d-3)\delta_z^A \Phi^z=0\,, \ee
which can be represented as
\beq
\label{06092008-06} && (z^2(\Box+\partial_z^2) +(1-d)z\partial_z + d - 1
)\Phi^a=0\,,
\\[5pt]
\label{06092008-07} && (z^2(\Box+\partial_z^2) +(1-d)z\partial_z + 2d -
4)\Phi^z =0\,. \qquad
\eeq
Introducing the canonically normalized field $\phi^A$,
\be \label{06092008-08} \Phi^A  =  z^{\frac{d-1}{2}}\phi^A \,, \ee
and using the identification $\phi^z=\phi$ we make sure that equations
\rf{06092008-06} and \rf{06092008-07} amount to the respective equations
\rf{15052008-02} and \rf{15052008-03}, while modified Lorentz gauge condition
\rf{06092008-03} takes the form given in \rf{15052008-01}.

Equations of motion \rf{06092008-01} are invariant under the gauge
transformations
\be \label{06092008-09} \delta \Phi^A = \hat{\partial}^A \Xi \,.\ee
Making the rescaling
\be \label{06092008-10} \Xi = z^{\frac{d-3}{2}}\xi\,,\ee
we check that gauge transformations \rf{06092008-09} lead to the ones given
in \rf{15052008-05}, \rf{15052008-06}.

{\bf Spin-2 field}. Einstein equations of motion for the massless spin-2
field in $AdS_{d+1}$ can be represented as
\beq \label{12092008-03}
&& \DD^2 h^{AB} - \DD^A \DD^C h^{CB} - \DD^B \DD^C h^{CA} + \DD^A \DD^B h
\nonumber\\[3pt]
&& \hspace{1.5cm} + 2 h^{AB} - 2 \eta^{AB} h =0\,,
\\[3pt]
&& \hspace{2.3cm} h\equiv h^{AA}\,,
\eeq
where field with the flat indices, $h^{AB}$, is related with the field
carrying the base manifold indices in a standard way $h^{AB} = e_\mu^A
e_\nu^B h^{\mu\nu}$. Gauge transformations of $h^{AB}$ take the form
\be  \label{21052008-01}\delta h^{AB} = \DD^A \Xi^B + \DD^B \Xi^A\,.\ee

In terms of $h^{AB}=h^{ab},h^{za},h^{zz}$, our modified de Donder gauge
condition is defined to be
\be \label{12092008-01} \DD^B h^{AB} - \half \DD^A h + 2 h^{zA} - \eta^{zA} h
= 0 \,.\ee
In the Poincar\'e coordinates, this gauge condition can be represented as
\be \label{12092008-02} z\partial^B h^{AB} - \half z\partial^A h + (1-d)
h^{zA} = 0   \ee
Introducing the canonically normalized fields $\widetilde{\phi}^{AB}$,
\be h^{AB} = z^{\frac{d-1}{2}} \widetilde{\phi}^{AB} \,,\ee
and using \rf{12092008-02} we represent equations \rf{12092008-03} as
\beq
\label{13052008-01} && (\Box + \partial_z^2 -  \frac{d^2-1)}{4z^2})
\tilde{\phi}^{ab} - \frac{2}{z^2} \eta^{ab} \tilde{\phi}^{zz} = 0 \,,
\\[5pt]
\label{13052008-07} && (\Box + \partial_z^2 - \frac{(d-1)(d-3)}{4z^2})
\tilde{\phi}^{za} = 0 \,,
\\[5pt]
\label{13052008-08} && (\Box + \partial_z^2  - \frac{(d-3)(d-5)}{4z^2} )
\tilde{\phi}^{zz} = 0 \,.
\eeq
From these equations, we see that the modified de Donder gauge itself does
not lead automatically to decoupled equations. In order to get the decoupled
equations we introduce our fields $\phi^{ab}$, $\phi^a$, $\phi$ defined by
\beq \label{12092008-04}
&& \phi^{ab} = \widetilde{\phi}^{ab} + \frac{1}{d-2}\eta^{ab}
\widetilde{\phi}^{zz}\,,
\\[5pt]
\label{12092008-05} && \phi^a =  \widetilde{\phi}^{za}\,,
\\[5pt]
\label{12092008-06} && \phi = \frac{1}{2} \sigg \widetilde{\phi}^{zz}\,,
\eeq
where $\sigg$ is defined in \rf{080423-02(add)}. In terms of our fields
\rf{12092008-04}-\rf{12092008-06}, the gauge-fixed equations of motion
\rf{13052008-01}-\rf{13052008-08} take the decoupled form given in
\rf{15052008-31}-\rf{15052008-33}.

Gauge transformations we use in the Section \ref{sec10} are obtained from
\rf{21052008-01} by introducing
\be \Xi^A = z^{\frac{d-3}{2}} \xi^A\,,\ee
and making the identification for the $so(d-1,1)$ algebra scalar mode $\xi
\equiv \xi^z$.

{\bf Arbitrary spin field}. For massless arbitrary spin-$s$ field in
$AdS_{d+1}$, we define our modified de Donder gauge condition as follows.
Consider totally symmetric double-traceless $so(d,1)$ algebra tensor field
$\Phi^{A_1\ldots A_s}$, $\Phi^{AA BB A_5\ldots A_s} =0$. The modified de
Donder gauge condition, found in Ref.\cite{Metsaev:2008ks}, is defined as
\beq \label{12092008-09}
&& \DD^B \Phi^{A_1\ldots A_{s-1}B} - \frac{s-1}{2} \DD^{(A_1}
\Phi^{A_2A_3\ldots A_{s-1} ) BB}
\nonumber\\[5pt]
&& + 2\Phi^{A_1\ldots A_{s-1}z} - (s-1)\eta^{z (A_1} \Phi^{A_2\ldots A_{s-1})
BB} = 0\,, \qquad \ \ \
\eeq
where the symmetrization of the indices $A_1\ldots A_{s-1}$ is normalized as
$(A_1\ldots A_n) = \frac{1}{n!}(A_1\ldots A_n + (n!-1)$terms). Note however
that gauge condition \rf{12092008-09} itself does not lead automatically to
decoupled equations. One needs to make transformation similar to the one in
\rf{12092008-04}-\rf{12092008-06}. Discussion of the transformation and the
field variables which lead to decoupled equations of motion in Section
\ref{sec11} may be found in Ref.\cite{Metsaev:2008ks}.

\section{ Modified Lorentz and de Donder gauge conditions
in conformal flat space} \label{appen03new}

We now generalize of the modified Lorentz and de Donder gauge conditions to
the case of massless arbitrary spin fields propagating in conformal flat
space.

Line element of conformal flat space takes the form
\be \label{15102008-01}
ds^2 = \frac{1}{Z^2} dx^A dx^A\,,
\ee
where conformal factor $Z=Z(x)$ depends on coordinates $x^A$. For
parametrization of conformal space \rf{15102008-01}, vielbein $e^A=e^A_\mu
dx^\mu$ and Lorentz connection $\omega_\mu^{BC}$ are given by
\be
\label{15102008-02}
e_\mu^A = \frac{1}{Z}\delta_\mu^A\,,\qquad
\omega_\mu^{BC} = \frac{1}{Z}(\delta_\mu^C Z^B - \delta_\mu^B Z^C)\,,
\ee
\be
Z^A \equiv \partial^A Z\,.  \ee
We note that $AdS_{d+1}$ space is obtained by requiring the conformal factor
$Z$ to satisfy the equation
\beq
&& Z \partial^A \partial^B Z = \half \eta^{AB}( Z^C Z^C-1)\,,\qquad
d>1\,,\qquad
\\[3pt]
&& Z \partial^A \partial^A Z =  Z^A Z^A - 1\,,\hspace{2cm} d=1\,.
\eeq
With choice made in \rf{15102008-02}, the covariant derivative takes the form
\beq
&& \DD^A \Phi^B = \hat{\partial}^A \Phi^B + \omega^{ABC}\Phi^C\,,
\\
&& \hat{\partial}^A = Z\partial^A \,,\qquad  \omega^{ABC} = \eta^{AC} Z^B -
\eta^{AB} Z^C\,. \qquad \eeq

We note that various conformal flat geometries are specialized by appropriate
choice of the conformal factor $Z$. This is to say that the Poincar\'e
parametrization of $AdS_{d+1}$ space with coordinates $x^A=x^a$, $z$,
$a=0,1,\ldots,d-1$, is specialized by
\be \label{16102008-01} Z(x) = z\,. \ee
Also we note that stereographic parametrization of $AdS_{d+1}$ space with
coordinates $x^A$, $A=0,1,\ldots, d$, is specialized by
\be Z(x) = 1 - \frac{1}{4}x^Ax^A\,.  \ee
Famous $AdS_{d+1}\times S^{d+1}$ space is also conformal flat. This is to say
that the $AdS_{d+1}\times S^{d+1}$ space can be described by coordinates $x^A
= x^a,x^M$, $a=0,1,\ldots,d-1$, $M=d,\ldots, 2d+1$, with conformal factor
given by
\be Z(x) = \sqrt{ x^M x^M }\,.  \ee

Now let us describe modified gauge conditions for massless fields in
conformal flat space.

For massless spin-1 field, our modified Lorentz gauge condition takes the
form
\be  \label{15102008-03} \DD^A\Phi^A + 2Z^A\Phi^A =0\,,  \ee
while for massless spin-2 field the modified de Donder gauge is defined to be
\be \label{15102008-06} \DD^B h^{AB} - \half \DD^A h + 2 Z^B h^{AB} - Z^A h =
0 \,.\ee

For massless arbitrary spin-$s$ field propagating in conformal flat space,
the modified de Donder gauge condition takes the form
\beq \label{15102008-07}
&&\hspace{-0.7cm} \DD^B \Phi^{A_1\ldots A_{s-1}B} - \frac{s-1}{2} \DD^{(A_1}
\Phi^{A_2A_3\ldots A_{s-1} ) BB}
\\[5pt]
&&\hspace{-0.7cm} + 2 Z^B \Phi^{A_1\ldots A_{s-1}B} - (s-1) Z^{(A_1}
\Phi^{A_2\ldots A_{s-1}) BB} = 0\,.
\nonumber
\eeq

It is easy to see that by choosing $Z$ corresponding to Poincar\'e
parametrization \rf{16102008-01} gauge conditions \rf{15102008-03},
\rf{15102008-06}, \rf{15102008-07} reduce to the respective gauge conditions
given in \rf{06092008-03}, \rf{12092008-01}, \rf{12092008-09}.

\section{ Matching of conformal boost symmetries }
\label{appen04}

We now demonstrate matching of the improved $K^a$-transformations of the
non-normalizable bulk $AdS$ modes and the conformal boost transformations of
the boundary shadow fields. Matching of conformal boost symmetries of bulk
normalizable $AdS$ modes and boundary currents can be demonstrated in a quite
similar way.

Improved $K^a$-transformations of $AdS$ field take the form
\be \label{14092008-01} K_{impr}^a\phik  = K_{_{AdS}}^a\phik + G_{_{AdS}}
|\xi^{K^a}\rangle\,, \ee
where the compensating gauge transformation parameter $|\xi^{K^a}\rangle$
corresponding to the non-normalizable solution is given in \rf{20052008-33}.
The generic generator of $K^a$-symmetries, denoted by $K_{_{AdS}}^a$ in this
Appendix,  is given in \rf{conalggenlis04ads}, while the gauge transformation
operator $G_{_{AdS}}$ can be read from \rf{18052008-15},
\be
\label{14092008-02} G_{_{AdS}} \equiv \alpar - e_1  -
\frac{\alpha^2}{2s+d-6-2N_z}\eb_1\,.
\ee
Now we are going to demonstrate that the improved $K^a$-transformations of
the non-normalizable massless spin-$s$ $AdS_{d+1}$ modes become
$K^a$-transformations of the shadow field. Thus, we are going to prove the
following relation
\be \label{14092008-03} K_{impr}^a |\phi_{non-norm}\rangle = U_{-\nu}
K_{_{CFT}}^a \phishk \,, \ee
where $K_{_{CFT}}^a$ stands for representation of the conformal boost
generator on space of the shadow field given in \rf{conalggenlis04}.

To prove relation \rf{14092008-03} we represent the operator $K_{_{AdS}}^a$
as
\beq
\label{14092008-04}  && K_{_{AdS}}^a = K_{\Delta_{AdS}}^a + R_\smone^a +
M^{ab} x^b + R_\smzero^a \,,
\\[3pt]
&& \label{14092008-05} \quad\qquad K_{\Delta_{AdS}}^a \equiv -\half
x^2\partial^a + x^a D_{_{AdS}} \,, \eeq
where $D_{_{AdS}}$ takes the form given in \rf{conalggenlis03ads}, while
operators $R_\smzero^a$, $R_\smone^a$ are given in
\rf{14092008-07},\rf{14092008-08}. Then, we note the relations
\beq
\label{14092008-09} && \hspace{-0.7cm}(K_{\Delta_{AdS}}^a +
R_\smone^a)|\phi_{non-norm}\rangle = U_{-\nu} K_{\Delta_{sh}}^a \phishk\,,
\qquad
\\[7pt]
\label{14092008-10} && \hspace{-0.7cm}(M^{ab}x ^b + R_\smzero^a)|
\phi_{non-norm}\rangle + G_{_{AdS}}|\xi_{non-norm}^{K^a} \rangle
\nonumber\\[5pt]
&& \hspace{1.7cm} = U_{-\nu} ( M^{ab}x^b + R_{sh}^a )\phishk\,,
\eeq
where
\be K_{\Delta_{sh}}^a \equiv -\half x^2\partial^a + x^a D_{sh}\,,
\ee
and $D_{sh}$ takes the form given in \rf{conalggenlis03} with $\Delta$ in
\rf{08092008-04}, while $R_{sh}^a$ takes the form given in \rf{20052008-16}.
Using \rf{14092008-09},\rf{14092008-10}, we see that relation
\rf{14092008-03} holds.

We now make comment on the derivation of relations
\rf{14092008-09},\rf{14092008-10}. These relations are obtained by using the
following general formulas
\beq
\label{14092008-11} && (K_{\Delta_{AdS}}^a + R_\smone^a) U_{-\nu} = U_{-\nu}
(K_{\Delta_{sh}}^a + x^a z\partial_z )
\nonumber \\[3pt]
&& \hspace{2cm}  -  q^{\nu - \frac{3}{2}} \partial^a (\partial_q Z_{-\nu}
(qz))z\partial_z\,,
\\[9pt]
\label{14092008-12} && (M^{ab}x ^b + R_\smzero^a) U_{-\nu} +  G_{_{AdS}}
(zU_{-\nu +1} \bar{C}_\perp^a)
\nonumber\\[5pt]
&& \hspace{2.7cm} \approx U_{-\nu} ( M^{ab}x^b + R_{sh}^a )\,,
\eeq
where $q$ is defined in \rf{15052008-12} and we use the notation $Z_\nu(z)
\equiv \sqrt{z} J_\nu(z)$. In \rf{14092008-12} and in some relations given
below, the signs $\approx$ indicate that these relations are valid by
applying to the ket-vector $\phishk$ subject to differential constraint
\rf{17052008-05sh}. We now see that by applying relations \rf{14092008-11}
and \rf{14092008-12} to $\phishk$ we obtain the respective relations
\rf{14092008-09} and \rf{14092008-10}.

Finally, we note the helpful formulas for deriving relation \rf{14092008-12},
\beq
&& M^{ab}x ^b U_{-\nu} \approx  U_{-\nu}  M^{ab}x^b
\nonumber\\[5pt]
&& \qquad \qquad -  z U_{-\nu+1} ( G_{sh} \bar{C}_\perp^a + c_2 \Cwt^a + 2
c_1 \bar\alpha^a \Box )\,,
\\[7pt]
&& R_\smzero^a U_{-\nu} =  - zU_{-\nu+1}\eb_{1,1}\Cwt^a + z U_{ -\nu - 1}
e_{1,1}\bar\alpha^a\,,
\\[7pt]
&& G_{_{AdS}} (zU_{-\nu +1} \bar{C}_\perp^a) \approx zU_{-\nu+1} G_{sh}
\bar{C}_\perp^a -U_{-\nu} 2e_{1,1}\bar{C}_\perp^a \,. \qquad
\nonumber\\
&&
\eeq
These formulas can be obtained by using differential constraint
\rf{17052008-05sh} and relations for the operator $U_\nu$ given in
\rf{20052008-20}-\rf{20052008-27}. Also, to derive $R_{sh}^a$-term in
\rf{14092008-12} we use the formula $z U_{-\nu-1} + z\Box U_{-\nu+1} = -2 \nu
U_{-\nu}$.

\end{document}